\documentclass[11pt]{article}
\usepackage{cite}
\usepackage{amsmath,amscd,amsfonts,mathtools,amssymb}
\usepackage[small,bf,hang]{caption}
\usepackage{slashed}
\usepackage{mathabx}
\usepackage{latexsym,epsfig}
\usepackage{stmaryrd}
\usepackage{float}
\usepackage{accents}

\usepackage[vcentermath]{youngtab}

\usepackage[usenames,dvipsnames]{color}
\usepackage{amssymb}
\usepackage{breqn}
\usepackage{tikz}
\usetikzlibrary{shapes, shadows, arrows, fit, positioning, calc}

\def\hybrid{
        \topmargin -20pt
        \oddsidemargin 0pt
        \headheight 0pt \headsep 0pt
        \textwidth 6.25in 
        \textheight 9.5in 
        \marginparwidth .875in
        \parskip 5pt plus 1pt \jot = 1.5ex}

\hybrid

 \csname
@addtoreset\endcsname{equation}{section}

\newcommand{\ut}[1]{\underaccent{\tilde}{#1}}

\def\moth{\mathsurround=0pt}
\newdimen\zo \zo=0pt

\def\tick{\leaders\hrule height 0.5ex depth 0pt \hskip 0.5pt}
\def\upboxfill{$\moth \setbox\zo\hbox{\tick}%
  \hskip 3pt\hbox to 0pt{$\tick$\hss}\hrulefill \hbox to 7.5pt{$\tick$\hss}$}

\def\dtick{\leaders\hrule height .34pt depth 0.5ex \hskip 0.5pt}
\def\downboxfill{$\moth \setbox\zo\hbox{\dtick}%
  \hskip 2pt\hbox to 0pt{$\dtick$\hss}\hrulefill \hbox to 2pt{$\dtick$\hss}$}

\def\bec{\begin{center}}
\def\ec{\end{center}}

\def\nn{\nonumber}

\def\be{\begin{equation}}
\def\ee{\end{equation}}
\newcommand{\beq}{\begin{equation}\begin{aligned}}
\newcommand{\eeq}{\end{aligned}\end{equation}}
\def\bea{\begin{eqnarray}}
\def\eea{\end{eqnarray}}
\def\ba{\begin{array}}
\def\ea{\end{array}}

\allowdisplaybreaks[2]

\begin{document}

\begin{titlepage}

\begin{center}
\vskip 1.6cm
{\Large \bf {The generalized Bergshoeff-de Roo identification II}}\\
 \vskip 2.0cm
{\large {W. Baron$^{1}$ and D. Marques$^2$}}
\vskip 0.5cm

$^1$ {\it Instituto de F\'isica La Plata (CONICET-UNLP)}\\ {\it Calle 49 y 115, La Plata, Argentina} \\[1ex]

$^2$ {\it Instituto de Astronom\'ia y F\'isica del Espacio (CONICET-UBA)}\\ {\it Ciudad Universitaria, Buenos Aires, Argentina} \\[1ex]

\vskip 0.5cm

{\small \verb"wbaron@fisica.unlp.edu.ar , diegomarques@iafe.uba.ar"}

\vskip 1cm
{\bf Abstract}	
\end{center}

\vskip 0.2cm

\noindent
\begin{narrower}

{\small We recently introduced a T-duality covariant mechanism to compute all-order higher- derivative interactions in the heterotic string. Here we extend the formalism to account for a two-parameter family of corrections that also include the bosonic string and HSZ theory. We use our result to compute the full second order Double Field Theory (DFT) for generic values of the parameters, including the generalized Green-Schwarz transformation and its invariant action.}

\end{narrower}

\vskip 1.5cm

\end{titlepage}

\tableofcontents

\vspace{0.5cm}

\section{Introduction}

Kaluza-Klein reductions of supergravity and its higher derivatives give rise to lower dimensional field theories with continuous global symmetries. When certain interactions in the higher dimensional theory are unknown, they could be constrained by demanding the emergence of such global symmetries after compactification. Alternatively, one may try to formulate the parent theory in the framework of Double (or Exceptional) Field Theory \cite{Siegel, DFT} (for reviews see \cite{Reviews}), in which the duality symmetries are manifest prior to compactifying.

The last years have witnessed progress in constraining higher-derivative interactions through dualities. There are methods based on explicit reductions, such as cosmological \cite{Meissner, Cosmological}, circle \cite{Ortin, Kaloper, Circle} and intermediate \cite{Intermediate} compactifications. There is also a duality covariant sigma-model approach to higher derivatives \cite{Bonezzi}. Here we will focus on higher derivatives in DFT, for which originally there were two alternatives. 

In one approach the corrections were accounted for through enlarging the duality group structure by adding higher-derivative interactions in the extra directions of the generalized tangent space \cite{Approach1a, Approach1b} (see also \cite{Lee}). The local symmetries and the action remain unchanged, but the duality structure is deformed. This method was only worked out for the heterotic string to first order in $\alpha'$, and has the disadvantage that the deformations are not manifestly duality covariant, so duality covariance has to be checked explicitly.

There is a second approach in which the duality structure remains unmodified (namely the duality group is still the continuous $O(D,D)$), and higher-derivatives enter through deformations of the local symmetries. In some cases it is generalized diffeomorphisms that are deformed \cite{HSZ, HSZcont, DFTalphaprime}, and in others the double Lorentz symmetries \cite{MarquesNunez, OddStory}. The distinction  between both cases is discussed in \cite{BckgndInd}. Two parameters $a$ and $b$ control the deformations, and depending on how they are chosen the framework accounts for the first order corrections to the heterotic string (one of the parameters vanishes), the bosonic string ($a = b$) or other duality symmetric theories such as a Lorentz deformed version of HSZ theory ($a = -b$).
 
More recently, a general framework was proposed \cite{GenBdR} in which the two approaches described above were shown to be equivalent in the heterotic case. The idea is to start with an extended duality group $O(D + p,D + q)$ as in the heterotic formulation of DFT \cite{HeteroticDFT}, and then perform an $O(D,D)$ decomposition along the lines of \cite{HSZheterotic}. We will discuss this extensively later, but for the moment let us state that $p$ and $q$ count the number of negative and positive eigenvalues of the Killing metric of the gauge group, respectively.
\begin{table}[H]
\begin{center}
\begin{tabular}{|l|l|l|} \hline
& {\bf Extended}& {\bf Double} \\ \hline
Duality group & $O(D + p,D + q)$  & $O(D,D) $ \\ \hline
Lorentz group & $\underline{O(D-1,1)} \times  \overline{O(1 + p, D + q - 1)} $  & $\underline {O(D-1,1)} \times  \overline {O(1, D - 1)}$ \\ \hline
Fields &  Generalized frame ${\cal E}_{\cal M}{}^{\cal A}$ ,  &  Generalized frame  $E_M{}^A$ ,  \\
& Dilaton $d$& Vectors ${\cal E}^{\tilde \mu}{}_{\underline a}$ , Dilaton $d$\\ \hline
Other symm. & Extended gen. diffeos. & Double gen. diffeos. $\times$ $\cal K$ \\ \hline
\end{tabular}
    \end{center}
\end{table}
The result is a DFT coupled to $k = p + q$ extra vectors ${\cal E}^{\tilde \mu}{}_{\underline a}$ that transform under a certain gauge group $\cal K$ as generalized connections. One then has a generalized connection in the {\it double} picture with respect to the gauge group $\cal K$ (which in turn descends from the generalized diffeomorphism in the {\it extended} picture). On the other hand there is a generalized spin connection ${\cal F}_{\underline a \overline {\cal B C}}$ in the {\it extended} picture with respect to the Lorentz factor $\overline {O(1 + p, D+ q - 1)}$. The idea in \cite{GenBdR} was then to {\it identify} these two independent symmetries
\begin{equation}
    {\cal K} \ \ \ \leftrightarrow \ \ \ \overline { O(1 + p,D+q-1)} \ ,
\end{equation}
and match the independent degrees of freedom ${\cal E}^{\tilde \mu}{}_{\underline a}$ with the composite degrees of freedom  ${\cal F}_{\underline {a} \overline { \cal B C}}$ through the generators of the resulting group $(t_{\tilde \mu})_{\overline {\cal B C}}$
\begin{equation}
     -g  \, {\cal E}^{\tilde\mu}{}_{ \underline a} \, (t_{\tilde\mu})_{\overline {\cal B C}} \ =\ {\cal F}_{\underline a \overline {\cal B C}} \ .
\end{equation}
After this identification, when the formalism is seen from an $O(D + p,D+q)$ perspective the first approach described above is recovered, and when scrutinized after its $O(D,D)$ decomposition it reproduces the second approach, thus proving their equivalence. This procedure is the duality covariant version of that in \cite{BdR} and was then referred to as the {\it Generalized Bergshoeff-de Roo identification} in \cite{GenBdR}. The advantage of this generalized identification is that it is exact, and generates an infinite tower of higher derivatives in the heterotic string. The reason for this is that the identification requires the symmetry group to be infinite dimensional, as will be reviewed soon.

As mentioned, the approach in which the double Lorentz symmetry is deformed admits a two-parameter $(a,b)$ extension, the heterotic string being a particular choice in parameter space. {\bf The first original result} in this paper is an extension of the generalized Bergshoeff-de Roo identification that captures this bi-parametric freedom. Let us briefly anticipate the result by showing how the discussion above is modified. Here we further extend the extended duality group in a more symmetric fashion to $O(D+k,D+k)$ with $k = p  + q$, and again realize an $O(D,D)$ decomposition.
\begin{table}[H]
\begin{center}
\begin{tabular}{|l|l|l|} \hline
& {\bf Extended}& {\bf Double} \\ \hline
Duality group & $O(D+k,D + k)$  & $O(D,D)$ \\ \hline
Lorentz group & $\begin{matrix}\underline{O(D+ q -1,1 + p)} \\ \times  \overline{O(1+ p, D + q - 1)} \end{matrix} $  & $\underline {O(D-1,1)} \times  \overline {O(1, D - 1)} $ \\ \hline
Fields & Generalized frame ${\cal E}_{\cal M}{}^{\cal A}$ , & Generalized frame  $E_M{}^A$ ,  Dilaton $d$
\\
&   Dilaton $d$ & Vectors ${\cal E}^{\tilde \mu}{}_{\underline a}$ , Vectors ${\cal E}^{\undertilde \mu}{}_{\overline a}$ , Scalars $\Omega_{\tilde \mu \ut \nu}$\\ \hline
Other symmetries & Extended gen. diffeos. & Double gen. diffeos. $\times$ $\cal K$ \\ \hline
\end{tabular}
    \end{center}
\end{table}
The result is a DFT coupled to $2k$ extra vectors and $k^2$ scalars, that jointly populate the following components of the extended generalized frame ${\cal E}^{\tilde \mu}{}_{\underline {\cal A}}$  and ${\cal E}^{\undertilde \mu}{}_{\overline {\cal A}}$, which transform under the gauge group $\cal K$ as generalized connections.  On the other hand there are two generalized spin connections ${\cal F}_{\overline {\cal A} \underline {\cal B C}}$ and ${\cal F}_{\underline {\cal A} \overline {\cal B C}}$ in the {\it extended} picture with respect to the Lorentz factors $\underline {O(D+ q - 1, 1 + p)}$ and $\overline {O(1 + p, D + q - 1)}$, respectively. The idea here is  to {\it identify} the symmetries
\begin{equation}
    {\cal K} \ \ \ \leftrightarrow \ \ \ \underline {O(D+q - 1, 1 + p)} \times \overline { O(1 + p,D+q-1)}\ ,
\end{equation}
by matching the independent degrees of freedom ${\cal E}^{\undertilde \mu}{}_{\overline {\cal A}}$ and ${\cal E}^{\tilde \mu}{}_{\underline {\cal A}}$ with the composite degrees of freedom  ${\cal F}_{\overline {\cal A} \underline {\cal B C}}$ and ${\cal F}_{\underline {\cal A} \overline {\cal B C}}$ through the generators of each factor of the resulting group $(t_{\tilde \mu})_{\overline {\cal B C}}$ and $(t_{\undertilde \mu})_{\underline {\cal B C}}$ 
\begin{eqnarray}\label{ExactIdentification}
\begin{aligned}
     -\; g_1 \; {\cal E}^{ {\ut\mu}}{}_{\overline{\cal A}} \; (t_{ {\ut\mu}})_{\underline{\cal B C}} &= {\cal F}_{\overline {\cal A} \underline {\cal B C}}\, , \\ 
     -\; g_2 \; {\cal E}^{ {\tilde \mu}}{}_{\underline{\cal A}} \; (t_{ {\tilde \mu}})_{\overline{\cal B C}} &= {\cal F}_{\underline {\cal A} \overline {\cal B C}}  \ .
\end{aligned}
\end{eqnarray}

The couplings $g_1$ and $g_2$ are related to the parameters $a$ and $b$. We will explain in detail how to extract perturbative results in powers of $a$ and $b$ from (\ref{ExactIdentification}). It is amazing that these can be obtained systematically from the standard two-derivative action, equations of motion, gauge transformations, etc. in the extended setup. Schematically, the resulting perturbative action in the double space is the sum of terms of the form ${\cal R}^{(m,n)}$, where the supra-label indicates that each term scales like $a^n b^m$. 

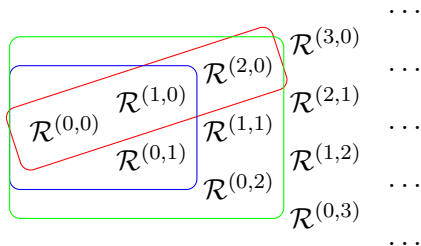
\tikzstyle{block} = [text width=2.1em, minimum height = 1em , text centered, node distance = 3em]
\begin{center}
\begin{tikzpicture}\label{PerturbativeRiccis} 
\node [block] (R00) {${\cal R}^{(0,0)}$};
\node [block, right of = R00, yshift = 1 em] (R10) {${\cal R}^{(1,0)}$};
\node [block, right of = R00, yshift = - 1 em] (R01) {${\cal R}^{(0,1)}$};
\node [block, right of = R10, yshift = 1 em] (R20) {${\cal R}^{(2,0)}$};
\node [block, right of = R10, yshift = -1 em] (R11) {${\cal R}^{(1,1)}$};
\node [block, right of = R01, yshift = - 1 em] (R02) {${\cal R}^{(0,2)}$};
\node [block, right of = R20, yshift = 1 em] (R30) {${\cal R}^{(3,0)}$};
\node [block, right of = R20, yshift = -1 em] (R21) {${\cal R}^{(2,1)}$};
\node [block, right of = R02, yshift = 1 em] (R12) {${\cal R}^{(1,2)}$};
\node [block, right of = R02, yshift = - 1 em] (R03) {${\cal R}^{(0,3)}$};
\node [block, right of = R30, yshift =  1 em] (R40) {$\dots$};
\node [block, right of = R03, yshift =  -1 em] (R04) {$\dots$};
\node [block, right of = R21, yshift =  1 em] (R31) {$\dots$};
\node [block, right of = R12, yshift =  1 em] (R22) {$\dots$};
\node [block, right of = R12, yshift =  -1 em] (R13) {$\dots$};
\node [draw=red, rounded corners, rotate fit=18, fit=(R00) (R10) (R20) ] {};
\node [draw=blue, rounded corners, fit=(R00) (R10) (R01) ] {};
\node [draw=green, rounded corners, fit=(R00) (R10) (R01) (R20) (R11) (R02) ] {};
\end{tikzpicture}
\end{center}

The term ${\cal R}^{(0,0)}$ is the standard two-derivative generalized Ricci scalar of DFT \cite{Siegel,DFT}. It is invariant under generalized diffeomorphisms, and double Lorentz transformations to lowest order. However, the double Lorentz symmetry receives higher derivative corrections. To first order they take the form of a generalized Green-Schwarz transformation, under which ${\cal R}^{(0,0)}$ is not invariant, and then first order corrections $a {\cal R}^{(0,1)} + b {\cal R}^{(1,0)}$ are induced in the action. This is pictured in the blue box, the results were introduced in \cite{MarquesNunez} and cast in a Gauged DFT form in \cite{OddStory}. It turns out that the algebra of the bi-parametric generalized Green-Schwarz transformation only closes to first order, and then higher corrections are required. For the heterotic case the second order corrections ${\cal R}^{(2,0)}$, contained here in the red box, are completely determined by the symmetry transformations introduced in \cite{GenBdR}. {\bf The second original result} in this paper is the computation of the full symmetries and action of DFT to second order in the bi-parametric case, which corresponds to the green box. This includes the second order corrections to the generalized Green-Schwarz transformation.

{\bf The third original result} is to show that to second order this extension accounts for the bi-parametric Green-Schwarz transformation of the Kalb-Ramond field (a two parameter generalization of the original deformation \cite{GreenSchwarz})
\be
\delta b_{\mu \nu} = \frac a 2 \partial_{[\mu} \Lambda^{a b} \widehat \omega^{(-)}_{\nu]ab} - \frac b 2 \partial_{[\mu} \Lambda^{a b} \widehat \omega^{(+)}_{\nu]ab}  \ , \nn
\ee
when the field redefinitions required to connect with the supergravity fields in the Bergshoeff-de Roo scheme are implemented. The spin connections include the three-form field strength as torsion, which in turn is sourced by Lorentz Chern-Simons three-forms. As expected, to second order (and presumably to all orders) the Lorentz transformation on the frame field and dilaton in supergravity remain uncorrected.  This confirms the expectations that the deformations due to the parameters $a$ and $b$ induce the full tower of corrections contained in the Chern-Simons terms that source the three-form curvature, plus all the corrections connected to these by T-duality.

The paper is organized as follows. Section \ref{SecMonoparam} is devoted to set the notation by reviewing the mono-parametric identification for the  heterotic case introduced in \cite{GenBdR}. In Section \ref{SecBiparam} we present the bi-parametric generalized Bergshoeff-de Roo identification, work it perturbatively to second order,  extract from it the second order corrections to the generalized Green-Schwarz transformation, analyze its closure and present the invariant action. Finally in Section \ref{SecSugra} we perform the minimal field redefinitions that  trivialize the Lorentz transformation of the vielbein and dilaton to second order, and show that the resulting transformation for the Kalb-Ramond field is the expected bi-parametric Lorentz Green-Schwarz transformation in the Bergshoeff-de Roo scheme.

\section{The heterotic generalized BdR identification} \label{SecMonoparam}
\subsection{The extended space}

Our starting point is the gauged extension of DFT \cite{HeteroticDFT, GaugedDFT} in the frame formulation \cite{Siegel, FrameDFT, Jeon, Coimbra}. We begin with a brief review of some basics that will be useful in the forthcoming sections, and will serve in addition to set the notation and conventions followed here. 

The idea is to start with an extended tangent space, acted on by the rigid action of some split orthogonal group $\cal G$, that includes $G = O(D,D)$ as a subgroup.  The dof are a generalized dilaton $d$ and a generalized frame ${\cal E}_{\cal M}{}^{\cal A}$, constrained by demanding that the $\cal G$-invariant metric $\eta$ is preserved by the generalized frame 
\begin{eqnarray}
\eta_{\cal M N}={\cal E}_{\cal M}{}^{\cal A} \; \eta_{\cal A B}\;  {\cal E}_{\cal N}{}^{\cal B}\ .
\end{eqnarray}

The local symmetries include generalized diffeomorphisms and gauge symmetries in a duality covariant way generated by a $\cal G$-vector $\xi$, in addition to the extended local Lorentz transformations with respect to a group $\cal H$, parameterized by $\Gamma$ in the adjoint of $\cal H$. Infinitesimally they take the form
\begin{eqnarray}
\delta d &=& \xi^{\cal N} \partial_{N} d - \frac 1 2 \partial_{\cal N} \xi^{\cal N} \ , \nonumber\\ 
\delta {\cal E}_{\cal M}{}^{\cal A}&=&\xi^{\cal N}\partial_{\cal N}{\cal E}_{\cal M}{}^{\cal A} 
+  \left( \partial_{\cal M}\xi^{\cal N} - \partial^{\cal N}\xi_{\cal M} \right) {\cal E}_{\cal N}{}^{\cal A} 
+{\hat f}_{\cal M N}{}^{\cal P} \xi^{\cal N} {\cal E}_{\cal P}{}^{\cal A} 
+ {\cal E}_{\cal M}{}^{\cal B} \Gamma_{\cal B}{}^{\cal A} \;. \label{TransformationExtended}
\end{eqnarray}
The consistency of these transformations requires the imposition of linear and quadratic constraints on the gaugings $ {\hat f}_{\cal M N}{}^{\cal P}$,
 \begin{eqnarray} \label{quadraticconstraints}
 \hat{f}_{\cal M N P} = \hat{f}_{[{\cal M N P}]} \ , \ \ \ \ \hat{f}_{[{\cal M N}}{}^{\cal K} \hat{f}_{{\cal P}]{\cal K}}{}^{\cal L} = 0\ . \label{LinearQuadratic}
 \end{eqnarray}
Together with the the strong constraint
\begin{eqnarray} \label{strongconstraint}
\eta^{\cal M N} \partial_{\cal M}\otimes \partial_{\cal N} =0\;,\;\;\;\;\;
\hat{f}_{\cal M N}{}^{\cal P} \partial_{\cal P}=0 \;,
\end{eqnarray}
they  guaranty the closure of the algebra
\begin{eqnarray}
[\delta_{(\Gamma_1,\xi_1)},\delta_{(\Gamma_2,\xi_2)}]=-\delta_{(\Gamma_{12},\xi_{12})}\;,\label{closure}
\end{eqnarray} 
defining the following brackets
\begin{eqnarray}
\xi_{12}^{\cal M}&=&2\xi^{\cal N}_{[1}\partial_{\cal N}\xi^{\cal M}_{2]} 
+ \partial^{\cal M}\xi^{\cal N}_{[1}\xi_{2]{\cal N}} 
+ \hat{f}_{\cal N P}{}^{\cal M}\xi^{\cal N}_{1} \xi^{\cal P}_{2}\;,\\
\Gamma_{12{\cal A B}}&=&2 \xi^{\cal N}_{[1}\partial_{\cal N}\Gamma_{2]{\cal A B}} + \Gamma_{1 {\cal A}}{}^{\cal C} \Gamma_{2] {\cal B C}}\;.\label{CloParam}
\end{eqnarray}

In the frame or flux formulation \cite{Siegel, FrameDFT}, the main characters are the generalized fluxes
\begin{eqnarray}
{\cal F}_{\cal A}&=& 2 \;{\cal D}_{\cal A}d -\Omega_{\cal B A}{}^{\cal B}\, , \\
{\cal F}_{\cal A B C}&=& 
3\;\Omega_{[{\cal A B C}]} 
+ {\hat f}_{\cal M N P} {\cal E}^{\cal M}{}_{\cal A} {\cal E}^{\cal N}{}_{\cal B} {\cal E}^{\cal P}{}_{\cal C} \ , \label{ExtendedFluxes}
\end{eqnarray}
defined in terms of $\Omega_{\cal A B C}$ which is named the generalized Weitzenb\"ock connection
\begin{eqnarray}
\Omega_{\cal A B C}= {\cal D}_{\cal A}{\cal E}^{\cal N}{}_{\cal B} {\cal E}^{\cal P}{}_{\cal C} \; \eta_{\cal N P} \ ,
\end{eqnarray}
and we have introduced the flat derivative ${\cal D}_{\cal A}\,= {\cal E}^{\cal M}{}_{\cal A} \partial_{\cal M}$ .
The generalized fluxes behave as scalars under generalized diffeomorphisms but transform non covariantly under the extended Lorentz transformations
\begin{eqnarray}
\delta {\cal F}_{\cal A}&=&\xi^{\cal M}\partial_{\cal M}{\cal F}_{\cal A} + \Gamma^{\cal B}{}_{\cal A}\; {\cal F}_{\cal B} 
- {\cal D}_{\cal B}\Gamma^{\cal B}{}_{\cal A}\;,\\
\delta {\cal F}_{\cal A B C}&=&\xi^{\cal M}\partial_{\cal M}{\cal F}_{\cal A B C} \;
+ 3\;\left(\vphantom{\frac12}\Gamma^{D}{}_{[\cal A}\; {\cal F}_{\cal B C] D}- {\cal D}_{[\cal A} \Gamma_{\cal BC]}\right)\,. \label{TransfFluxes}
\end{eqnarray}
The strong constraint (\ref{strongconstraint}) together with the quadratic constraints (\ref{quadraticconstraints}) imply the following generalized Bianchi identities
\begin{eqnarray}
{\cal D}_{[\cal A} {\cal D}_{\cal B]} &=& \frac12 {\cal F}_{\cal A B}{}^{\cal C} {\cal D}_{\cal C}\;,\label{BI1} \\
{\cal D}_{[\cal A} {\cal F}_{\cal B C D]}&=&\frac34 {\cal F}_{[\cal AB}{}^{\cal E} {\cal F}_{\cal C D] E}\;,\label{BI2}\\
{\cal D}^{\cal C} {\cal F}_{\cal C AB} &=& {\cal F}^{\cal C} {\cal F}_{\cal C A B}- 2 {\cal D}_{[\cal A} {\cal F}_{\cal B]}\label{BI3}\;.
\end{eqnarray}
It is also useful to rewrite some other conditions that follow from the strong constraint in terms of flat derivatives
\begin{eqnarray}
{\cal D}_{\cal A} f \;{\cal D}^{\cal A} g &=& 0\;,\label{SC1}\\
{\cal D}^{\cal A} {\cal D}_{\cal A} f - {\cal F}^{\cal A} {\cal D}_{\cal A} f &=& 0\;,\label{SC2}\\
{\cal F}^{\cal A B C} {\cal D}_{\cal A} f \;{\cal D}_{\cal B}\, g\; {\cal D}_{\cal C} h &=&0\;\label{SC3},
\end{eqnarray}
for any function $f,g,h$.

\subsection{The double space and the identification}

We have just considered a generic scenario in which  the double space is extended in order to introduce gaugings in a duality covariant way. We will now discuss a concrete realization of this extension. We begin with the following extended duality group $\cal G$ and the extended double Lorentz symmetry group $\cal H$
\begin{eqnarray}
{\cal G}= O(D + p , D+ q)\;\;,\;\;\;\;\;\;\;\;\;\;\;\;{\cal H}=\underline{O(D-1,1)}\times \overline{O(1 + p,D+ q-1)}\;\;.
\end{eqnarray}
The extension is characterized by the quantity
\begin{equation}
    k = p + q \ ,
\end{equation}
which is the dimension of the gauge group produced by the gaugings. In Table \ref{NotationMono} we clarify the notation adopted for the groups and indices in this section. The same notation extends to other sections, though some of the groups will get enhanced later. 

The idea is to perform a $G$ and $H$ decomposition of $\cal G$ and $\cal H$, respectively. Every $\cal G$-vector, such as derivatives or parameters, splits in $G$-vectors and internal components, $\partial_{\cal M} = (\partial_{M} , \partial_{\tilde\mu})$ and $\xi^{\cal M} = (\xi^{M},\xi^{\tilde\mu})$. Only the internal components  ${\hat f}_{\tilde\mu \tilde \nu \tilde \rho}$ of the gaugings are non vanishing, and then from the double space point of view the parameters $\xi^{M}$ generate double generalized diffeomorphisms, and the parameters $\xi^{\tilde\mu}$ generate gauge transformations with respect to some group ${\cal K}$ of dimension $k$ with structure constants ${\hat f}_{\tilde\mu \tilde \nu \tilde \rho}$.  This requires that no fields or parameters in the theory depend on the internal coordinates, while the dependence on the double space is strong constrained as usual
\begin{eqnarray}
\partial_{\tilde\mu}=0 \;,\;\;\;\;\;\;\;\;\eta^{M N}\partial_{M}\otimes \partial_{N} =0\;.
\end{eqnarray}

\begin{table}[H]
\begin{center}
\begin{tabular}{|l|l|l|l|} \hline
{\bf Name}& {\bf Group}  & {\bf Indices}  &  {\bf Metric} \\ \hline
$G$ $\begin{matrix} \\ ~\end{matrix}$  & $O(D,D)$ & $M$ & $\eta_{M N}$ \\ \hline
$g$ $\begin{matrix} \\ ~\end{matrix}$  & $O(p , q)$ & $\tilde\mu$ & $\kappa_{\tilde\mu \tilde\nu}$ \\ \hline
$\cal G$ & $O(D + p , D+ q)$ & ${\cal M} = (M,\,\tilde\mu)$ &
$\eta_{\cal M N} = \left(\begin{matrix}\eta_{M N} & 0 \\ 0 & \kappa_{\tilde\mu \tilde\nu}\end{matrix}\right)$ \\ \hline
$\underline {\cal H}$ = $\underline H$ $\begin{matrix} \\ ~\end{matrix}$ & $\underline{O(D-1,1)}$ & $\underline{\cal A} = \underline a$ & $P_{\underline {\cal  A} \underline {\cal B}} = P_{\underline a \underline b}$ \\ \hline
$\overline H$ $\begin{matrix} \\ ~\end{matrix}$  & $\overline{O(1,D-1)}$ & $\overline a$ & $\bar P_{\overline {a b}}$ \\ \hline
$\overline h$ $\begin{matrix} \\ ~\end{matrix}$ & $\overline{O(p , q)}$ & $\overline \alpha$ & $\kappa_{\overline {\alpha \beta}}$ \\ \hline
$\overline {\cal H}$ & $\overline{O(1 + p,D+ q-1)}$ & $\overline{\cal A} = (\overline a,\, \overline \alpha)$ & $\bar P_{\overline {\cal A} \overline {\cal B}} = \left(\begin{matrix} \bar P_{\overline {a b}} & 0 \\ 0 & \kappa_{\overline {\alpha \beta}}\end{matrix}\right)$ \\ \hline
$\cal H$ & $\underline{{\cal H}} \times \overline{{\cal H}}$ & ${\cal A} = (\underline{{\cal A}},\,\overline{{\cal A}}) = (\underline a,\, \overline a ,\, \overline \alpha)$ & $\eta_{\cal A B} = \left(\begin{matrix}P_{\underline {\cal A B}} & 0 \\ 0 & \bar P_{\overline {\cal A B}} \end{matrix} \right)$  \\ \hline
$H$ & $\underline{H}\times \overline{H}$ & $A = (\underline a,\, \overline a)$ & $\eta_{A B} = \left(\begin{matrix}P_{\underline a \underline b}& 0 \\ 0 & \bar P_{\overline {ab}}\end{matrix}\right)$\\ \hline
\end{tabular}
    \end{center}
    \caption{Groups, metrics and index structure for the extended space relevant to heterotic DFT. For those familiar with \cite{GenBdR}, let us note two differences with the table shown there. First note a small change in the notation: we find it more convenient to use $\tilde\mu$  instead of $\alpha$ for the curved internal index, as it has a more natural extension to the bi-parametric case. Second, we are now writing explicitly the split signature $(p,q)$ of the internal metric $\kappa_{\tilde \mu \tilde \nu}$, which by abuse of notation was omitted in \cite{GenBdR}.}\label{NotationMono}
\end{table}

The extended $\cal G$-valued generalized frame ${\cal E}_{\cal M}{}^{\cal A}$ admits a $G$ and $H$ covariant parameterization in terms of the double generalized frame $E_{M}{}^{A}$, $k$ vectors $A_{M}{}^{\tilde\nu}$ and $k(k-1)/2$ scalars $e_{\tilde\mu}{}^{\overline\alpha}$
\begin{eqnarray}
{\cal E}_M{}^A &=& (\chi^{\frac 1 2 }){}_M{}^N\, E_N{}^A\,, \nonumber\\
{\cal E}_M{}^{\overline \alpha} &=& - {\cal A}_M{}^{ {\tilde\mu}} \, e_{ {\tilde
\mu}}{}^{\overline \alpha}\,, \label{FrameParameterizationMono}\\
{\cal E}_{ {\tilde\mu}}{}^A &=& {\cal A}^M{}_{ {\tilde\mu}}\, E_M{}^A\,, \nonumber\\
{\cal E}_{ {\tilde\mu}}{}^{\overline\alpha} &=& (\Box^{\frac 1 2}){}_{ {\tilde\mu}}{}^{ {\tilde\nu}}\, e_{ {\tilde\nu}}{}^{\overline\alpha}\, ,\nonumber
\end{eqnarray}
where 
\begin{eqnarray}
\chi_{M N} = \eta_{M N} - {\cal A}_M{}^{ {\tilde\mu}}\, {\cal A}_{N {\tilde\mu}} \ , \ \ \ \ \ \Box_{ {\tilde\mu}  {\tilde\nu}} = \kappa_{ {\tilde\mu}  {\tilde\nu}} - {\cal A}_{M  {\tilde\mu}} \, {\cal A}^M{}_{ {\tilde\nu}} \ ,
\end{eqnarray}
and all indices are implicitly raised and lowered with the double invariant metrics $\eta_{A B}$ or $\eta_{M N}=E_{M}{}^{A}\; \eta_{A B} \; E_{N}{}^{B}$ and the Killing metric of the gauge group $\kappa_{\overline{\alpha \beta}}$ or $\kappa_{\tilde\mu \tilde\nu}=e_{\tilde\mu}{}^{\overline\alpha}\;\kappa_{\overline{\alpha \beta}}\; e_{\tilde\nu}{}^{\overline\beta}$.

The extended generalized frame $\cal E$ parameterizes the coset ${\cal G}/{\cal H}$  and so carries $D(D+k)$ physical degrees of freedom (dof). They are contained in the double generalized frame $E$ which parameterizes the coset $G/H$ and so carries $D^2$ dof. The remaining $Dk$ dof are captured by ${\cal E}^{\tilde\nu}{}_{\underline a}$. The rest of the components are gauge dof and can be eliminated by the action of $\cal H$, which can be used to implement the following gauge fixing
\begin{equation} \label{gaugefixing}
{\cal E}^{\tilde\nu}{}_{\overline a}=E^{M}{}_{\overline a} \, {\cal A}_{M}{}^{\tilde\nu}=0 \ , \ \ \  e_{\tilde\mu}{}^{\overline\alpha}=constant \,. 
\end{equation}
The reason why we can freeze the scalars is that they are pure gauge dof because the coset $g/h$ is trivial. This gauge fixing breaks the group $\cal H$ down to $H$. In fact, freezing the components (\ref{gaugefixing}) implies locking their gauge transformations $\delta{\cal E}^{\tilde\nu}{}_{\overline a}=\delta e_{\tilde\mu}{}^{\overline\alpha} =0$, which fixes the following components of the parameters of $\cal H$
\begin{eqnarray}
\Gamma_{\overline {\alpha a}} &=& e^{\tilde\mu}{}_{\overline \alpha} \, (\Box^{- \frac 1 2}){}_{\tilde\mu}{}^{\tilde\nu}\, \partial_M \xi_{\tilde\nu} \, E^M{}_{\bar a}\ , \label{GaugedFixedParametersMono}\\
\Gamma_{\overline {\alpha \beta}} &=&  e^{\tilde\mu}{}_{[\overline \alpha} \, e^{\tilde\nu}{}_{\overline \beta]} \, (\Box^{-\frac 1 2}){}_{\tilde\mu}{}^{\tilde\rho} \left(\delta (\Box^{\frac 1 2}){}_{\tilde\rho \tilde\nu} - {\cal A}^{M}{}_{\tilde\nu}  \, \partial_{M} \xi_{\tilde\rho}- g\, f_{ \tilde\rho \tilde\sigma}{}^{\tilde\tau} \, \xi^{\tilde\sigma} \,(\Box^{\frac 1 2}){}_{\tilde\tau \tilde\nu} \right)
\ , \nonumber
\end{eqnarray}
where we have explicitly introduced the gauge coupling constant $g$ and the dimensionless structure constants $f_{\tilde\mu \tilde\nu}{}^{\tilde\rho} =g^{-1} \hat{f}_{\tilde\mu \tilde\nu}{}^{\tilde\rho}$.

Let us now discuss the so-called {\it generalized BdR identification} in this heterotic scenario. For more details see \cite{GenBdR}. There are two gauge groups in the theory, and both have generalized connections. One is the group $\cal K$, the connection being the projected field ${\cal E}_{\tilde \mu \underline a}=E^{M}{}_{\underline a}\, {\cal A}_{M \tilde\mu}$. When the generalized Lie derivative (\ref{TransformationExtended}) is reduced to its components, it yields
\begin{eqnarray}
\delta {\cal E}_{\tilde\mu \underline a } = \widehat {\cal L}_\xi {\cal E}_{\tilde\mu \underline a} - {\cal D}_{\underline a} \xi_{\tilde\mu} + g f_{\tilde\mu \tilde\nu}{}^{\tilde\rho} \xi^{\tilde\nu} {\cal E}_{\tilde\rho \underline a} + {\cal E}_{\tilde\mu \underline d} \Gamma^{\underline d}{}_{\underline {a}} \ . \label{TransformationA}
\end{eqnarray}
The other is the Lorentz group $\overline {\cal H}$, the connection being a certain projection of the generalized fluxes ${\cal F}_{\underline a {\overline {\cal BC}}}$, which according to (\ref{TransfFluxes}) transform as
\begin{eqnarray}
\delta {\cal F}_{\underline a \overline {\cal B C}} = \widehat {\cal L}_\xi {\cal F}_{\underline a \overline {\cal B C}} - {\cal D}_{\underline a} \Gamma_{\overline{\cal B C}} + 2 {\cal F}_{\underline a  \overline {\cal D} [\overline{\cal C}} \Gamma^{\overline{\cal D}}{}_{\overline{\cal B}]} + {\cal F}_{\underline d  \overline {\cal B C}} \Gamma^{\underline d}{}_{\underline {a}} \ . \label{TransformationFProjected}
\end{eqnarray}
The former are independent physical dof, while the later are composite dof, yet as different as they are, they both transform in the same way with respect to different groups. Then, if we {\it choose} these groups to coincide
\begin{eqnarray}
{\cal K} = \overline {\cal H} \ ,
\end{eqnarray}
we can express the connection ${\cal E}_{\tilde \mu \underline a}$ and the parameters $\xi_{\tilde \mu}$ in terms of the adjoint indices of $\overline {\cal H}$ through its generators $(t^{\tilde\mu})_{\overline {\cal B C}}$
\begin{eqnarray}
{\cal E}_{\underline a \overline {\cal B C}} = -g  \, {\cal E}_{\tilde\mu \underline a} \, (t^{\tilde\mu})_{\overline {\cal B C}} \ , \ \ \ \xi_{\overline {\cal {BC}}} = - g  \, \xi_{\tilde \mu} \, (t^{\tilde\mu})_{\overline {\cal B C}}\ , \label{IndexRelation}
\end{eqnarray}
in which case (\ref{TransformationA}) takes the form 
\begin{eqnarray}
 \delta {\cal E}_{\underline a \overline{\cal B C}}= \widehat {\cal L}_\xi {\cal E}_{\underline a \overline{\cal B C}} - {\cal D}_{\underline a} \xi_{\overline{\cal B C}} +  2 {\cal E}_{\underline a \overline{\cal D} [\overline{\cal C}}\, \xi^{\overline{\cal D}}{}_{\overline{\cal B}]} +  {\cal E}_{\underline d \overline{\cal B C}}\, \Gamma^{\underline{d}}{}_{\underline a} \ . \label{TransformationAH}
\end{eqnarray}
Now, the comparison between (\ref{TransformationFProjected}) and (\ref{TransformationAH}) establishes a way to {\it lock} the gauge vectors in terms of the generalized fluxes 
\begin{eqnarray}
\boxed{\begin{aligned}
\xi_{\overline {\cal A B}} &= -g  \, \xi_{\tilde\mu} \, (t^{\tilde\mu})_{\overline {\cal A B}} \ =\ \Gamma_{\overline {\cal A B}}\ , \\
{\cal E}_{\underline a \overline {\cal B C}} &= -g  \, {\cal E}_{\tilde\mu \underline a} \, (t^{\tilde\mu})_{\overline {\cal B C}} \ =\ {\cal F}_{\underline a \overline {\cal B C}}  \ . \end{aligned}}\label{LockingMono}
\end{eqnarray}
This is  the {\it generalized Bergshoeff-de Roo identification}. It has the appearance of being impossible because the dimensions  dim$({\cal K}) = k$ and dim$(\overline {\cal H}) = (D+k)(D+k-1)/2$ are different for any finite $k$. The only way out is that these are connections of an infinite dimensional orthogonal group. This is somehow expected, because this identification is exact (by this we mean that the transformations (\ref{TransformationFProjected}) and (\ref{TransformationAH}) are identical) and then we expect it to generate an infinite tower of higher derivatives, as opposed to the original identification in \cite{BdR} which only held to first order.

\subsection{The perturbative expansion}\label{sectionMono}

After the identification the expected remaining dof are the standard $G$-valued generalized frame $E_M{}^A$ and the dilaton $d$. They inherit their transformation properties from those of the extended space (\ref{TransformationExtended}) after insertion of the parameterization (\ref{FrameParameterizationMono}) and the identification (\ref{LockingMono}). The vectorial components {\it induce} gauge transformations to the generalized frame with respect to $\cal K$, which after the identification become higher-derivative corrections to the Lorentz transformations. These corrections can be extracted perturbatively in powers of $\alpha'$ order by order. To first order they were shown in \cite{GenBdR} to reproduce the first order generalized Green-Schwarz transformation introduced in \cite{MarquesNunez}. The perturbative expansion proceeds as follows. The identification relates $\tilde\mu$ with $\overline{\cal A B} = (\overline{a b} , \overline{a \beta} , \overline{\alpha b} , \overline{\alpha \beta} )$ through the generators $(t^{\tilde\mu})_{\overline {\cal A B}}$. The indices $\overline\alpha$ can then be curved back to $\tilde \mu$ through $e_{\tilde\mu}{}^{\overline\alpha}$ in (\ref{FrameParameterizationMono}). This triggers a never ending iteration that permits to compute every order in the derivative expansion.

The exact transformation of the generalized frame after the identification is given by
\begin{eqnarray}
\delta E_M{}^{\overline a} &=& \widehat {\cal L}_\xi E_M{}^{\overline a}\, + \,E_M{}^{\overline b}\, \Lambda_{\overline b}{}^{\overline a} - \frac{1}{g^2 X_R} E_{M}{}^{\underline c} \;(\chi^{-\frac12})_{\underline c}{}^{\underline b} \;{\cal F}_{\underline b \overline{\cal C}\overline{\cal D}} \;D^{\overline a} \Gamma^{\overline{\cal C}\overline{\cal D}}\;,  \label{deltaEoverline2} \\
\delta E_M{}^{\underline a} &=& \widehat {\cal L}_\xi E_M{}^{\underline a} + E_M{}^{\underline b}\, \Lambda_{\underline b}{}^{\underline a} + \frac{1}{g^2 \, X_R}\partial_{\overline M}
														 \Gamma^{\overline{\cal C}\overline{\cal D}}
														\;(\chi^{-\frac12})^{\underline a}{}_{\underline b}\;{\cal F}^{\underline b}{}_{\overline{\cal C}\overline{\cal D}}\ , \nn
\end{eqnarray}
where we introduced the Dynkin index $X_R$ and redefined the Lorentz parameters of the double space 
\begin{eqnarray}
\Lambda_{\underline {a b}} &=& \Gamma_{\underline {a b}} - E^M{}_{[\underline a} E^N{}_{\underline b]} (\chi^{-\frac 1 2}){}_M{}^P \left(\delta (\chi^{\frac 1 2})_{P N} - \partial_P \xi^\alpha {\cal A}_{N \alpha}\right)\ , \label{LorentzParamterRedefinition}\\
\Lambda_{\overline {a  b}} &=& \Gamma_{\overline {a b}}  \ . \nn
\end{eqnarray}

To trigger the perturbative expansion we first split coordinates $\overline {\cal A}\to(\overline a, \overline\alpha)$ in the $\overline{\cal C D}$ contraction between the extended fluxes and Lorentz parameters, then replace by the different components of the fluxes
\begin{eqnarray} 
{\cal F}_{\underline a\overline{b c} }&=& (\chi^{\frac12})_{\underline a}{}^{\underline e}\;
F_{\underline e\overline{b c} }\ ,\label{FabcGF}\\
{\cal F}_{\underline a \overline{b \gamma}}&=& - \left[(\chi^{\frac12})_{\underline a}{}^{\underline e}
\left(
{\cal E}_{\tilde \mu}{}^{ \underline d} F_{\overline b\underline{d e}} + D_{\overline b} {\cal E}_{\tilde \mu \underline e}\right)-D_{\overline b}(\Box^{\frac12})_{\tilde \mu}{}^{\tilde \nu}\; {\cal E}_{\tilde \nu \underline a}\right] e^{\tilde \mu}{}_{\bar \gamma}\ ,\label{FabgammaGF}\\
{\cal F}_{\underline a\overline{\alpha\beta} }&=&g\; f_{\tilde \mu \tilde \nu}{}^{\tilde \lambda} \; {\cal E}_{\tilde \lambda\,\underline a}\;(\Box^{\frac12})^{\tilde \mu}{}_{\tilde \rho} (\Box^{\frac12})^{\tilde \nu}{}_{\tilde \sigma}\;e^{\tilde \rho}{}_{\overline{\alpha}} \,e^{\tilde \sigma}{}_{\overline \beta}\label{FaalpbetGF} \\
&&+ (\chi^{\frac12})_{\underline a}{}^{\underline b}\; {\cal E}_{\tilde \mu}{}^{\underline c}\;
 e^{\tilde \mu}{}_{[\overline \alpha}\, e^{\tilde \nu}{}_{\overline \beta]}
 \left[ F_{\underline{bcd}} {\cal E}_{\tilde \nu}{}^{\underline d}
+ (2\, D_{\underline c} {\cal E}_{\tilde \nu\, \underline b}-D_{\underline b} {\cal E}_{\tilde \nu\, \underline c})\right]\nn \\
&&+  e^{\tilde \mu}{}_{[\overline \alpha}\, e^{\tilde\nu}{}_{\overline \beta]}\;  D_{\underline b}(\Box^{\frac12})^{\tilde \rho}{}_{\tilde\mu} \left[
(\chi^{\frac12})_{\underline a}{}^{\underline b}(\Box^{\frac12})_{\tilde \nu \tilde \rho} +{\cal E}_{\tilde \rho \underline a}\; {\cal E}_{\tilde \nu}{}^{\underline b}
 \right]\ ,
\nn
\end{eqnarray}
and the extended Lorentz components, through (\ref{GaugedFixedParametersMono}) and (\ref{LorentzParamterRedefinition}).  Here we introduced double generalized fluxes  and flat derivatives 
\begin{equation}
    F_{A B C} = 3\, D_{[A} E^M{}_B E^N{}_{C]} \, \eta_{M N} \ , \ \ \ \  D_A = E^M{}_A \partial_M \, . \label{DoubleFluxes}
\end{equation}

All the replacements above are exact. They depend on ${\cal E}_{\tilde\mu\underline a}$ though, but at a higher order in a $g^{-1}$ expansion, except for (\ref{FaalpbetGF}) whose only effect is to renormalize the leading contribution (responsible for the $b$ parameter), as we will discuss later. Hence, repeating recursively this procedure leads to a derivative expansion of the Lorentz transformation.  Up to second order one finds \cite{GenBdR}
\small
\begin{eqnarray}
\boxed{\begin{aligned}
\delta E_{M}{}^{\overline a}=& \ \widehat {\cal L}_\xi E_M{}^{\overline a} + E_M{}^{\overline b}\, \Lambda_{\overline b}{}^{\overline a} - \frac b 2  \, E_M{}^{\underline d} F_{\underline d \overline {b c}} \, D^{\overline a}  \Lambda^{\overline {bc}}  \\
&-\frac12 b^2 E_M{}^{\underline b}\left[
														D^{\overline{a}}D^{\overline c}\Lambda^{\overline{e f}}\left(F_{\overline c\underline{d b}}  F^{\underline d}{}_{\overline{e f}} + D_{\overline c} F_{\underline b \overline{e f}}\right)	
														\right.
-F_{\underline b \overline{e f}} F_{\underline c \overline d}{}^{\overline f}\left(
F^{\underline c}{}^{\overline{h d}} D^{\overline a}\Lambda_{\overline h}{}^{\overline e}
-F^{\underline c}{}^{\overline{h e}} D^{\overline a}\Lambda_{\overline h}{}^{\overline d}
\right) \\
&\ \ \ \ \ \ \ \ \   +\;\left. F^{\underline c}{}_{\overline{e f}}\;D^{\overline a}\Lambda^{\overline{e}}{}_{\overline g} \left(F_{\underline {b c d}} F^{\underline d \overline{g f}}
-D_{\underline b} F_{\underline c}{}^{\overline{g f}} + 2\; D_{\underline c} F_{\underline b}{}^{ \overline{g f}}
\right)
+F_{\underline b \overline{e f}}D^{\overline a}\left(D^{\underline c}\Lambda^{\overline{e d}} F_{\underline c \overline{d}}{}^{\overline f}\right)						
							\right] ,\,\end{aligned}} \nn
\end{eqnarray}\normalsize

\begin{equation}
\label{Second1}
\end{equation}
and \small
\begin{eqnarray}
\boxed{\begin{aligned}
\delta E_M{}^{\underline a} =& \ \widehat {\cal L}_\xi E_M{}^{\underline a} + E_M{}^{\underline b}\, \Lambda_{\underline b}{}^{\underline a} + \frac b 2 \partial_{\overline M} \Lambda^{\overline {b c}} \, F^{\underline a}{}_{\overline {b c}} \\
& +\frac12 b^2\, E_M{}^{\overline b}\left[
D_{\overline{b}}D^{\overline c}\Lambda^{\overline{e f}}\left(F_{\overline c\underline d}{}^{\underline a} F^{\underline d}{}_{\overline{e f}} +
D_{\overline c} F^{\underline a}{}_{\overline{e f}}\right)	\right.
-F^{\underline a}{}_{\overline{e f}} F_{\underline c \overline d}{}^{\overline f}\left(
F^{\underline c}{}^{\overline{h d}} D_{\overline b}\Lambda_{\overline h}{}^{\overline e}
-F^{\underline c}{}^{\overline{h e}} D_{\overline b}\Lambda_{\overline h}{}^{\overline d}
\right) \\
&\ \ \ \ \ \ \ \ \   + \left. F^{\underline c}{}_{\overline{e f}}\;D_{\overline b}\Lambda^{\overline{e}}{}_{\overline g} \left(F^{\underline a}{}_{\underline{c d}} F^{\underline d \overline{g f}}
-D^{\underline a} F_{\underline c}{}^{\overline{g f}} + 2\; D_{\underline c} F^{\underline a  \overline{g f}}\right)
+F^{\underline a}{}_{\overline{e f}}D_{\overline b}\left(D^{\underline c}\Lambda^{\overline{e d}} F_{\underline c \overline{d}}{}^{\overline f}\right)\right],\, \end{aligned}}\nn
\end{eqnarray}\normalsize
\begin{equation}
\label{Second2}
\end{equation}

where 
\begin{eqnarray}
b = \frac{2}{g^2 (-1+X_R)}\ .\label{abmatch}
\end{eqnarray}

Let us briefly point out how this parameter forms. Consider the contractions in (\ref{deltaEoverline2}), which are schematically of the form
\begin{equation}
    \Psi_{\overline {\cal A B}} \Phi^{\overline {\cal A B}} = \Psi_{\overline {a b}} \Phi^{\overline {a b}} + \underset {\text {Higher\ order}}
    { \underline{ \Psi_{\overline {a \beta}} \Phi^{\overline {a \beta}} + \Psi{}_{\overline {\alpha b} } \Phi^{\overline {\alpha b}}  } } + \Psi_{\overline {\alpha \beta}} \Phi^{\overline {\alpha \beta}} \ . \label{IndexDecomp1}
\end{equation}
The off-diagonal part is of higher order because its leading order already contains vector fields, which are identified with the generalized fluxes that carry derivatives. The last term, corresponding to the purely internal part, happens to obey the following relations due to the identification
\begin{equation}
      \Psi_{\overline {\alpha \beta}} \Phi^{\overline {\alpha \beta}} = \frac 1 {X_R} \, \Psi_{\overline {\cal A B}} \Phi^{\overline {\cal A B}} + {\text {Higher\ order} } \ . \label{IndexDecomp2}
\end{equation}
This tells us on the one hand that the purely internal sector (where $\overline h$ acts) starts at the same order than the purely external sector (where $\overline H$ acts). On the other hand, interestingly the internal $\overline h$ contraction can be re-expressed up to higher orders in terms of the full $\overline {\cal H}$ contraction, by use of the identification. Then,    
combining (\ref{IndexDecomp1}) with (\ref{IndexDecomp2}) permits to eliminate the internal contraction
\begin{equation}
    \frac 1 { g^2\, X_R} \Psi_{\overline {\cal A B}} \Phi^{\overline {\cal A B}} = \frac b 2 \, \Psi_{\overline {a b}} \Phi^{\overline {a b}} + \text{Higher\ order} \ ,
\label{FirstOrderContractions} \end{equation}
generating at the same time the parameter $b$ defined in (\ref{abmatch}).  Proceeding forward towards more derivatives requires keeping track of the higher order terms, which interestingly can again undergo this procedure. The only non-straightforward step for higher orders is that, in general, the structures that obey these cyclic relations are not single contractions as in (\ref{IndexDecomp1}), but consist of sums of terms with more that two indices contracted. It then happens that the identities above fail to apply to independent terms in the sum, but hold for the full summation. To clarify this point it is instructive to discuss a concrete example. 

When implementing this procedure for the  last term in (\ref{deltaEoverline2}), we get
\begin{eqnarray}
\frac 1 { g^2\, X_R} (\chi^{-\frac12})_{\underline c}{}^{\underline b} {\cal F}_{\underline b \overline {\cal C D}} D^{\overline a}\Gamma^{\overline {\cal C D}} &=& \frac b 2 F_{\underline c\overline {c d}} D^{\overline a}\Lambda^{\overline {c d}} + \text{Higher\ order} \ ,\label{example1}
\end{eqnarray}
which is simply a concrete realization of (\ref{FirstOrderContractions}), and explains the ${\cal O}(b)$ contribution to the generalized Green-Schwarz transformation in (\ref{Second1}). Keeping track of the higher order contributions, one can identify among them the following combination
\begin{eqnarray}
\delta E_M{}^{\overline a}&\supset& \;\frac{2}{g_2^4\,X_{R}(-1+X_{R})} \,E_{\underline M}{}^{\underline b}\, \;{\cal F}_{\underline b}{}_{\overline{\cal E F}} {\cal F}_{\underline c}{}_{\overline{\cal G}}{}^{\overline{\cal F}}\left(
{\cal F}^{\underline c \overline{\cal C G}} D^{\overline a} \Gamma_{\overline{\cal C}}{}^{\overline{\cal E} }
 -  {\cal F}^{\underline c \overline{\cal C E}} D^{\overline a}\Gamma_{\overline{\cal C}}{}^{\overline{\cal G} }\right)\;.
 \label{example2}
\end{eqnarray}
Note that the difference now is that there are {\it two} terms with  a {\it four}-index  $\overline {\cal H}$-contraction on $\overline {\cal C E F G}$ (while in (\ref{IndexDecomp1}) we started with {\it one} term with a {\it two}-index contraction). We now perform the $\overline h \times \overline H$ splitting as in (\ref{IndexDecomp1}) for these terms 
\begin{eqnarray}
{\cal F}_{\underline b}{}_{\overline{\cal E F}} {\cal F}_{\underline c}{}_{\overline{\cal G}}{}^{\overline{\cal F}}
{\cal F}^{\underline c \overline{\cal C G}} D^{\overline a} \Gamma_{\overline{\cal C}}{}^{\overline{\cal E} }
\;=\; F_{\underline b \overline{e f}} F_{\underline c \overline d}{}^{\overline f}
F^{\underline c}{}^{\overline{h d}} D^{\overline a}\Lambda_{\overline h}{}^{\overline e}
\;+\; F_{\underline b \overline{\alpha \beta}} F_{\underline c \overline \gamma}{}^{\overline \beta}
F^{\underline c}{}^{\overline{\delta \gamma}} D^{\overline a}\Lambda_{\overline \delta}{}^{\overline \alpha}\;+\;\dots\; ,
\label{IndexDecomp1B1} \\ 
{\cal F}_{\underline b}{}_{\overline{\cal E F}} {\cal F}_{\underline c}{}_{\overline{\cal G}}{}^{\overline{\cal F}}  
{\cal F}^{\underline c \overline{\cal C E}} D^{\overline a}\Gamma_{\overline{\cal C}}{}^{\overline{\cal G} }\;=\; F_{\underline b \overline{e f}} F_{\underline c \overline d}{}^{\overline f} F^{\underline c}{}^{\overline{h e}} D^{\overline a}\Lambda_{\overline h}{}^{\overline d}
\;+\; F_{\underline b \overline{\alpha \beta}} F_{\underline c \overline \gamma}{}^{\overline \beta} F^{\underline c}{}^{\overline{\delta \alpha}} D^{\overline a}\Lambda_{\overline \delta}{}^{\overline \gamma}
\;+\; \dots\; ,\label{IndexDecomp1B2}
\end{eqnarray}
where the dots stand for higher orders.  The subtlety arises when studying the realization of (\ref{IndexDecomp2}) in this case. We find that  
\begin{eqnarray}
F_{\underline b \overline{\alpha \beta}} F_{\underline c \overline \gamma}{}^{\overline \beta}
F^{\underline c}{}^{\overline{\delta \gamma}} D^{\overline a}\Lambda_{\overline \delta}{}^{\overline \alpha}&=&
\frac{1}{X_R} \; {\cal F}_{\underline b}{}_{\overline{\cal E F}} {\cal F}_{\underline c}{}_{\overline{\cal G}}{}^{\overline{\cal F}} 
{\cal F}^{\underline c \overline{\cal C G}} D^{\overline a} \Gamma_{\overline{\cal C}}{}^{\overline{\cal E} } 
\;+\; \Delta_{\underline b}{}^{\overline a} \, ,
\label{example2B} \\
F_{\underline b \overline{\alpha \beta}} F_{\underline c \overline \gamma}{}^{\overline \beta}
F^{\underline c}{}^{\overline{\delta \alpha}} D^{\overline a}\Lambda_{\overline \delta}{}^{\overline \gamma}&=&
\frac{1}{X_R} \; {\cal F}_{\underline b}{}_{\overline{\cal E F}} {\cal F}_{\underline c}{}_{\overline{\cal G}}{}^{\overline{\cal F}} 
{\cal F}^{\underline c \overline{\cal C E}} D^{\overline a} \Gamma_{\overline{\cal C}}{}^{\overline{\cal G} } 
\;+\; \Delta_{\underline b}{}^{\overline a}  \  , \label{example2B2}
\end{eqnarray}
where the anomalous factor is given by 
\begin{eqnarray}
\Delta_{\underline b}{}^{\overline a}&=&
-2\; {\cal F}_{\underline b}{}_{\overline{\cal E F}} {\cal F}_{\underline c}{}_{\overline{\cal G}}{}^{\overline{\cal F}}
{\cal F}^{\underline c \overline{\cal C G}} D^{\overline a} \Gamma_{\overline{\cal C}}{}^{\overline{\cal E} } -
2\; {\cal F}_{\underline b}{}_{\overline{\cal G F}} {\cal F}_{\underline c}{}_{\overline{\cal E}}{}^{\overline{\cal F}}
{\cal F}^{\underline c \overline{\cal C G}} D^{\overline a} \Gamma_{\overline{\cal C}}{}^{\overline{\cal E} } 
+ \; {\cal F}_{\underline b}{}_{\overline{\cal E F}} {\cal F}_{\underline c}{}^{\overline{\cal E F}}
{\cal F}^{\underline c \overline{\cal C G}} D^{\overline a} \Gamma_{\overline{\cal C G}} \cr &&
\; {\cal F}_{\underline b}{}_{\overline{\cal E F}} {\cal F}_{\underline c}{}_{\overline{\cal G H}}
{\cal F}^{\underline c \overline{\cal F H}} D^{\overline a} \Gamma^{\overline{\cal E G} }  
- \; {\cal F}_{\underline b}{}_{\overline{\cal E F}} {\cal F}_{\underline c}{}_{\overline{\cal G H}}
{\cal F}^{\underline c \overline{\cal G H}} D^{\overline a} \Gamma^{\overline{\cal E F} }\;
 - \; {\cal F}_{\underline b}{}_{\overline{\cal E F}} {\cal F}_{\underline c}{}_{\overline{\cal G H}}
{\cal F}^{\underline c \overline{\cal E F}} D^{\overline a} \Gamma^{\overline{\cal G H} }  \cr &&
+ \; {\cal F}_{\underline b}{}_{\overline{\cal E F}} {\cal F}_{\underline c}{}_{\overline{\cal G H}}
{\cal F}^{\underline c \overline{\cal E G}} D^{\overline a} \Gamma^{\overline{\cal F H} } + {\rm Higher \ order} \, .
\end{eqnarray}
It is quite remarkable that exactly the same anomaly appears in (\ref{example2B}) and (\ref{example2B2}), and that it cancels for the particular combination (\ref{example2}), leading to 
\begin{eqnarray}
\frac{2}{g_2^4\,X_{R}(-1+X_{R})} \,E_{\underline M}{}^{\underline b}\, \;{\cal F}_{\underline b}{}_{\overline{\cal E F}} {\cal F}_{\underline c}{}_{\overline{\cal G}}{}^{\overline{\cal F}}\left(
{\cal F}^{\underline c \overline{\cal C G}} D^{\overline a} \Gamma_{\overline{\cal C}}{}^{\overline{\cal E} }
 -  {\cal F}^{\underline c \overline{\cal C E}} D^{\overline a}\Gamma_{\overline{\cal C}}{}^{\overline{\cal G} }\right)
 \;=\;\;\;\;\;\;\;\;\;\;\;\;\;\;\;\cr\cr
\;=\;\frac  {b^2} 2\,E_{\underline M}{}^{\underline b}\, \;{\cal F}_{\underline b}{}_{\overline{e f}} {\cal F}_{\underline c}{}_{\overline{d}}{}^{\overline{f}}\left(
{\cal F}^{\underline c \overline{g d}} D^{\overline a} \Gamma_{\overline{g}}{}^{\overline{e} }
 -  {\cal F}^{\underline c \overline{g e}} D^{\overline a}\Gamma_{\overline{g}}{}^{\overline{d} }\right) \;+\; \text{Higher\ order}
 \cr\cr  
\;=\;\frac {b^2} 2 \,E_{\underline M}{}^{\underline b}\, \;{F}_{\underline b}{}_{\overline{e f}} {F}_{\underline c}{}_{\overline{d}}{}^{\overline{f}}\left(
{F}^{\underline c \overline{g d}} D^{\overline a} \Lambda_{\overline{g}}{}^{\overline{e} }
 -  {F}^{\underline c \overline{g e}} D^{\overline a}\Lambda_{\overline{g}}{}^{\overline{d} }\right) \;+\; \text{Higher\ order}  \ .
\end{eqnarray}
These are the last pair of terms in the second line in the generalized Green-Schwarz transformation (\ref{Second1}). All the other ${\cal O}(b^2)$ terms and higher can be treated analogously.

Although it certainly looks like this is the case, let us remark that we do not have a proof that the parameter $b$ will form to all orders, nor that the recursive relations required to completely remove the gauge dof will converge at all orders. However, if we {\it assume} that the steps leading to the formation of the $b$ parameter can be repeated over and over, it is then possible to implement a {\it systematic} procedure to compute order by order in the perturbative expansion, that can be built into a computer program. It should follow a precise route in order to succeed. The first step requires switching all the $g$-fundamental indices $\tilde\mu$  at a given order into $\overline {\cal H}$-adjoint indices $\overline {\cal A B}$. This is readily implemented by replacing
\begin{eqnarray}
{\cal E}_{\tilde\mu}{}^{\underline a} = \frac{1}{X_R}\,{\cal F}_{\underline{a} \overline{\cal A}}{}^{\overline{\cal B}} \; (t_{\tilde\mu})_{\overline{\cal B}}{}^{\overline{\cal A}} \;,\;\;\;
\xi^{\tilde\mu} = \frac{1}{X_R} \Gamma_{\overline{\cal A}}{}^{\overline{\cal B}} \; (t^{\tilde\mu})_{\overline{\cal B}}{}^{\overline{\cal A}} \;,\;\;\;
f_{\tilde \mu \tilde\nu}{}^{\tilde \rho} = -\frac{2}{X_R} (t_{\tilde\mu})_{\overline{\cal A}}{}^{\overline{\cal B}} 
\; (t_{\tilde\nu})_{\overline{\cal B}}{}^{\overline{\cal C}} \;
(t^{\tilde\rho})_{\overline{\cal C}}{}^{\overline{\cal A}}\; ,  \ \ \ \ 
\end{eqnarray}
and then by eliminating the generators through (\ref{GenPr2}).  The next step consists in splitting indices $\overline{\cal A}$ into $\overline{a}, \overline{\alpha}$ in the previous expression. The terms with generalized fluxes or Lorentz parameters containing mixed $\overline H$ and $\overline h$ contractions must be separated as they are higher order. Those with pure $\overline h$ contractions should be replaced and expanded by their expressions in the Appendix and the gauge fixing conditions (\ref{GaugedFixedParametersMono}). Once this is done, the leading terms of such an expansion will combine with the pure $\overline H$ contraction to form the parameter $b$, the rest must be separated as it is higher order. Finally, one is left with generalized fluxes and Lorentz transformations in the extended space with pure $\overline H$ contractions (now properly weighted with the parameter $b$), which should now be replaced by the expressions in the Appendix and the redefinitions of the Lorentz parameter (\ref{LorentzParamterRedefinition}), in terms of the fluxes and parameters in the double space, plus higher orders. This isolates the relevant contribution to a given order, which is now properly weighted with the parameter $b$, and separates the higher order contributions, which further admit an identical treatment.   

The same algorithm could be adapted to the full bi-parametric deformation of DFT to be discussed below.  One can also adapt this algorithm to find higher orders in the invariant action. The main issue here is the optimization of the algorithm as the number of couplings grows exponentially as we move to higher orders. Of course, since the whole algorithm is based on an assumption, in the end one should check if the result is correct and consistent. This is typically a difficult task as Bianchi identities can be responsible for the equality between seemingly different terms.

\section{The bi-parametric generalized BdR identification} \label{SecBiparam}

We now move to the bi-parametric case, where both $a$ and $b$ can be turned on simultaneously. This requires a further extension of the mono-parametric setup, consisting in a double extended space with a duality group ${\cal G} = O(D+k,D+k)$, which is now a more symmetric scenario, as expected.  We show in Table \ref{NotationBi} the implications of this extension for the relevant symmetries, and the notation that we will adopt from now on.
\begin{table}[H]
\begin{center}
\begin{tabular}{|l|l|l|l|} \hline
{\bf Name}& {\bf Group}  & {\bf Indices}  &  {\bf Metric} \\ \hline
$G$ $\begin{matrix} \\ ~\end{matrix}$  & $O(D,D)$ & $M$ & $\eta_{M N}$ \\ \hline
$g$ $\begin{matrix} \\ ~\end{matrix}$  & $O(k,k) = O(p + q', q + p')$ & $\hat \mu = (\ut \mu ,\, \tilde \mu)$ & $\kappa_{\hat \mu \hat \nu} = \left(\begin{matrix}- \kappa_{\ut \mu \ut \nu} & 0 \\ 0 & \kappa_{\tilde\mu \tilde\nu} \end{matrix} \right)$ \\ \hline
$\cal G$ & $\begin{matrix}O(D+k,D+k) \ \ \ \ \ \ \ \ \ \ \ \ \ \ \ \ \\ = O(D + p + q', D+q + p')\end{matrix}$ & ${\cal M} = (M,\,\hat \mu)$ &
$\eta_{\cal M N} = \left(\begin{matrix}\eta_{M N} & 0 \\ 0 & \kappa_{\hat\mu \hat\nu}\end{matrix}\right)$ \\ \hline
 $\underline H$ $\begin{matrix} \\ ~\end{matrix}$ & $\underline{O(D-1,1)}$ & $ \underline a$ & $P_{\underline a \underline b}$ \\ \hline
$\overline H$ $\begin{matrix} \\ ~\end{matrix}$  & $\overline{O(1,D-1)}$ & $\overline a$ & $\bar P_{\overline {a b}}$ \\ \hline
$H$ & $\underline{H}\times \overline{H}$ & $A = (\underline a,\, \overline a)$ & $\eta_{A B} = \left(\begin{matrix}P_{\underline a \underline b}& 0 \\ 0 & \bar P_{\overline {ab}}\end{matrix}\right)$\\ \hline
$h$ $\begin{matrix} \\ ~\end{matrix}$ & $\underline h \times \overline h = \underline{O(q',p')} \times \overline{O(p,q)}$ & $\hat \alpha = (\underline \alpha,\, \overline \alpha)$ & $\kappa_{\hat \alpha \hat \beta} = \left(\begin{matrix} - \kappa_{\underline{\alpha \beta}} & 0 \\ 0 & \kappa_{\overline {\alpha \beta}}\end{matrix}\right)$ \\ \hline
$\underline {\cal H}$ & $\underline{O(D+q'-1,1 + p')}$ & $\underline{\cal A} = (\underline a,\, \underline \alpha)$ & $P_{\underline {\cal A} \underline {\cal B}} = \left(\begin{matrix}  P_{\underline {a b}} & 0 \\ 0 & - \kappa_{\underline {\alpha \beta}}\end{matrix}\right)$ \\ \hline
$\overline {\cal H}$ & $\overline{O(1+ p,D+q-1)}$ & $\overline{\cal A} = (\overline a,\, \overline \alpha)$ & $\bar P_{\overline {\cal A} \overline {\cal B}} = \left(\begin{matrix} \bar P_{\overline {a b}} & 0 \\ 0 & \kappa_{\overline {\alpha \beta}}\end{matrix}\right)$ \\ \hline
$\cal H$ & $\underline{{\cal H}} \times \overline{{\cal H}}$ & ${\cal A} = (\underline{{\cal A}},\,\overline{{\cal A}})$ & $\eta_{\cal A B} = \left(\begin{matrix}P_{\underline {\cal A B}} & 0 \\ 0 & \bar P_{\overline {\cal A B}} \end{matrix} \right)$  \\ \hline
\end{tabular}
    \end{center}
    \caption{Groups, metrics and index structure for the extended space relevant to the bi-parametric case. In the bi-parametric case the relevant choice is $q' = q$, $p' = p$ and $k = p + q$. If the prime quantities were independent and set to zero, this table then reproduces Table \ref{NotationMono}. }\label{NotationBi}
\end{table}

The counting of dof is now a little different than before. The extended frame ${\cal E}_{\cal M}{}^{\cal A}$ parameterizes the coset ${\cal G} / {\cal H}$, now containing $(D  + k)^2$ physical dof. We obviously accommodate $D^2$ of them into a double generalized frame $E_M{}^A$ parameterizing the coset $G/H$. There are other $k^2$ physical dof that are captured by a scalar frame $e_{\hat \mu}{}^{\hat \alpha}$, parameterizing the coset $g/h$, which is now non-trivial as opposed to the mono-parametric case. Also there is now a {\it pair} of projected vectors ${\cal E}^{\tilde \mu}{}_{\underline a}{}$ and ${\cal E}^{\ut \mu}{}_{\overline a}$, each containing $Dk$ dof. Compared to the heterotic case, there are then extra vector fields and scalars, that will have to be identified.

Because now the duality group is enhanced, the generalized diffeomorphisms can accomodate a gauge group $\cal K$ of dimension $2k$. We then take it to be a direct product ${\cal K}=\underline{\cal K}\times \overline{\cal K}$, where $\underline{\cal K}$ and $\overline{\cal K}$ are two independent $k$-dimensional gauge groups. The only non vanishing components of the extended gaugings are then $\hat{f}_{ {\ut\mu} {\ut\nu}}{}^{{\ut\rho}}$ and $\hat{f}_{ {\tilde\mu} {\tilde\nu}}{}^{ {\tilde\rho}}$ and the consistency of the deformation then requires that each pair of gaugings must satisfy the linear and quadratic constraints independently. The strong constraint further requires $\partial_{ {\ut\mu}}=0=\partial_{ {\tilde\mu}}$ . 

The source of the two parameters $(a,b)$ are the two gauge couplings $g_1$ and $g_2$ of $\underline{\cal K}$ and $\overline{\cal K}$ respectively, whose (dimensionless) structure constants are given by $f_{ {\ut\mu} {\ut\nu}}{}^{{\ut\rho}}=g_1^{-1}\, \hat{f}_{ {\ut\mu} {\ut\nu}}{}^{{\ut\rho}}$ and $f_{ {\tilde\mu} {\tilde\nu}}{}^{ {\tilde\rho}}=g_2^{-1}\, \hat{f}_{ {\tilde\mu} {\tilde\nu}}{}^{ {\tilde\rho}}$, respectively. Inspired by the heterotic case we now plan to identify the groups
\begin{equation}\underline{\cal K}= \underline {\cal H} \ , \ \ \ \overline{\cal K}  = \overline {\cal H} \ ,
\end{equation}
with generators $(t_{ {\ut\mu}})_{\underline {\cal AB}}$ and $(t_{ {\tilde\mu}})_{\overline {\cal AB}}$, which  satisfy the following algebraic relations
\begin{eqnarray}
\left(t^{ {\ut\mu}}\right)_{\underline{\cal A} \underline{\cal B}} \left(t_{{\ut\nu}}\right)^{\underline{\cal A} \underline{\cal B}}&=& X_{R_1} \,\delta^{ {\ut\mu}}_{{\ut\nu}}\ ,\;\;\;\;\;\;\;\;\;\;\;\;\;\;\;\;\;\;\;
\left(t^{ {\tilde\mu}}\right)_{\overline{\cal A} \overline{\cal B}} \left(t_{ {\tilde\nu}}\right)^{\overline{\cal A} \overline{\cal B}}= X_{R_2} \,\delta^{ {\tilde\mu}}_{ {\tilde\nu}}\ ,
\label{GenPr1}\\
\left(t^{ {\ut\mu}}\right)_{\underline{\cal A} \underline{\cal B}} (t_{ {\ut\mu}})^{\underline{\cal C} \underline{\cal D}}&=&X_{R_1} \;\delta_{\underline{\cal A} \underline{\cal B}}^{\underline{\cal C} \underline{\cal D}}
\ ,\;\;\;\;\;\;\;\;\;\;\;\;\;\;\;\;
\left(t^{ {\tilde\mu}}\right)_{\overline{\cal A} \overline{\cal B}} (t_{ {\tilde\mu}})^{\overline{\cal C} \overline{\cal D}}=X_{R_2} \;\delta_{\overline{\cal A} \overline{\cal B}}^{\overline{\cal C} \overline{\cal D}}\ ,
\ \label{GenPr2}
\end{eqnarray}
$X_{R_i}$ being the Dynkin index of each representation. Here we used the killing metrics $\kappa_{ {\ut\mu}{\ut\nu}}$ and $\kappa_{ {\tilde\mu} {\tilde\nu}}$ to rise and lower indices in the algebra, $e.g.$ $t^{ {\ut\mu}}=\kappa^{ {\ut\mu}{\ut\nu}} t_{\ut\nu}$, $f_{ {\ut\mu}{\ut\nu}{\ut\rho}}=\kappa_{{\ut\rho}{\ut\sigma}} \, f_{ {\ut\mu}{\ut\nu}}{}^{{\ut\sigma}}$ and similarly for tilded indices. This should be contrasted with the frame components, $e.g.$ ${\cal E}_{ {\ut\mu}{\cal A}}=\eta_{ {\ut\mu}{\cal N}} \,{\cal E}^{\cal N}{}_{\cal A}=-\kappa_{ {\ut\mu}{\ut\nu}}\, {\cal E}^{{\ut\nu}}{}_{\cal A}$. 

As before, by use of the generators, we can cast certain components of the extended generalized frame in the same structure as the generalized fluxes in the extended space
\begin{eqnarray}
{\cal E}_{\overline{\cal A}{\underline{\cal B C}}}&=&-\; g_1 \; {\cal E}^{ {\ut\mu}}{}_{\overline{\cal A}} \; (t_{ {\ut\mu}})_{\underline{\cal B C}}\,,\;\;\;\;\;\;\;\;\;\;\;\;\;\;\;\;\;\;\;\;\;
\xi_{\underline{\cal B C}}=-\; g_1 \; \xi^{ {\ut\mu}}\; (t_{ {\ut\mu}})_{\underline{\cal B C}}\,,\cr
{\cal E}_{\underline{\cal A}{\overline{\cal B C}}}&=&-\; g_2 \; {\cal E}^{ {\tilde\mu}}{}_{\underline{\cal A}} \; (t_{ {\tilde\mu}})_{\overline{\cal B C}}\,,\;\;\;\;\;\;\;\;\;\;\;\;\;\;\;\;\;\;\;\;\;
\xi_{\overline{\cal B C}}=-\; g_2 \; \xi^{ {\tilde\mu}}\; (t_{ {\tilde\mu}})_{\overline{\cal B C}}\,.\label{EABC}
\end{eqnarray}
On the right we have done the same thing with the gauge components of the parameters that generate generalized diffeomorphisms.  Written in this form, their transformation with respect to local symmetries reads 
\begin{eqnarray}
\delta {\cal E}_{\underline{\cal A}{\overline{\cal B C}}}&=& \widehat {\cal L}_{\xi}{\cal E}_{\underline{\cal A}{\overline{\cal B C}}}+ {\cal E}_{\underline{\cal D}{\overline{\cal B C}}}\; \Gamma^{\underline{\cal D}}{}_{\underline{\cal A}}-{\cal D}_{\underline {\cal A}}\xi_{\overline{\cal B C}} + 2\; {\cal E}_{\underline{\cal A}{\overline{\cal D}[\overline {\cal B}}}\, \xi_{\overline{\cal C}]}{}^{\overline{\cal D}}\,,\cr
\delta {\cal E}_{\overline{\cal A}{\underline{\cal B C}}}&=&  \widehat {\cal L}_{\xi}{\cal E}_{\overline{\cal A}{\underline{\cal B C}}} +  {\cal E}_{\overline{\cal D}{\underline{\cal B C}}}\; \Gamma^{\overline{\cal D}}{}_{\overline{\cal A}}-{\cal D}_{\overline {\cal A}}\xi_{\underline{\cal B C}} + 2\; {\cal E}_{\overline{\cal A}{\underline{\cal D}[\underline {\cal B}}}\, \xi_{\underline{\cal C}]}{}^{\underline{\cal D}}\,.\label{deltaFab}
\end{eqnarray}
Written as such, they happen to transform in exactly the same way as the extended generalized fluxes (\ref{TransfFluxes})
\begin{eqnarray}
\delta {\cal F}_{\underline{\cal A}{\overline{\cal B C}}}&=& \widehat {\cal L}_{\xi}{\cal F}_{\underline{\cal A}{\overline{\cal B C}}}+ {\cal F}_{\underline{\cal D}{\overline{\cal B C}}}\; \Gamma^{\underline{\cal D}}{}_{\underline{\cal A}}-{\cal D}_{\underline {\cal A}}\Gamma_{\overline{\cal B C}} + 2\; {\cal F}_{\underline{\cal A}{\overline{\cal D}[\overline {\cal B}}}\, \Gamma_{\overline{\cal C}]}{}^{\overline{\cal D}}\,,\cr
\delta {\cal F}_{\overline{\cal A}{\underline{\cal B C}}}&=&  \widehat{\cal L}_{\xi}{\cal F}_{\overline{\cal A}{\underline{\cal B C}}} +  {\cal F}_{\overline{\cal D}{\underline{\cal B C}}}\; \Gamma^{\overline{\cal D}}{}_{\overline{\cal A}}-{\cal D}_{\overline {\cal A}}\Gamma_{\underline{\cal B C}} + 2\; {\cal F}_{\overline{\cal A}{\underline{\cal D}[\underline {\cal B}}}\, \Gamma_{\underline{\cal C}]}{}^{\underline{\cal D}}\,,
\end{eqnarray}
which readily suggests
\bea
\boxed{\begin{aligned}
\;\;\xi_{\overline {\cal A B}} &=&\Gamma_{\overline {\cal A B}}\,,\;\;\;\;\;\;\;\;\;\;\;\;\;\;\;\;\;
{\cal E}_{\underline{\cal A} \overline {\cal B C}} = {\cal F}_{\underline{\cal A} \overline {\cal B C}}\,,\;\;\cr
\;\;\xi_{\underline {\cal A B}} &=& \Gamma_{\underline {\cal A B}}\,,\;\;\;\;\;\;\;\;\;\;\;\;\;\;\;\;
{\cal E}_{\overline{\cal A} \underline {\cal B C}} = {\cal F}_{\overline{\cal A} \underline {\cal B C}}\,.\;\;
  \end{aligned}}\label{LockingBi}
\eea
This is the {\it generalized  Bergshoeff-de Roo identification} in the full bi-parametric case. Again, it is exact in the sense that both transformations match identically under this identification.

The extended frame admits a parameterization identical in structure to that of the mono-parametric case (\ref{FrameParameterizationMono})
\bea
{\cal E}_M{}^A &=& (\chi^{\frac 1 2 }){}_M{}^N\, E_N{}^A\,, \nn\\
{\cal E}_M{}^{\hat \alpha} &=& - {\cal A}_M{}^{ {\hat\mu}} \, e_{ {\hat\mu}}{}^{\hat \alpha}\,, \label{FrameParameterizationA}\\
{\cal E}_{ {\hat\mu}}{}^A &=& {\cal A}^M{}_{ {\hat\mu}}\, E_M{}^A\,, \nn\\
{\cal E}_{ {\hat\mu}}{}^{\hat \alpha} &=& (\Box^{\frac 1 2}){}_{ {\hat\mu}}{}^{ {\hat\nu}}\, e_{ {\hat\nu}}{}^{\hat \alpha}\,, \nn
\eea
where we now redefined the  quantities
\bea
\chi_{M N} = \eta_{M N} - {\cal A}_M{}^{ {\hat\mu}}\, {\cal A}_{N {\hat\mu}} \ , \ \ \ \ \ \Box_{ {\hat\mu}  {\hat\nu}} = \eta_{ {\hat\mu}  {\hat\nu}} - {\cal A}_{M  {\hat\mu}} \, {\cal A}^M{}_{ {\hat\nu}} \ ,
\eea
that satisfy the useful identity
\be
{\cal A}_M{}^{ {\hat\mu}}\, f(\Box){}_{ {\hat\mu}}{}^{ {\hat\nu}} = f(\chi){}_M{}^N \, {\cal A}_N{}^{ {\hat\nu}} \ ,
\ee
for any function $f$.
As opposed to the mono-parametric situation,  $e_{ {\hat\mu}}{}^{\hat \alpha}$ is now $g =  O(k,k)$-valued, so it is convenient to further parameterize it as 
\bea
e_{ {\tilde\mu}}{}^{\overline\alpha} &=& ({\Pi}^{\frac 1 2 })_{ {\tilde\mu}}{}^{ {\tilde\nu}}\, \overline{e}_{ {\tilde\nu}}{}^{\overline\alpha}\,, \nn\\
e_{ {\tilde\mu}}{}^{\underline\alpha} &=& - \Omega_{ {\tilde\mu}}{}^{{\ut\nu}} \, \underline{e}_{{\ut\nu}}{}^{\underline \alpha}\,, \label{FrameParameterizationB}\\
e_{ {\ut\mu}}{}^{\overline\alpha} &=&  \Omega^{ {\tilde\nu}}{}_{ {\ut\mu}} \, \overline{e}_{ {\tilde\nu}}{}^{\overline \alpha}\,, \nn\\
e_{ {\ut\mu}}{}^{\underline\alpha} &=&({\Pi}^{\frac 1 2 })_{ {\ut\mu}}{}^{{\ut\nu}}\, \underline{e}_{{\ut\nu}}{}^{\underline\alpha}\,, \nn
\eea
where $\overline{e}_{ {\tilde\mu}}{}^{\overline\alpha}$ and  $\underline{e}_{ {\ut\mu}}{}^{\underline\alpha}$ are independent $O(p,q)$ and $O(q',p')$ matrices respectively and  
\bea
{\Pi}_{ {\tilde\mu} {\tilde\nu}} = \eta_{ {\tilde\mu} {\tilde\nu}} - \Omega_{ {\tilde\mu}}{}^{{\ut\rho}}\,\Omega_{ {\tilde\nu}}{}_{{\ut\rho}} \ , \;\;\;\;\;\;\;\;\;\;\; {\Pi}_{ {\ut\mu}{\ut\nu}} = \eta_{ {\ut\mu}{\ut\nu}} - \Omega^{ {\tilde\rho}}{}_{ {\ut\mu}}\,\Omega_{ {\tilde\rho} {\ut\nu}} \ .
\eea
Note that in this parameterization the counting of dof exhausts the dim$(g) = 2 k^2 - k$, of which $k^2 - k$ are contained in $\overline e$ and $\underline e$, and the other $k^2$ in $\Omega$.

Due to the original $\cal H$ symmetry, there are many non-physical gauge dof. It will then turn out to be convenient to perform a gauge fixing to remove some of them
\be
E^{M}{}_{\overline a}\, {\cal A}_{M}{}^{ {\tilde\mu}} = 0\,,\;\;\;\;E^{M}{}_{\underline a}\, {\cal A}_{M}{}^{ {\ut\mu}} = 0\,,\;\;\;\;\overline{e}_{ {\tilde\mu}}{}^{\overline\alpha}={\rm constant}\,,\;\;\;\;\; \underline{e}_{ {\ut\mu}}{}^{\underline\alpha}={\rm constant}\,.\label{GF}
\ee
Demanding that these constraints are gauge invariant $\delta {\cal E}_{ {\tilde\mu}}{}^{\overline a} = \delta {\cal E}_{ {\ut\mu}}{}^{\underline a} = \delta\overline{e}_{ {\tilde\mu}}{}^{\overline\alpha} = \delta\underline{e}_{ {\ut\mu}}{}^{\underline\alpha}=0$ freezes the following components of the ${\cal H}$ parameters
\bea
\Gamma_{\underline {\alpha a}} &=& \underline{e}^{ {\ut\mu}}{}_{\underline \alpha} \, ({\Sigma}^{-\frac12}){}_{ {\ut\mu}}{}^{{\ut\nu}}\, {\cal E}^P{}_{\underline a}\, \partial_P \xi_{{\ut\nu}}\,, \nn \\
\Gamma_{\overline {\alpha a}} &=& \overline{e}^{ {\tilde\mu}}{}_{\overline \alpha} \, ({\Sigma}^{-\frac12}){}_{ {\tilde\mu}}{}^{ {\tilde\nu}}\, {\cal E}^P{}_{\overline a}\, \partial_P \xi_{ {\tilde\nu}}\,, \label{GaugedFixedParameters}\\
\Gamma_{\underline {\alpha \beta}} &=&  \underline{e}^{ {\ut\mu}}{}_{[\underline \alpha} \, \underline{e}^{{\ut\nu}}{}_{\underline \beta]} \, ({\Sigma}^{-\frac12}){}_{ {\ut\mu}}{}^{{\ut\rho}} \left(\delta ({\Sigma}^{\frac12}){}_{{\ut\rho}{\ut\nu}} - ({\Pi}^{\frac12})_{{\ut\nu}}{}^{{\ut\sigma}}\, {\cal A}^{M}{}_{{\ut\sigma}} \partial_M \xi_{{\ut\rho}}\, - g_1\, f_{{\ut\rho} {\ut\sigma}}{}^{{\ut\lambda}} \, \xi^{{\ut\sigma}} \,({\Sigma}^{\frac12}){}_{{\ut\lambda} {\ut\nu}} \right)\ , \nn\\
\Gamma_{\overline {\alpha \beta}} &=&  \overline{e}^{ {\tilde\mu}}{}_{[\overline \alpha} \, \overline{e}^{ {\tilde\nu}}{}_{\overline \beta]} \, ({\Sigma}^{-\frac12}){}_{ {\tilde\mu}}{}^{ {\tilde\rho}} \left(\delta ({\Sigma}^{\frac12}){}_{ {\tilde\rho} {\tilde\nu}} - ({\Pi}^{\frac12})_{ {\tilde\nu}}{}^{ {\tilde\sigma}}\, {\cal A}^{M}{}_{ {\tilde\sigma}} \partial_M \xi_{ {\tilde\rho}}\, - g_2\, f_{ {\tilde\rho}  {\tilde\sigma}}{}^{ {\tilde\lambda}} \, \xi^{ {\tilde\sigma}} \,({\Sigma}^{\frac12}){}_{ {\tilde\lambda}  {\tilde\nu}} \right)\,,\nn
\eea
where we have introduced $ \Sigma^{\frac12}=\Box^{\frac12}\cdot{\Pi}^{\frac12}$.

\subsection{The perturbative expansion}\label{SecPerturbative}

The generalized transformations in the extended setup (\ref{TransformationExtended}), the proposed parameterization in terms of $G$-covariant components (\ref{FrameParameterizationA}) and the generalized BdR identification (\ref{LockingBi}) lead to an exact, yet implicit, double Lorentz $H$-transformation for the double generalized frame
\begin{eqnarray}
\delta E_M{}^{\underline a} &=& E_M{}^{\underline b} {\Lambda}_{\underline b}{}^{\underline a} + \frac{1}{g_1^2 X_{R_1}} E_{\overline M}{}^{\overline b} (\chi^{-\frac12})_{\overline b}{}^{\overline c} {\cal F}_{\overline c \underline{\cal CD}} D^{\underline a}\Gamma^{\underline{\cal CD}}+\frac{1}{g_2^2 X_{R_2}} \partial_{\overline M}\Gamma^{\overline{\cal CD}} (\chi^{-\frac12})^{\underline a}{}_{\underline b} {\cal F}^{\underline b}{}_{\overline{\cal C D}}\; ,\label{exactGGS} \\
\delta E_M{}^{\overline a} &=& E_M{}^{\overline b} {\Lambda}_{\overline b}{}^{\overline a}- \frac{1}{g_2^2 X_{R_2}} E_{\underline M}{}^{\underline b} (\chi^{-\frac12})_{\underline b}{}^{\underline c} {\cal F}_{\underline c \overline{\cal CD}} D^{\overline a}\Gamma^{\overline{\cal CD}} -\frac{1}{g_1^2 X_{R_1}} \partial_{\underline M}\Gamma^{\underline{\cal CD}} (\chi^{-\frac12})^{\overline a}{}_{\overline b} {\cal F}^{\overline b}{}_{\underline{\cal C D}}\; , \ \ \ \ \ \ \ \ \ \ \ \  \nn
\end{eqnarray}
where we have redefined 
\begin{eqnarray}
 {\Lambda}_{\underline a\underline b}&=&(\chi^{\frac12})_{\underline a}{}^{\underline c}\,\Gamma_{\underline c}{}^{\underline d}\, (\chi^{-\frac12})_{\underline {d b}} + (\chi^{\frac12})_{\underline a}{}^{\underline c}\, \delta (\chi^{-\frac12})_{\underline{c b}}-{\cal E}_{ {\tilde\mu}\underline a} e^{ {\tilde\mu}\underline\gamma} \Gamma_{\underline\gamma}{}^{\underline c}(\chi^{-\frac12})_{\underline c \underline b} + D_{\underline a} \xi^{ {\tilde\mu}} {\cal E}_{ {\tilde\mu}}{}^{\underline c} (\chi^{-\frac12})_{\underline{c b}}\;,\label{Lambda} \\
  {\Lambda}_{\overline {a b}}&=& (\chi^{\frac12})_{\overline a}{}^{\overline c}\,\Gamma_{\overline c}{}^{\overline d}\, (\chi^{-\frac12})_{\overline {d b}} + (\chi^{\frac12})_{\overline a}{}^{\overline c}\, \delta (\chi^{-\frac12})_{\overline{c b}}-{\cal E}_{ {\ut\mu}\overline a} e^{ {\ut\mu}\overline\gamma} \Gamma_{\overline\gamma}{}^{\overline c}(\chi^{-\frac12})_{\overline {c b}} + D_{\overline a} \xi^{ {\ut\mu}} {\cal E}_{ {\ut\mu}}{}^{\overline c} (\chi^{-\frac12})_{\overline{c b}}  \ . \ \ \ \ \ \ \ \  \nn
\end{eqnarray}
The transformation (\ref{exactGGS}) hides an infinite expansion in terms of $G$-covariant fields and parameters, named the {\it generalized Green-Schwarz transformation}. To compute this transformation perturbatively in integer powers of $g_1^{-2}$ and $g_2^{-2}$ requires taking into account the following four actions:
\begin{enumerate}
    \item First one should perform an $\underline h \times \underline H$ decomposition of $\underline {\cal H}$ by splitting indices  $\underline {\cal A} = (\underline a, \underline \alpha)$, and an $\overline h \times \overline H$ decomposition of $\overline {\cal H}$ by splitting indices $\overline {\cal A} = (\overline a, \overline \alpha)$.
    \item Identify the following components of the extended Lorentz parameters $\Gamma_{\overline {a \beta}}$, $\Gamma_{\overline {\alpha \beta}}$, $\Gamma_{\underline {a \beta}}$, $\Gamma_{\underline {\alpha \beta}}$, and replace them by the gauge fixing conditions (\ref{GaugedFixedParameters}). 
    \item Redefine the $\underline H \times \overline H$ components of the Lorentz parameters  $\Gamma_{\underline {a b}} \to \Lambda_{\underline {ab}}$ and $\Gamma_{\overline {a b}} \to \Lambda_{\overline {ab}}$, through (\ref{Lambda}).
    \item  Rewrite the components of fluxes in the extended space $\cal F$ (\ref{ExtendedFluxes}), in terms of the fluxes in the double space $F$ (\ref{DoubleFluxes}) and  the internal components of the extended generalized frame. We write these expressions explicitly in the Appendix (\ref{F1abcGF})-(\ref{F1alpbetgamGF}) in order to lighten the notation here.
\end{enumerate}
 Up to four derivatives one gets the following transformation for the $\overline H$ projection of the double generalized frame
{\small \begin{eqnarray}
\delta E_M{}^{\overline a} &=& E_M{}^{\overline b} {\Lambda}_{\overline b}{}^{\overline a}-\frac{1}{g_2^2}\frac{1}{(-1+X_{R_2})}\; F_{\underline M \overline{c d}} \;D^{\overline a} {\Lambda}^{\overline{c d}} 
-\frac{1}{g_1^2}\frac{1}{(-1+X_{R_1})} \partial_{\underline M}{\Lambda}^{\underline{cd}}\, F^{\overline a}{}_{\underline{cd}}\\ &&+\;\frac{2}{g_2^4\,X_{R_2}(-1+X_{R_2})} \,E_{\underline M}{}^{\underline b}\, \left[ \;{\cal F}_{\underline b}{}_{\overline{\cal E F}} {\cal F}_{\underline c}{}_{\overline{\cal G}}{}^{\overline{\cal F}}\left(
{\cal F}^{\underline c \overline{\cal C G}} D^{\overline a} \Gamma_{\overline{\cal C}}{}^{\overline{\cal E} }
 -  {\cal F}^{\underline c \overline{\cal C E}} D^{\overline a}\Gamma_{\overline{\cal C}}{}^{\overline{\cal G} }\right)
\right.\nn\\
&&\;\;\;\;-\;\left(D^{\overline a}\Gamma^{\overline{\cal EG}}\right) \left(F_{\underline{ b  c d}}{\cal F}^{\underline c}{}_{\overline{\cal E F}} {\cal F}^{\underline d}{}_{\overline{\cal G}}{}^{\overline{\cal F}}
+ D_{\underline b}{\cal F}^{\underline c}{}_{\overline{\cal  EF}} {\cal F}_{\underline c}{}_{\overline{\cal G}}{}^{\overline{\cal F}}  - 2\; D_{\underline c}  {\cal F}_{\underline b}{}_{\overline{\cal E F}}
{\cal F}^{\underline c}{}_{\overline{\cal G}}{}^{\overline{\cal F}}   \right)\nn\\
 && \;\;\;\; -\left.\,{\cal F}_{\underline  b\overline{\cal E F} }D^{\overline a} \left(D^{\underline c}\Gamma^{\overline{\cal E G}}{\cal F}_{\underline c \overline{\cal G}}{}^{\overline{\cal F}} \right)-\; D^{\overline a} D^{\overline c}  \Gamma^{\overline{\cal E F}}\left(F_{\overline c\underline{d b}} {\cal F}^{\underline d}{}_{\overline{\cal  EF}} + D_{\overline{c}} {\cal F}_{\underline b}{}_{\overline{\cal  EF}}         \right) \right] \nn\\&&
 -\frac{1}{g_2^2}\frac{1}{(-1+X_{R_2})}\;\frac{1}{g_1^2}\frac{1}{X_{R_1}}\;\left[
 F_{\underline M\overline{ef}}\, D^{\overline a}\left(D^{\overline e}\Gamma^{\underline{\cal{CD}}}\; {\cal F}^{\overline f}{}_{\underline{\cal{CD}}}\right)
\right.\nn\\&&
\;\;\;\;\;\;\;\;\;\;\;\;\;\;\;\;\;\;\;\;\;\;\;\;\;\;\;\;\;\; \left.+ \left( F_{\underline{M}\overline{eh}}\,{\cal F}^{\overline h\underline{\cal{CD}}} \,{\cal F}_{\overline f\underline{\cal{CD}}}- \partial_{\underline M} {\cal F}_{\overline e}{}^{\underline{\cal{CD}}}\;  {\cal F}_{\overline f\underline{\cal{CD}}}\right) D^{\overline a}{\Lambda}^{\overline{ef}}\right]\nn \\&&
+ \frac{1}{g_1^2}\frac{1}{(-1+X_{R_1})}\;\frac{1}{g_2^2}\frac{1}{X_{R_2}}\;\left[
\, \partial_{\underline M}\left( D^{\underline c}\Gamma^{\overline{\cal{CD}}}\, {\cal F}^{\underline d}{}_{\overline{\cal{CD}}}\right)F^{\overline a}{}_{\underline{cd}}\right. \nn\\&&
\;\;\;\;\;\;\;\;\;\;\;\;\;\;\;\;\;\;\; \;\;\;\;\;\;\;\;\;\;\left. \,+\, \partial_{\underline M}{\Lambda}^{\underline{cd}}\left(\,F^{\overline a}{}_{\underline{ce}}
 {\cal F}^{\underline e\overline{\cal{CD}}} {\cal F}_{\underline d\overline{\cal{CD}}}
 -D^{\overline a}{\cal F}_{\underline c}{}^{\overline{\cal{CD}}} {\cal F}_{\underline d\overline{\cal{CD}}}\right)
  \right] 
 \nn \\&&
 +\frac{1}{g_1^4}\frac{2}{X_{R_1}(-1+X_{R_1})} \left[ \partial_{\underline M}\left(D^{\underline c}\Gamma^{\underline{\cal CD}}\right)\left( F_{\underline c \,\overline b}{}^{\overline a}{\cal F}^{\overline b}{}_{\underline{\cal{CD}}} + D_{\underline c}{\cal F}^{\overline a}{}_{\underline{\cal CD}} \right)\right. \nn \\ &&
 +\left(\partial_{\underline M}\Gamma^{\underline{\cal CE}}\right)\left(
 2\,{\cal F}_{\overline c}{}^{\underline{\cal{D}}}{}_{\underline{\cal{E}}} {\cal F}^{\overline a}{}_{[\underline{\cal D}}{}^{\underline{\cal{F}}} \, {\cal F}^{\overline c}{}_{\underline{\cal C}] \underline{\cal{F}}}\,+ 
 F^{\overline a}{}_{\overline{cd}}\, {\cal F}^{\overline d}{}_{\underline{\cal CD}}\,
  {\cal F}^{\overline c}{}^{\underline{\cal D}}{}_{\underline{\cal{E}}}
  \,+\,{\cal F}^{\overline c}{}_{\underline{\cal CD}}\, D^{\overline a}{\cal F}_{\overline c}{}^{\underline{\cal D}}{}_{\underline{\cal E}}\right.\nn\\&&
\;\;\;\;\;\;\;\;\;\;\;\;\;\;\;\;\;\;\left. \left.\,+\, 2\,\vphantom{1^{\frac12}}
 D^{\overline c}{\cal F}^{\overline a}{}_{\underline{\cal CD}}\,
 {\cal F}_{\overline c}{}^{\underline{\cal D}}{}_{\underline{\cal{ E}}}\
 \right)\,+\,
 \partial_{\underline M}\left( D^{\overline{c}}\Gamma^{\underline{\cal{CD}}} {\cal F}_{\overline{c} \underline{\cal{CE}}}\right)\,{\cal F}^{\overline a\underline{\cal E}}{}_{\underline{\cal D}}\;\right]
 \;\;+\, {\cal O}\left(g_i^{-6}\right) \ .\nn
\end{eqnarray}}
The factors $(-1+X_i)^{-1}$ are a consequence of the cyclic relations (explained at the end of Section \ref{sectionMono}) necessary to cast the $h$-covariant contractions in terms of $H$-covariant ones. Note that there are no $h$-covariant indices $\underline \alpha$ and $\overline \alpha$ in this expression. The two derivative part of the Lorentz transformations in the first line above is fully expressed in terms of the $H$-covariant indices of the double space. The higher derivative terms of order $g_i^{-4}$  are written in terms of the extended fields and parameters. We must then repeat these steps once again for these terms in order to get the complete four derivative transformations (we drop here all contributions of order $g_i^{-6}$ and higher) 

{\small \bea
\boxed{\begin{aligned}
& \delta E_M{}^{\overline a} = E_M{}^{\overline b} {\Lambda}_{\overline b}{}^{\overline a}-\;\frac a 2\, \partial_{\underline M}{\Lambda}^{\underline{cd}}\, F^{\overline a}{}_{\underline{cd}}- \;\frac b 2 \, F_{\underline M \overline{c d}} \;D^{\overline a} {\Lambda}^{\overline{c d}} 
\\ & \ \ \ \ \ \ \  -\, \frac {b^2} 2 \; E_M{}^{\underline b}\left[
														D^{\overline{a}}D^{\overline c}{\Lambda}^{\overline{e f}}\left(F_{\overline c\underline{d b}} F^{\underline d}{}_{\overline{e f}} + D_{\overline c} F_{\underline b \overline{e f}}\right)	
														\right.
-F_{\underline b \overline{e f}} F_{\underline c \overline d}{}^{\overline f}\left(
F^{\underline c}{}^{\overline{h d}} D^{\overline a}{\Lambda}_{\overline h}{}^{\overline e}
-F^{\underline c}{}^{\overline{h e}} D^{\overline a}{\Lambda}_{\overline h}{}^{\overline d}
\right)\nonumber \\
&\ \ \ \ \ \ \ \ \ \ \ \ \ \ \  +\;\left. F^{\underline c}{}_{\overline{e f}}\;D^{\overline a}{\Lambda}^{\overline{e}}{}_{\overline g} \left(F_{\underline {b c d}} F^{\underline d \overline{g f}}
-D_{\underline b} F_{\underline c}{}^{\overline{g f}} + 2\; D_{\underline c} F_{\underline b}{}^{ \overline{g f}}
\right)
+F_{\underline b \overline{e f}}D^{\overline a}\left(D^{\underline c}{\Lambda}^{\overline{e d}} F_{\underline c \overline{d}}{}^{\overline f}\right)					
							\right]\nonumber\\& \ \ \ \ \ \ \ 
 -\frac {a\,b} 4 \; E_M{}^{\underline b}\,\left[D^{\overline a}{\Lambda}^{\overline{ef}}
 \left( F_{\underline{b}\overline{eh}}\,F^{\overline h\underline{cd}} \,F_{\overline f\underline{cd}}- D_{\underline b} F_{\overline e}{}^{\underline{cd}}\;  F_{\overline f\underline{cd}}\right) 
+F_{\underline b\overline{ef}}\, D^{\overline a}\left(D^{\overline e}{\Lambda}^{\underline{cd}}\; F^{\overline f}{}_{\underline{cd}}\right) \right.\nn\\&
\;\;\;\;\;\;\;\;\;\;\;\;\;\;\;\;\;\;
\left.-\, D_{\underline b}{\Lambda}^{\underline{ef}}\left(F^{\overline a}{}_{\underline{eh}}
 F^{\underline h\overline{cd}} F_{\underline f\overline{cd}}
 -D^{\overline a}F_{\underline e}{}^{\overline{cd}} F_{\underline f\overline{cd}}\right) -\,F^{\overline a}{}_{\underline{ef}}\, D_{\underline b}\left( D^{\underline e}{\Lambda}^{\overline{cd}}\, F^{\underline f}{}_{\overline{cd}}\right)\,\right]\nn \\&
 \ \ \ \ \ \ \  +\frac {a^2} 2 \; E_M{}^{\underline b}\left[
														D_{\underline{b}}D^{\underline c}{\Lambda}^{\underline{e f}}\left(F_{\underline c\overline{d}}{}^{\overline{a}} F^{\overline d}{}_{\underline{e f}} + D_{\underline c} F^{\overline a}{}_{\underline{e f}}\right)	
														\right.
-F^{\overline{a}}{}_{\underline{e f}} F_{\overline c \underline d}{}^{\underline f}\left(
F^{\overline c}{}^{\underline{h d}} D_{\underline{b}}{\Lambda}_{\underline h}{}^{\underline e}
-F^{\overline c}{}^{\underline{h e}} D_{\underline{b}}{\Lambda}_{\underline h}{}^{\underline d}
\right)\nonumber \\
&\ \ \ \ \ \ \ \ \ \ \ \ \ \ \  +\;\left. F^{\overline c}{}_{\underline{e f}}\;D_{\underline{b}}{\Lambda}^{\underline{e}}{}_{\underline g} \left(F^{\overline a}{}_{\overline{c d}} F^{\overline d \underline{g f}}
-D^{\overline a} F_{\overline c}{}^{\underline{g f}} + 2\; D_{\overline c} F^{ \overline{a}\underline{g f}}
\right)
+F^{\overline a}{}_{\underline{e f}}D_{\underline{b}}\left(D^{\overline c}{\Lambda}^{\underline{e d}} F_{\overline c \underline{d}}{}^{\underline f}\right)						
							\right]\ . \nn 
\end{aligned}} 
\eea} 
\be \label{VarEup} \ee
We see once again that $g_i$ and $X_{R_i}$ arrange themselves into the combination
\begin{eqnarray}
a=\frac{1}{g_1^2}\frac{2}{(-1+X_{R_1})}\;,\;\;\;\;\;\;\;\;\;\;\;b=\frac{1}{g_2^2}\frac{2}{(-1+X_{R_2})}\ .
\end{eqnarray}
Repeating  the computations analogously for the transformation of the $\underline H$ projection of the double generalized frame leads to
{\small \bea
\boxed{\begin{aligned}
& \delta E_M{}^{\underline a} = E_M{}^{\underline b} {\Lambda}_{\underline b}{}^{\underline a}+\;\frac b 2\, \partial_{\overline M}{\Lambda}^{\overline{cd}}\, F^{\underline a}{}_{\overline{cd}}+\;\frac a 2\, F_{\overline M \underline{c d}} \;D^{\underline a} {\Lambda}^{\underline{c d}} 
\\ & \ \ \ \ \ \ \ -\, \frac {a^2} 2\;  E_M{}^{\overline b}\left[
														D^{\underline{a}}D^{\underline c}{\Lambda}^{\underline{e f}}\left(F_{\underline c\overline{d b}} F^{\overline d}{}_{\underline{e f}} + D_{\underline c} F_{\overline b \underline{e f}}\right)	
														\right.
-F_{\overline b \underline{e f}} F_{\overline c \underline d}{}^{\underline f}\left(
F^{\overline c}{}^{\underline{h d}} D^{\underline a}{\Lambda}_{\underline h}{}^{\underline e}
-F^{\overline c}{}^{\underline{h e}} D^{\underline a}{\Lambda}_{\underline h}{}^{\underline d}
\right)\nonumber \\
&\ \ \ \ \ \ \ \ \ \ \ \ \ \ \  +\;\left. F^{\overline c}{}_{\underline{e f}}\;D^{\underline a}{\Lambda}^{\underline{e}}{}_{\underline g} \left(F_{\overline {b c d}} F^{\overline d \underline{g f}}
-D_{\overline b} F_{\overline c}{}^{\underline{g f}} + 2\; D_{\overline c} F_{\overline b}{}^{ \underline{g f}}
\right)
+F_{\overline b \underline{e f}}D^{\underline a}\left(D^{\overline c}{\Lambda}^{\underline{e d}} F_{\overline c \underline{d}}{}^{\underline f}\right)						
							\right]\nonumber\\& \ \ \ \ \ \ \ 
 -\frac {a \, b}4\; E_M{}^{\overline b}\,\left[D^{\underline a}{\Lambda}^{\underline{ef}}
 \left( F_{\overline{b}\underline{eh}}\,F^{\underline h\overline{cd}} \,F_{\underline f\overline{cd}}- D_{\overline b} F_{\underline e}{}^{\overline{cd}}\;  F_{\underline f\overline{cd}}\right) 
+F_{\overline b\underline{ef}}\, D^{\underline a}\left(D^{\underline e}{\Lambda}^{\overline{cd}}\; F^{\underline f}{}_{\overline{cd}}\right) \right.\nn\\&
\;\;\;\;\;\;\;\;\;\;\;\;\;\;\;\;\;\;
\left.-\, D_{\overline b}{\Lambda}^{\overline{ef}}\left(F^{\underline a}{}_{\overline{eh}}
 F^{\overline h\underline{cd}} F_{\overline f\underline{cd}}
 -D^{\underline a}F_{\overline e}{}^{\underline{cd}} F_{\overline f\underline{cd}}\right) -\,F^{\underline a}{}_{\overline{ef}}\, D_{\overline b}\left( D^{\overline e}{\Lambda}^{\underline{cd}}\, F^{\overline f}{}_{\underline{cd}}\right)\,\right]\nn \\&
 \ \ \ \ \ \ \ +\frac {b^2} 2\;  E_M{}^{\overline b}\left[
D_{\overline{b}}D^{\overline c}{\Lambda}^{\overline{e f}}\left(F_{\overline c\underline d}{}^{\underline a} F^{\underline d}{}_{\overline{e f}} +
D_{\overline c} F^{\underline a}{}_{\overline{e f}}\right)	\right.
-F^{\underline a}{}_{\overline{e f}} F_{\underline c \overline d}{}^{\overline f}\left(
F^{\underline c}{}^{\overline{h d}} D_{\overline b}{\Lambda}_{\overline h}{}^{\overline e}
-F^{\underline c}{}^{\overline{h e}} D_{\overline b}{\Lambda}_{\overline h}{}^{\overline d}
\right)\nonumber \\
&\ \ \ \ \ \ \ \ \ \ \ \ \ + \left. F^{\underline c}{}_{\overline{e f}}\;D_{\overline b}{\Lambda}^{\overline{e}}{}_{\overline g} \left(F^{\underline a}{}_{\underline{c d}} F^{\underline d \overline{g f}}
-D^{\underline a} F_{\underline c}{}^{\overline{g f}} + 2\; D_{\underline c} F^{\underline a  \overline{g f}}\right)
+F^{\underline a}{}_{\overline{e f}}D_{\overline b}\left(D^{\underline c}{\Lambda}^{\overline{e d}} F_{\underline c \overline{d}}{}^{\overline f}\right)\right]\ .\nn
\end{aligned}} 
\eea} 
\be \label{VarEdown} \ee
It can be checked that these transformations preserve the $G$-valuedness of the double generalized frame, and also close together with the generalized diffeomorphisms into transformations produced by the following corrected brackets\footnote{\label{typo}We thank F. Hassler and A. Gitsis for identifying a typo in Equation (\ref{brackets}).}
{\small \bea
\xi_{12}^M&=& 2 \xi_{[1}^{P} \partial_{P} \xi_{2]}^{M} + \partial^{M} \xi_{[1}^{P} \xi_{2]{P}}
+ \frac{a}{2}\; {\Lambda}_{[1}^{\underline{cd}} \; \partial^M {\Lambda}_{2]}{}_{\underline{cd}}
- \frac{b}{2}\; {\Lambda}_{[1}^{\overline{cd}} \; \partial^M {\Lambda}_{2]}{}_{\overline{cd}} \label{brackets}\\&& 
- \frac{a\,b}{2}\,\left[ \partial^M{\Lambda}_{[1}{}^{\underline{ef}}\; D_{\underline{f}}{\Lambda}_{2]}{}_{\overline{cd}}  \;F_{\underline e}{}^{\overline{cd}} 
+ \partial^M{\Lambda}_{[1}{}^{\overline{ef}}\; D_{\overline{f}}{\Lambda}_{2]}{}_{\underline{cd}}  \;F_{\overline e}{}^{\underline{cd}}\right] \nn \\&&
 +\;a^2\left[ \partial^M{\Lambda}_{[1}^{\underline{ef}}\; D^{\overline c}{\Lambda}_{2]\underline e}{}^{\underline d} F_{\overline c\underline{df}}\;+\;\frac12 \partial^M\left(D^{\underline c}{\Lambda}_{[1}^{\underline{ef}}\right)D_{\underline{c}} {\Lambda}_{2]}{}_{\underline{ef}}   \right] \nn\\&&
 +\;b^2\left[ \partial^M{\Lambda}_{[1}^{\overline{ef}}\; D^{\underline c}{\Lambda}_{2]\overline e}{}^{\overline d} F_{\underline c\overline{df}}\;+\;\frac12 \partial^M\left(D^{\overline c}{\Lambda}_{[1}^{\overline{ef}}\right)D_{\overline{c}} {\Lambda}_{2]}{}_{\overline{ef}}   \right]\ , \nn \\&&\nn \\
{\Lambda}_{12}^{\overline {ab}}&=& 2\;\xi_{[1}^N\partial_N{\Lambda}_{2]}^{\overline{ab}}\,-\,2\;{\Lambda}_{[1}^{\overline{ac}}\,{\Lambda}_{2]}{}_{\overline c}{}^{\overline b}
- \,a\; D^{\overline a}{\Lambda}_{[1}^{\underline{cd}}  \;D^{\overline b}{\Lambda}_{2]}{}_{\underline{cd}}
+ \,b\; D^{\overline a}{\Lambda}_{[1}^{\overline{cd}}  \;D^{\overline b}{\Lambda}_{2]}{}_{\overline{cd}}\nn \\&&
+\,a\,b\,\left[F^{\underline g}{}_{\overline{cd}}\; D^{[\overline a}{\Lambda}_{[1}{}^{\overline{cd}}\;
F^{\overline b]}{}_{\underline{ef}}\; D_{\underline g}{\Lambda}_{2]}^{\underline{ef}}\;
-\; D^{[\overline a}{\Lambda}_{[1}{}_{\underline{ef}}\;
D^{\overline b]}\left(F^{\underline{e}}{}_{\overline{cd}}D^{\underline f}{\Lambda}_{2]}{}^{\overline{cd}}
\;\right)\right.\nonumber\cr&&
\left.\;\;\;\;\;\;\;\;\;\;\;\;\;\;\;\;\;\;\;\;\;\;\;\;\;\;\;\;\;\;\;\;\;\;\;\;\;\;\;\;\;\;\;\;\;\;\;\;\;\;\;\;\;\;\;\;\;\;\;-\; 
D^{[\overline a}{\Lambda}_{[1}{}_{\overline{ef}}\;
D^{\overline b]}\left(F^{\overline{e}}{}_{\underline{cd}}D^{\overline f}{\Lambda}_{2]}{}^{\underline{cd}}
\;\right)\right] \nn \\&& 
\;+\;  a^2 \left[ \frac12 D^{\underline c}{\Lambda}_{[1}^{\underline{ef}}\;  D_{\underline c}{\Lambda}_{2]}^{\underline{gh}} F^{\overline a}{}_{\underline{ef}}\; F^{\overline b}{}_{\underline {gh}}  
+ D^{\overline a} D^{\underline e}{\Lambda}_{[1}^{\underline{cd}}\;  D^{\overline b} D_{\underline e}{\Lambda}_{2]}{}_{\underline{cd}}
+\,2\;  D^{[\overline a}\left(D^{\overline c}{\Lambda}_{[1}^{\underline{ed}}\; F_{\overline c\underline d}{}^{\underline f} \right)\;  D^{\overline b]}{\Lambda}_{2]}{}_{\underline{ef}} \right]\nn \\&& 
\;+\;  b^2 \left[\frac12 D^{\overline a}{\Lambda}_{[1}^{\overline{cd}}\;  D^{\overline b}{\Lambda}_{2]}^{\overline{ef}} F^{\underline g}{}_{\overline{cd}}\; F_{\underline g\overline {ef}} 
 + D^{\overline a} D^{\overline e}{\Lambda}_{[1}^{\overline{cd}}\;  D^{\overline b} D_{\overline e}{\Lambda}_{2]}{}_{\overline{cd}}
+\,2\;  D^{[\overline a}\left(D^{\underline c}{\Lambda}_{[1}^{\overline{ed}}\; F_{\underline c\overline d}{}^{\overline f} \right)\;  D^{\overline b]}{\Lambda}_{2]}{}_{\overline{ef}} \right] \ ,\nn \\ \nn \\
{\Lambda}_{12}^{\underline {ab}}&=& 2\;\xi_{[1}^N\partial_N{\Lambda}_{2]}^{\underline{ab}}\,-\,2\;{\Lambda}_{[1}^{\underline{ac}}\,{\Lambda}_{2]}{}_{\underline c}{}^{\underline b}
- \,a\; D^{\underline a}{\Lambda}_{[1}^{\underline{cd}}  \;D^{\underline b}{\Lambda}_{2]}{}_{\underline{cd}}
+ \,b\; D^{\underline a}{\Lambda}_{[1}^{\overline{cd}}  \;D^{\underline b}{\Lambda}_{2]}{}_{\overline{cd}}\nn \\ &&+\,a\,b\,\left[F^{\overline g}{}_{\underline{cd}}\; D^{[\underline a}{\Lambda}_{[1}{}^{\underline{cd}}\;
F^{\underline b]}{}_{\overline{ef}}\; D_{\overline g}{\Lambda}_{2]}^{\overline{ef}}\;
-\; D^{[\underline a}{\Lambda}_{[1}{}_{\overline{ef}}\;
D^{\underline b]}\left(F^{\overline{e}}{}_{\underline{cd}}D^{\overline f}{\Lambda}_{2]}{}^{\underline{cd}}
\;\right)\right.\nonumber\cr&&
\left.\;\;\;\;\;\;\;\;\;\;\;\;\;\;\;\;\;\;\;\;\;\;\;\;\;\;\;\;\;\;\;\;\;\;\;\;\;\;\;\;\;\;\;\;\;\;\;\;\;\;\;\;\;\;\;\;\;\;\;-\; 
D^{[\underline a}{\Lambda}_{[1}{}_{\underline{ef}}\;
D^{\underline b]}\left(F^{\underline{e}}{}_{\overline{cd}}D^{\underline f}{\Lambda}_{2]}{}^{\overline{cd}}
\;\right)\right] \nn \\&& 
\;+\;  a^2 \left[\frac12  D^{\underline a}{\Lambda}_{[1}^{\underline{cd}}\;  D^{\underline b}{\Lambda}_{2]}^{\underline{ef}} F^{ \overline g}{}_{\underline{cd}}\; F_{\overline g \underline {ef}}
+ D^{\underline a} D^{\underline e}{\Lambda}_{[1}^{\underline{cd}}\;  D^{\underline b} D_{\underline e}{\Lambda}_{2]}{}_{\underline{cd}}
+\,2\;  D^{[\underline a}\left(D^{\overline c}{\Lambda}_{[1}^{\underline{ed}}\; F_{\overline c\underline d}{}^{\underline f} \right)\;  D^{\underline b]}{\Lambda}_{2]}{}_{\underline{ef}} \right]\nn \\ &&
\;+\;  b^2 \left[ \frac12 D^{\overline c}{\Lambda}_{[1}^{\overline{ef}}\;  D_{\overline c}{\Lambda}_{2]}^{\overline{gh}} F^{\underline a}{}_{\overline{ef}}\; F^{\underline b}{}_{\overline {gh}}
+ D^{\underline a} D^{\overline e}{\Lambda}_{[1}^{\overline{cd}}\;  D^{\underline b} D_{\overline e}{\Lambda}_{2]}{}_{\overline{cd}}
+\,2\;  D^{[\underline a}\left(D^{\underline c}{\Lambda}_{[1}^{\overline{ed}}\; F_{\underline c\overline d}{}^{\overline f} \right)\;  D^{\underline b]}{\Lambda}_{2]}{}_{\overline{ef}} \right]\ . \nn
\eea}

The transformations (\ref{VarEup},\ref{VarEdown}) are the second order corrections to the generalized Green-Schwarz transformation. The first order reproduces the results in \cite{MarquesNunez} and the second order in the mono-parametric case reproduces (\ref{Second1},\ref{Second2}) originally found in \cite{GenBdR}.

\subsection{The bi-parametric action to second order}\label{SectBi-parDFT}

In the previous sections we introduced an exact generalized BdR identification (\ref{LockingBi}), and used it to obtain second-order four-derivative corrections to the generalized Green-Schwarz transformations (\ref{VarEup}-\ref{VarEdown}). We now exploit this identification further to get the full invariant four and six-derivative couplings in the action of DFT.

The starting point is the standard two-derivative $\cal G$-covariant action in the extended space
\begin{eqnarray}
{\cal S}=\int\;dX\; e^{-2d}\;{\cal R}\;.\label{ExtendAction}
\end{eqnarray}
It is useful to decompose it as a sum
\begin{eqnarray}
{\cal R}={\cal R}_0+{\cal R}_1+{\cal R}_2 \ ,
\end{eqnarray}
where ${\cal R}_0$ is a constant, ${\cal R}_1$ contains vectorial generalized fluxes and therefore the generalized dilaton dependence
\begin{eqnarray}
{\cal R}_1=2\left(D^{\overline{\cal A}}{\cal F}_{\overline{\cal A}}-
D^{\underline{\cal A}}{\cal F}_{\underline{\cal A}}\right)-\left({\cal F}^{\overline{\cal A}}{\cal F}_{\overline{\cal A}}- {\cal F}^{\underline{\cal A}}{\cal F}_{\underline{\cal A}}\right)\;,
\end{eqnarray}
and ${\cal R}_2$ includes the three-form fluxes
\begin{eqnarray}
{\cal R}_2=\frac12\left({\cal F}_{\cal{\underline A}\overline{BC}}\;{\cal F}^{\cal{\underline A}\overline{BC}} -
{\cal F}_{\cal{\overline A}\underline{BC}}\;{\cal F}^{\cal{\overline A}\underline{BC}} \right)
+\frac16\left( {\cal F}_{\cal \overline{ABC}}\;{\cal F}^{\cal \overline{ABC}}-
 {\cal F}_{\cal \underline{ABC}}\;{\cal F}^{\cal \underline{ABC}}\right)\ .
\end{eqnarray}
The action written in this form is democratic with respect to overline and underline projections. Using Bianchi identities it can be taken to a simpler and equivalent form in which this symmetry is broken \cite{FrameDFT}. When the action is gauged either explicitly \cite{HeteroticDFT} or through a generalized Scherk-Schwarz reductions \cite{GSS, GaugedDFT}, in certain cases the Bianchi Identities that connect the two ways of writing the action fail con coincide, and the one that properly connects with  gauged supergravities is one in which the overline-underline symmetry is broken \cite{DualityOrbits}. Here we use the symmetric version because since we are interested in the bi-parametric case, where parameters $a$ and $b$ interpolate between the two projections, it is useful to preserve the symmetry between them.

The perturbative expansion follows from the same procedure discussed in Section \ref{SecPerturbative}, the relevant steps here being 1 and 4. One first has to perform an $\underline h \times \underline H$ decomposition of $\underline {\cal H}$ by splitting indices  $\underline {\cal A} = (\underline a, \underline \alpha)$, and an $\overline h \times \overline H$ decomposition of $\overline {\cal H}$ by splitting indices $\overline {\cal A} = (\overline a, \overline \alpha)$. Then rewrite the components of the fluxes in the extended space $\cal F$ (\ref{ExtendedFluxes}), in terms of the fluxes in the double space $F$ (\ref{DoubleFluxes}) and  the internal components of the extended generalized frame. The exact expressions for these can be found in (\ref{F1abcGF})-(\ref{Falpha}) in the Appendix. The outcome of such a procedure is a lengthy action, and so Cadabra software has been of great help \cite{Peeters:2007wn}. 

We found the following action
\begin{eqnarray}
{\cal S}=\int\;dX\; e^{-2d}\; \left( {\cal R}^{(-1)} +  {\cal R}^{(0)} + {\cal R}^{(1)} + {\cal R}^{(2)} \right)\;,\label{FullAction}
\end{eqnarray}
where 
\begin{eqnarray}
 {\cal R}^{(-1)} = {\cal R}_0 + \frac{k}{6}\left(g_2^2-g_1^2\right),
\end{eqnarray}
is an arbitrary constant because ${\cal R}_0$ is not fixed by duality, so we choose it to vanish. The two-derivative part is obviously the standard DFT generalized Ricci scalar 
\begin{eqnarray}
{\cal R}^{(0)}=2 D^{\overline{a}} F_{\overline{ a}} - F^{\overline{ a}} F_{\overline{ a}} + \frac12 F_{{\underline a}\overline{bc}}\; F^{{\underline a}\overline{bc}}
+\frac16 F_{ \overline{abc}}\; F^{ \overline{abc}} -  \left(\{\overline{a}, \overline{b}, \overline{c}, ... \}\leftrightarrow\{\underline{a}, \underline{b}, \underline{c}, ...\}\right)\; .
\end{eqnarray}
The first order decomposes as
\be
{\cal R}^{(1)} = a\; {\cal R}^{(0, 1)} + b\; {\cal R}^{(1, 0)}=  a \;{\cal R}^{(0,1)} +  \left(\begin{matrix} a\leftrightarrow b \cr \{\overline{a}, \overline{b}, \overline{c}, ... \}\leftrightarrow\{\underline{a}, \underline{b}, \underline{c}, ...\}\end{matrix}\right) \;,
\ee
with
\begin{eqnarray}
{\cal R}^{(0, 1)}&=&-F_{\overline{a}} F_{\overline{b}} F^{\overline{a} \underline{a} \underline{b}} F^{\overline{b}}{}_{\underline{a} \underline{b}} + 2D_{\overline{a}}F_{\overline{b}} F^{\overline{a} \underline{a} \underline{b}} F^{\overline{b}}{}_{\underline{a} \underline{b}} + 2D^{\overline{a}}F_{\overline{a}}{}^{\underline{a} \underline{b}} F_{\overline{b}} F^{\overline{b}}{}_{\underline{a} \underline{b}} + 2D_{\overline{a}}F^{\overline{b} \underline{a} \underline{b}} F_{\overline{b}} F^{\overline{a}}{}_{\underline{a} \underline{b}}
\cr&&
-  1/2 D_{\overline{a}}D_{\overline{b}}F^{\overline{a} \underline{a} \underline{b}} F^{\overline{b}}{}_{\underline{a} \underline{b}} - 3/2D^{\overline{a}}D_{\overline{b}}F^{\overline{b} \underline{a} \underline{b}} F_{\overline{a} \underline{a} \underline{b}} -  D^{\overline{a}}F_{\overline{a}}{}^{\underline{a} \underline{b}} D^{\overline{b}}F_{\overline{b} \underline{a} \underline{b}}
 +  1/2 D^{\underline{a}}F^{\overline{a} \underline{c} \underline{b}} D_{\underline{a}}F_{\overline{a} \underline{c} \underline{b}}  \cr&& -  1/2 D^{\overline{a}}F^{\overline{b} \underline{a} \underline{b}} D_{\overline{a}}F_{\overline{b} \underline{a} \underline{b}}  -  3/2 D^{\underline{a}}F^{\overline{a} \underline{c} \underline{b}} F^{\overline{b}}{}_{\underline{c} \underline{b}} F_{\overline{a} \overline{b} \underline{a}} -4D^{\overline{a}}F^{\overline{b} \underline{a} \underline{c}} F_{\overline{a} \underline{a}}{}^{\underline{b}} F_{\overline{b} \underline{c} \underline{b}}\cr&& 
 +  1/2 D^{\overline{a}}F^{\overline{c} \underline{a} \underline{b}} F^{\overline{b}}{}_{\underline{a} \underline{b}} F_{\overline{a} \overline{c} \overline{b}}  +  4/3 F_{\overline{a}}{}^{\underline{a} \underline{c}} F_{\overline{c} \underline{a}}{}^{\underline{b}} F_{\overline{b} \underline{c} \underline{b}} F^{\overline{a} \overline{c} \overline{b}}-F^{\overline{a} \underline{a} \underline{c}} F_{\overline{a} \underline{a}}{}^{\underline{d}} F^{\overline{b}}{}_{\underline{c}}{}^{\underline{b}} F_{\overline{b} \underline{d} \underline{b}}  \cr&&
 + F^{\overline{a} \underline{a} \underline{c}} F^{\overline{b}}{}_{\underline{a}}{}^{\underline{d}} F_{\overline{b} \underline{c}}{}^{\underline{b}} F_{\overline{a} \underline{d} \underline{b}} + F^{\overline{a} \underline{a} \underline{c}} F^{\overline{c}}{}_{\underline{a} \underline{c}} F_{\overline{a}}{}^{\overline{b} \underline{b}} F_{\overline{c} \overline{b} \underline{b}}\,.
\end{eqnarray}
It obviously coincides with the four-derivative action found in \cite{MarquesNunez}, and later rewritten in \cite{OddStory} in terms of generalized fluxes and flat derivatives.  

The previous actions were known, and now we present a new result: the six-derivative action. It decomposes as follows
\bea
{\cal R}^{(2)}&=& a^2 {\cal R}^{(0,2)}+ a b {\cal R}^{(1,1)} + b^2 {\cal R}^{(2,0)} \\
&=&  a^2 {\cal R}^{(0,2)} + a b \tilde{\cal R}^{(1,1)}-  \left(\begin{matrix} a\leftrightarrow b \cr \{\overline{a}, \overline{b}, \overline{c}, ... \}\leftrightarrow\{\underline{a}, \underline{b}, \underline{c}, ...\}\end{matrix}\right)\;.  \nn
\eea
In the last identity we cast the Lagrangian in a form that exhibits the symmetry with respect to the exchange of $a$ and $b$. It is convenient to split the contributions between those coming from ${\cal R}_1$ and ${\cal R}_2$, namely those with dilaton dependence and without.  We then write ${\cal R}^{(0,2)} = {\cal R}_{\Phi}^{(0,2)}+{\cal R}_{\not\Phi}^{(0,2)}$, where
{\footnotesize \begin{eqnarray}
 {\cal R}_{\Phi}^{(0,2)} &=&
-2D^{\underline{b}}F^{\overline{d} \underline{e} \underline{f}} D^{\overline{c}}D_{\underline{b}}F_{\overline{c} \underline{e} \underline{f}} F_{\overline{d}}-2D^{\underline{b}}F^{\overline{a} \underline{e} \underline{f}} D_{\overline{a}}D_{\underline{b}}F^{\overline{d}}{}_{\underline{e} \underline{f}} F_{\overline{d}}-2D^{\underline{b}}F^{\overline{a} \underline{e} \underline{f}} D_{\underline{b}}F^{\overline{d}}{}_{\underline{e} \underline{f}} D_{\overline{a}}F_{\overline{d}} \\ && 
-4D^{\overline{c}}D^{\overline{e}}F_{\overline{c}}{}^{\underline{c} \underline{e}} F^{\overline{d}} F_{\overline{d} \underline{c}}{}^{\underline{f}} F_{\overline{e} \underline{e} \underline{f}} + 2D^{\overline{b}}D_{\overline{b}}F^{\overline{e} \underline{c} \underline{e}} F^{\overline{d}} F_{\overline{d} \underline{c}}{}^{\underline{f}} F_{\overline{e} \underline{e} \underline{f}} + 2D^{\overline{b}}D^{\underline{f}}F_{\overline{b}}{}^{\underline{d} \underline{e}} F^{\overline{c}} F^{\overline{e}}{}_{\underline{d} \underline{e}} F_{\overline{c} \overline{e} \underline{f}}\cr&& 
-4D^{\overline{d}}F^{\overline{c}} D^{\overline{e}}F_{\overline{c}}{}^{\underline{c} \underline{e}} F_{\overline{d} \underline{c}}{}^{\underline{f}} F_{\overline{e} \underline{e} \underline{f}}-4D^{\overline{d}}F^{\overline{c}} D^{\overline{e}}F_{\overline{d}}{}^{\underline{c} \underline{e}} F_{\overline{c} \underline{c}}{}^{\underline{f}} F_{\overline{e} \underline{e} \underline{f}} + 2D^{\overline{b}}F^{\overline{c}} D_{\overline{b}}F^{\overline{e} \underline{c} \underline{e}} F_{\overline{c} \underline{c}}{}^{\underline{f}} F_{\overline{e} \underline{e} \underline{f}} \cr&& 
+ 2D^{\overline{d}}F^{\overline{b}} D_{\overline{b}}F^{\overline{e} \underline{c} \underline{e}} F_{\overline{d} \underline{c}}{}^{\underline{f}} F_{\overline{e} \underline{e} \underline{f}} + D^{\underline{b}}F^{\overline{c} \underline{e} \underline{f}} D_{\underline{b}}F^{\overline{d}}{}_{\underline{e} \underline{f}} F_{\overline{c}} F_{\overline{d}} + 2D^{\underline{f}}F^{\overline{a} \underline{d} \underline{e}} D_{\overline{a}}F^{\overline{c}} F^{\overline{e}}{}_{\underline{d} \underline{e}} F_{\overline{c} \overline{e} \underline{f}} \cr&& 
+ 2D^{\underline{f}}F^{\overline{a} \underline{d} \underline{e}} D_{\overline{a}}F^{\overline{e}}{}_{\underline{d} \underline{e}} F^{\overline{d}} F_{\overline{d} \overline{e} \underline{f}}-4D^{\overline{e}}D^{\overline{f}}F^{\overline{d} \underline{c} \underline{e}} F_{\overline{d}} F_{\overline{e} \underline{c}}{}^{\underline{f}} F_{\overline{f} \underline{e} \underline{f}} + 2D^{\overline{e}}D^{\overline{d}}F^{\overline{f} \underline{c} \underline{e}} F_{\overline{d}} F_{\overline{e} \underline{c}}{}^{\underline{f}} F_{\overline{f} \underline{e} \underline{f}} \cr&& 
+ 2D^{\overline{e}}D^{\underline{f}}F^{\overline{c} \underline{d} \underline{e}} F_{\overline{c}} F^{\overline{f}}{}_{\underline{d} \underline{e}} F_{\overline{e} \overline{f} \underline{f}}-2D^{\underline{f}}F^{\overline{c} \underline{d} \underline{e}} F^{\overline{b}} F_{\overline{c}} F^{\overline{e}}{}_{\underline{d} \underline{e}} F_{\overline{b} \overline{e} \underline{f}} + 4D^{\overline{c}}F_{\overline{c}}{}^{\underline{c} \underline{e}} D^{\overline{f}}F^{\overline{e}}{}_{\underline{c}}{}^{\underline{f}} F_{\overline{e}} F_{\overline{f} \underline{e} \underline{f}}\cr&& 
-2D^{\overline{c}}F_{\overline{c}}{}^{\underline{c} \underline{e}} D^{\overline{e}}F^{\overline{f}}{}_{\underline{c}}{}^{\underline{f}} F_{\overline{e}} F_{\overline{f} \underline{e} \underline{f}} + 8D^{\overline{d}}F^{\overline{e} \underline{c} \underline{e}} D^{\overline{f}}F_{\overline{d} \underline{c}}{}^{\underline{f}} F_{\overline{e}} F_{\overline{f} \underline{e} \underline{f}}-2D^{\overline{e}}F^{\overline{d} \underline{c} \underline{e}} D^{\overline{f}}F_{\overline{d} \underline{c}}{}^{\underline{f}} F_{\overline{e}} F_{\overline{f} \underline{e} \underline{f}}\cr&& 
-2D^{\overline{b}}F^{\overline{e} \underline{c} \underline{e}} D_{\overline{b}}F^{\overline{f}}{}_{\underline{c}}{}^{\underline{f}} F_{\overline{e}} F_{\overline{f} \underline{e} \underline{f}}-2D^{\overline{c}}F^{\overline{e} \underline{c} \underline{e}} F_{\overline{c}} F^{\overline{d}} F_{\overline{d} \underline{c}}{}^{\underline{f}} F_{\overline{e} \underline{e} \underline{f}} + 4D^{\overline{e}}F^{\overline{c} \underline{c} \underline{e}} F_{\overline{c}} F^{\overline{d}} F_{\overline{d} \underline{c}}{}^{\underline{f}} F_{\overline{e} \underline{e} \underline{f}}\cr&& 
 + 2D^{\underline{f}}F^{\overline{c} \underline{d} \underline{e}} D^{\overline{e}}F_{\overline{c}} F^{\overline{f}}{}_{\underline{d} \underline{e}} F_{\overline{e} \overline{f} \underline{f}} + 2D^{\underline{d}}F^{\overline{e} \underline{e} \underline{f}} D^{\overline{c}}F_{\overline{c}}{}^{\overline{f}}{}_{\underline{d}} F_{\overline{e}} F_{\overline{f} \underline{e} \underline{f}} + 2D^{\underline{d}}F^{\overline{a} \underline{e} \underline{f}} D_{\overline{a}}F^{\overline{e} \overline{f}}{}_{\underline{d}} F_{\overline{e}} F_{\overline{f} \underline{e} \underline{f}} \cr&& 
+ 2D^{\underline{f}}F^{\overline{d} \underline{d} \underline{e}} D^{\overline{e}}F^{\overline{f}}{}_{\underline{d} \underline{e}} F_{\overline{d}} F_{\overline{e} \overline{f} \underline{f}}-2D^{\overline{c}}F^{\overline{b}} F_{\overline{c}}{}^{\underline{c} \underline{e}} F^{\overline{f}}{}_{\underline{c}}{}^{\underline{f}} F^{\overline{g}}{}_{\underline{e} \underline{f}} F_{\overline{b} \overline{f} \overline{g}}-2D^{\overline{f}}F^{\overline{b}} F^{\overline{e} \underline{c} \underline{d}} F^{\overline{g}}{}_{\underline{c} \underline{d}} F_{\overline{f} \overline{g} \underline{f}} F_{\overline{b} \overline{e}}{}^{\underline{f}} \cr&& 
+ 4D^{\overline{c}}F^{\overline{b}} F_{\overline{c} \underline{c}}{}^{\underline{g}} F^{\overline{e}}{}_{\underline{e}}{}^{\underline{h}} F_{\overline{e} \underline{g} \underline{h}} F_{\overline{b}}{}^{\underline{c} \underline{e}}-4D^{\overline{e}}F^{\overline{b}} F^{\overline{d}}{}_{\underline{c}}{}^{\underline{g}} F_{\overline{d} \underline{e}}{}^{\underline{h}} F_{\overline{e} \underline{g} \underline{h}} F_{\overline{b}}{}^{\underline{c} \underline{e}} + D^{\overline{d}}F^{\overline{b}} F^{\overline{e}}{}_{\underline{c} \underline{d}} F_{\overline{d}}{}^{\underline{g} \underline{h}} F_{\overline{e} \underline{g} \underline{h}} F_{\overline{b}}{}^{\underline{c} \underline{d}}\cr&& 
-2D^{\overline{e}}F^{\overline{b}} F^{\overline{f}}{}_{\underline{c}}{}^{\underline{f}} F^{\overline{g}}{}_{\underline{e} \underline{f}} F_{\overline{b}}{}^{\underline{c} \underline{e}} F_{\overline{e} \overline{f} \overline{g}}-2D^{\overline{b}}F_{\overline{b}}{}^{\overline{f} \overline{g}} F^{\overline{e}} F_{\overline{e}}{}^{\underline{c} \underline{e}} F_{\overline{f} \underline{c}}{}^{\underline{f}} F_{\overline{g} \underline{e} \underline{f}}-2D^{\overline{b}}F_{\overline{b}}{}^{\overline{e} \underline{f}} F^{\overline{d}} F_{\overline{e}}{}^{\underline{d} \underline{e}} F^{\overline{g}}{}_{\underline{d} \underline{e}} F_{\overline{d} \overline{g} \underline{f}}\cr&& 
 + 4D^{\overline{e}}F^{\overline{f} \underline{c} \underline{e}} F^{\overline{c}} F_{\overline{c} \underline{c}}{}^{\underline{f}} F^{\overline{g}}{}_{\underline{e} \underline{f}} F_{\overline{e} \overline{f} \overline{g}}-8D^{\overline{d}}F^{\overline{e} \underline{c} \underline{e}} F^{\overline{c}} F_{\overline{c} \underline{c}}{}^{\underline{g}} F_{\overline{d} \underline{e}}{}^{\underline{h}} F_{\overline{e} \underline{g} \underline{h}} + 4D^{\overline{e}}F^{\overline{d} \underline{c} \underline{e}} F^{\overline{c}} F_{\overline{c} \underline{c}}{}^{\underline{g}} F_{\overline{d} \underline{e}}{}^{\underline{h}} F_{\overline{e} \underline{g} \underline{h}} \cr&& 
+ 4D^{\overline{b}}F_{\overline{b}}{}^{\underline{c} \underline{e}} F^{\overline{c}} F_{\overline{c} \underline{c}}{}^{\underline{g}} F^{\overline{e}}{}_{\underline{e}}{}^{\underline{h}} F_{\overline{e} \underline{g} \underline{h}} + D^{\overline{d}}F^{\overline{e} \underline{c} \underline{d}} F^{\overline{c}} F_{\overline{c} \underline{c} \underline{d}} F_{\overline{d}}{}^{\underline{g} \underline{h}} F_{\overline{e} \underline{g} \underline{h}} + 4D^{\overline{d}}F^{\overline{e} \underline{e} \underline{g}} F^{\overline{c}} F_{\overline{c}}{}^{\underline{f} \underline{h}} F_{\overline{d} \underline{e} \underline{f}} F_{\overline{e} \underline{g} \underline{h}}\cr&& 
-4D^{\overline{b}}F_{\overline{b}}{}^{\underline{e} \underline{g}} F^{\overline{c}} F_{\overline{c}}{}^{\underline{f} \underline{h}} F^{\overline{e}}{}_{\underline{e} \underline{f}} F_{\overline{e} \underline{g} \underline{h}} + D^{\overline{d}}F^{\overline{e} \underline{e} \underline{f}} F^{\overline{c}} F_{\overline{c}}{}^{\underline{g} \underline{h}} F_{\overline{d} \underline{e} \underline{f}} F_{\overline{e} \underline{g} \underline{h}} + D^{\overline{b}}F_{\overline{b}}{}^{\underline{e} \underline{f}} F^{\overline{c}} F_{\overline{c}}{}^{\underline{g} \underline{h}} F^{\overline{e}}{}_{\underline{e} \underline{f}} F_{\overline{e} \underline{g} \underline{h}}\cr&& 
-2D^{\overline{f}}F^{\overline{g} \underline{c} \underline{d}} F^{\overline{c}} F^{\overline{e}}{}_{\underline{c} \underline{d}} F_{\overline{c} \overline{e}}{}^{\underline{f}} F_{\overline{f} \overline{g} \underline{f}}-2D^{\overline{f}}F^{\overline{e} \underline{c} \underline{d}} F^{\overline{c}} F^{\overline{g}}{}_{\underline{c} \underline{d}} F_{\overline{c} \overline{e}}{}^{\underline{f}} F_{\overline{f} \overline{g} \underline{f}} + 4D^{\overline{d}}F^{\overline{f} \underline{c} \underline{e}} F^{\overline{c}} F_{\overline{d} \underline{c}}{}^{\underline{f}} F^{\overline{g}}{}_{\underline{e} \underline{f}} F_{\overline{c} \overline{f} \overline{g}}\cr&& 
-2D^{\overline{b}}F_{\overline{b}}{}^{\underline{c} \underline{e}} F^{\overline{c}} F^{\overline{f}}{}_{\underline{c}}{}^{\underline{f}} F^{\overline{g}}{}_{\underline{e} \underline{f}} F_{\overline{c} \overline{f} \overline{g}} + 2F^{\overline{a}} F^{\overline{b}} F_{\overline{a}}{}^{\underline{c} \underline{e}} F^{\overline{e}}{}_{\underline{c}}{}^{\underline{f}} F^{\overline{f}}{}_{\underline{e} \underline{f}} F_{\overline{b} \overline{e} \overline{f}}-2F^{\overline{a}} F^{\overline{b}} F_{\overline{a}}{}^{\underline{c} \underline{e}} F^{\overline{d}}{}_{\underline{e}}{}^{\underline{h}} F_{\overline{d} \underline{g} \underline{h}} F_{\overline{b} \underline{c}}{}^{\underline{g}} \cr&& 
+ 2F^{\overline{a}} F^{\overline{b}} F_{\overline{a}}{}^{\underline{e} \underline{g}} F^{\overline{d}}{}_{\underline{e} \underline{f}} F_{\overline{d} \underline{g} \underline{h}} F_{\overline{b}}{}^{\underline{f} \underline{h}} -  1/2 F^{\overline{a}} F^{\overline{b}} F_{\overline{a}}{}^{\underline{e} \underline{f}} F^{\overline{d}}{}_{\underline{e} \underline{f}} F_{\overline{d} \underline{g} \underline{h}} F_{\overline{b}}{}^{\underline{g} \underline{h}} + F^{\overline{a}} F^{\overline{b}} F^{\overline{e} \underline{c} \underline{d}} F^{\overline{f}}{}_{\underline{c} \underline{d}} F_{\overline{a} \overline{e}}{}^{\underline{f}} F_{\overline{b} \overline{f} \underline{f}}\cr&& 
-2D^{\overline{f}}F^{\overline{e} \overline{g} \overline{h}} F_{\overline{e}} F_{\overline{f}}{}^{\underline{c} \underline{e}} F_{\overline{g} \underline{c}}{}^{\underline{f}} F_{\overline{h} \underline{e} \underline{f}}-2D^{\overline{g}}F^{\overline{d} \overline{e} \underline{f}} F_{\overline{d}} F_{\overline{e}}{}^{\underline{d} \underline{e}} F^{\overline{h}}{}_{\underline{d} \underline{e}} F_{\overline{g} \overline{h} \underline{f}}-2D^{\overline{f}}F^{\overline{c} \underline{c} \underline{e}} F_{\overline{c}} F^{\overline{g}}{}_{\underline{c}}{}^{\underline{f}} F^{\overline{h}}{}_{\underline{e} \underline{f}} F_{\overline{f} \overline{g} \overline{h}} \cr&& 
+ 4D^{\overline{d}}F^{\overline{c} \underline{c} \underline{e}} F_{\overline{c}} F_{\overline{d} \underline{c}}{}^{\underline{g}} F^{\overline{f}}{}_{\underline{e}}{}^{\underline{h}} F_{\overline{f} \underline{g} \underline{h}}-4D^{\overline{f}}F^{\overline{c} \underline{c} \underline{e}} F_{\overline{c}} F^{\overline{e}}{}_{\underline{c}}{}^{\underline{g}} F_{\overline{e} \underline{e}}{}^{\underline{h}} F_{\overline{f} \underline{g} \underline{h}} + D^{\overline{e}}F^{\overline{c} \underline{c} \underline{d}} F_{\overline{c}} F^{\overline{f}}{}_{\underline{c} \underline{d}} F_{\overline{e}}{}^{\underline{g} \underline{h}} F_{\overline{f} \underline{g} \underline{h}} \ , \nn
\end{eqnarray}}
and the contribution of the three-form fluxes is given by
{\footnotesize 
}
On the other hand we also split the mixed $a b$ terms in those with dilaton dependence and without $\tilde{\cal R}^{(1,1)}= {\cal R}_{\Phi}^{(1,1)}+{\cal R}_{\not \Phi}^{(1,1)}$, finding on the one hand 
{\footnotesize \begin{eqnarray}
{\cal R}_{\Phi}^{(1,1)} & = &
 -D^{\underline{b}}D_{\underline{b}}F^{\overline{c} \underline{f} \underline{g}} F^{\underline{e}} F^{\overline{d}}{}_{\underline{f} \underline{g}} F_{\overline{c} \overline{d} \underline{e}}-D^{\overline{d}}D^{\overline{e}}F^{\overline{f} \underline{d} \underline{e}} F^{\underline{c}} F_{\overline{d} \underline{d} \underline{e}} F_{\overline{e} \overline{f} \underline{c}}-D^{\overline{d}}F^{\underline{a}} D^{\overline{e}}F^{\overline{f} \underline{d} \underline{e}} F_{\overline{d} \underline{d} \underline{e}} F_{\overline{e} \overline{f} \underline{a}}\\ && 
 -D^{\underline{b}}F^{\underline{c}} D_{\underline{b}}F^{\overline{c} \underline{f} \underline{g}} F^{\overline{d}}{}_{\underline{f} \underline{g}} F_{\overline{c} \overline{d} \underline{c}}-D^{\underline{g}}F^{\underline{b}} D_{\underline{b}}F^{\overline{c} \underline{e} \underline{f}} F^{\overline{d}}{}_{\underline{e} \underline{f}} F_{\overline{c} \overline{d} \underline{g}}-D^{\overline{c}}F_{\overline{c}}{}^{\underline{c} \underline{d}} D^{\overline{e}}F^{\overline{f}}{}_{\underline{c} \underline{d}} F^{\underline{e}} F_{\overline{e} \overline{f} \underline{e}} \cr && 
 -D^{\underline{h}}D^{\underline{e}}F^{\overline{c} \underline{f} \underline{g}} F_{\underline{e}} F^{\overline{d}}{}_{\underline{f} \underline{g}} F_{\overline{c} \overline{d} \underline{h}}-D^{\underline{f}}F^{\overline{b} \underline{g} \underline{h}} D^{\underline{e}}F_{\overline{b}}{}^{\overline{d}}{}_{\underline{e}} F_{\underline{f}} F_{\overline{d} \underline{g} \underline{h}}-D^{\underline{b}}F^{\overline{b} \underline{g} \underline{h}} D_{\underline{b}}F_{\overline{b}}{}^{\overline{d} \underline{f}} F_{\underline{f}} F_{\overline{d} \underline{g} \underline{h}}\cr && 
 -D^{\underline{g}}F^{\overline{c} \underline{e} \underline{f}} D^{\underline{h}}F^{\overline{d}}{}_{\underline{e} \underline{f}} F_{\underline{g}} F_{\overline{c} \overline{d} \underline{h}}-D^{\overline{d}}F^{\overline{e} \underline{e} \underline{f}} D^{\overline{f}}F_{\overline{d} \overline{e}}{}^{\underline{d}} F_{\underline{d}} F_{\overline{f} \underline{e} \underline{f}} -  \frac{1}{2} D^{\overline{b}}F^{\underline{a}} F^{\overline{f}}{}_{\underline{d} \underline{e}} F_{\overline{b}}{}^{\underline{d} \underline{e}} F^{\overline{g} \overline{h}}{}_{\underline{a}} F_{\overline{f} \overline{g} \overline{h}}\cr && 
  -  \frac{1}{2} D^{\overline{e}}F^{\overline{f} \overline{g} \overline{h}} F^{\underline{a}} F_{\overline{e}}{}^{\underline{d} \underline{e}} F_{\overline{f} \underline{d} \underline{e}} F_{\overline{g} \overline{h} \underline{a}} -  \frac{1}{2} D^{\overline{c}}F^{\overline{f} \underline{d} \underline{e}} F^{\underline{c}} F_{\overline{c} \underline{d} \underline{e}} F^{\overline{g} \overline{h}}{}_{\underline{c}} F_{\overline{f} \overline{g} \overline{h}} -  \frac{1}{2} D^{\overline{b}}F_{\overline{b}}{}^{\underline{d} \underline{e}} F^{\underline{c}} F^{\overline{f}}{}_{\underline{d} \underline{e}} F^{\overline{g} \overline{h}}{}_{\underline{c}} F_{\overline{f} \overline{g} \overline{h}} \cr && 
 -  \frac{1}{2} D^{\overline{d}}F^{\overline{f} \overline{g} \underline{b}} F_{\underline{b}} F_{\overline{d}}{}^{\underline{e} \underline{f}} F^{\overline{h}}{}_{\underline{e} \underline{f}} F_{\overline{f} \overline{g} \overline{h}} + D^{\overline{b}}F^{\underline{a}} F^{\overline{e}}{}_{\underline{d}}{}^{\underline{g}} F^{\overline{f}}{}_{\underline{f} \underline{g}} F_{\overline{b}}{}^{\underline{d} \underline{f}} F_{\overline{e} \overline{f} \underline{a}}-2D^{\overline{c}}F^{\overline{e} \underline{d} \underline{f}} F^{\underline{c}} F_{\overline{c} \underline{d}}{}^{\underline{g}} F^{\overline{f}}{}_{\underline{f} \underline{g}} F_{\overline{e} \overline{f} \underline{c}} \cr && 
 + D^{\overline{b}}F_{\overline{b}}{}^{\underline{d} \underline{f}} F^{\underline{c}} F^{\overline{e}}{}_{\underline{d}}{}^{\underline{g}} F^{\overline{f}}{}_{\underline{f} \underline{g}} F_{\overline{e} \overline{f} \underline{c}} + D^{\overline{d}}F^{\overline{e} \overline{f} \underline{b}} F_{\underline{b}} F_{\overline{d}}{}^{\underline{e} \underline{g}} F_{\overline{e} \underline{e}}{}^{\underline{h}} F_{\overline{f} \underline{g} \underline{h}} + D^{\overline{d}}D^{\underline{d}}F^{\overline{e} \overline{f} \underline{e}} F^{\underline{c}} F_{\overline{d} \underline{d} \underline{e}} F_{\overline{e} \overline{f} \underline{c}} \cr && 
 + D^{\underline{d}}F^{\overline{e} \overline{f} \underline{e}} D^{\overline{d}}F^{\underline{c}} F_{\overline{d} \underline{d} \underline{e}} F_{\overline{e} \overline{f} \underline{c}} + D^{\underline{c}}F^{\overline{e} \overline{f} \underline{d}} D^{\overline{d}}F_{\overline{d} \underline{c} \underline{d}} F^{\underline{e}} F_{\overline{e} \overline{f} \underline{e}} + D^{\underline{e}}F^{\overline{d} \overline{e} \underline{f}} D^{\overline{f}}F_{\overline{d} \overline{e}}{}^{\underline{d}} F_{\underline{d}} F_{\overline{f} \underline{e} \underline{f}} \cr && 
 +  \frac{1}{2} D^{\overline{b}}F^{\underline{a}} F^{\underline{e} \underline{f} \underline{g}} F_{\overline{b} \underline{e} \underline{f}} F^{\overline{e} \overline{f}}{}_{\underline{a}} F_{\overline{e} \overline{f} \underline{g}} +  \frac{1}{2} D^{\overline{d}}F^{\overline{e} \overline{f} \underline{c}} F^{\underline{b}} F_{\underline{c}}{}^{\underline{f} \underline{g}} F_{\overline{d} \underline{f} \underline{g}} F_{\overline{e} \overline{f} \underline{b}} +  \frac{1}{2} D^{\overline{b}}F_{\overline{b}}{}^{\underline{d} \underline{e}} F^{\underline{c}} F_{\underline{d} \underline{e}}{}^{\underline{g}} F^{\overline{e} \overline{f}}{}_{\underline{c}} F_{\overline{e} \overline{f} \underline{g}} \cr && 
 +  \frac{1}{2} D^{\overline{b}}F^{\underline{e} \underline{f} \underline{g}} F^{\underline{d}} F_{\overline{b} \underline{e} \underline{f}} F^{\overline{e} \overline{f}}{}_{\underline{d}} F_{\overline{e} \overline{f} \underline{g}} +  \frac{1}{2} D^{\overline{d}}F^{\overline{e} \overline{f} \underline{b}} F_{\underline{b}} F^{\underline{f} \underline{g} \underline{h}} F_{\overline{d} \underline{f} \underline{g}} F_{\overline{e} \overline{f} \underline{h}} + 2D^{\underline{g}}F^{\underline{b}} F^{\overline{c} \underline{e} \underline{f}} F^{\overline{e}}{}_{\underline{e} \underline{f}} F_{\overline{c}}{}^{\overline{f}}{}_{\underline{b}} F_{\overline{e} \overline{f} \underline{g}}\cr && 
 -2D^{\underline{b}}F^{\overline{c} \overline{e}}{}_{\underline{b}} F^{\underline{c}} F_{\overline{c}}{}^{\underline{f} \underline{g}} F^{\overline{f}}{}_{\underline{f} \underline{g}} F_{\overline{e} \overline{f} \underline{c}} + 2D^{\underline{g}}F^{\overline{c} \underline{e} \underline{f}} F^{\underline{d}} F^{\overline{e}}{}_{\underline{e} \underline{f}} F_{\overline{c}}{}^{\overline{f}}{}_{\underline{d}} F_{\overline{e} \overline{f} \underline{g}} + 2D^{\underline{g}}F^{\overline{e} \underline{e} \underline{f}} F^{\underline{d}} F^{\overline{c}}{}_{\underline{e} \underline{f}} F_{\overline{c}}{}^{\overline{f}}{}_{\underline{d}} F_{\overline{e} \overline{f} \underline{g}}\cr && 
 -2D^{\underline{h}}F^{\overline{c} \overline{e} \underline{c}} F_{\underline{c}} F_{\overline{c}}{}^{\underline{f} \underline{g}} F^{\overline{f}}{}_{\underline{f} \underline{g}} F_{\overline{e} \overline{f} \underline{h}}-D^{\overline{b}}F^{\underline{a}} F_{\overline{b}}{}^{\underline{d} \underline{e}} F^{\overline{e} \overline{g}}{}_{\underline{a}} F_{\overline{e}}{}^{\overline{h}}{}_{\underline{d}} F_{\overline{g} \overline{h} \underline{e}} + 2D^{\overline{d}}F^{\overline{e} \overline{g} \underline{c}} F^{\underline{b}} F_{\overline{d} \underline{c}}{}^{\underline{e}} F_{\overline{e}}{}^{\overline{h}}{}_{\underline{b}} F_{\overline{g} \overline{h} \underline{e}}\cr && 
 -D^{\overline{b}}F_{\overline{b}}{}^{\underline{d} \underline{e}} F^{\underline{c}} F^{\overline{e} \overline{g}}{}_{\underline{c}} F_{\overline{e}}{}^{\overline{h}}{}_{\underline{d}} F_{\overline{g} \overline{h} \underline{e}}-D^{\overline{d}}F^{\overline{e} \overline{g} \underline{b}} F_{\underline{b}} F_{\overline{d}}{}^{\underline{e} \underline{f}} F_{\overline{e}}{}^{\overline{h}}{}_{\underline{e}} F_{\overline{g} \overline{h} \underline{f}} + D^{\underline{d}}F^{\overline{c} \underline{f} \underline{g}} F_{\underline{d}} F^{\underline{e}} F^{\overline{d}}{}_{\underline{f} \underline{g}} F_{\overline{c} \overline{d} \underline{e}}\cr && 
 -F^{\underline{a}} F^{\underline{b}} F^{\overline{c} \underline{e} \underline{f}} F^{\overline{e}}{}_{\underline{e} \underline{f}} F_{\overline{c}}{}^{\overline{f}}{}_{\underline{a}} F_{\overline{e} \overline{f} \underline{b}} + D^{\overline{d}}F^{\overline{e} \underline{d} \underline{e}} F^{\underline{c}} F^{\overline{c}} F_{\overline{c} \underline{d} \underline{e}} F_{\overline{d} \overline{e} \underline{c}} +  \frac{1}{2} F^{\underline{a}} F^{\overline{a}} F_{\overline{a}}{}^{\underline{d} \underline{e}} F^{\overline{e}}{}_{\underline{d} \underline{e}} F^{\overline{f} \overline{g}}{}_{\underline{a}} F_{\overline{e} \overline{f} \overline{g}}\cr && 
 -F^{\underline{a}} F^{\overline{a}} F_{\overline{a}}{}^{\underline{d} \underline{f}} F^{\overline{e}}{}_{\underline{f} \underline{g}} F^{\overline{d}}{}_{\underline{d}}{}^{\underline{g}} F_{\overline{d} \overline{e} \underline{a}} \;,\nn
\end{eqnarray}}
and on the other
{\footnotesize 
}

Every term in the action is separately invariant under generalized diffeomorphisms and rigid $O(D,D)$ transformations. What fixes the couplings is the double Lorentz symmetry. Since this symmetry mixes different orders through the generalized Green-Schwarz transformation, invariance is achieved as follows
\begin{equation}
    \delta_\Lambda {\cal R} = \sum_{i  = 0}^{\infty}  \sum_{k  = 0}^i \delta_{\Lambda}^{(i-k)} {\cal R}^{(k)} = 0 \ . 
\end{equation}
Here we know the Lorentz transformation and the action to second order, and then invariance holds here to order ${\cal O}(\alpha'{}^2)$ only. Higher corrections require pursuing the perturbative expansion further.

\section{Supergravity, gauge fixing and field redefinitions} \label{SecSugra}

It is well known how to reduce the two-derivative DFT action to that of supergravity. This requires a $GL(D)$ decomposition of $O(D,D)$, a parameterization of the generalized fields, a gauge fixing of the double Lorentz transformations to its diagonal subgroup, and picking up a certain solution to the strong constraint. When higher derivatives are considered, these steps must be complemented with field redefintions. The reason is that the components of the generalized fields inherit the non-standard  Lorentz symmetry coming from the generalized Green-Schwarz transformation. In particular, one should seek a redefinition that renders the metric and dilaton Lorentz invariant. The two-form is different as it is expected to carry a Green-Schwarz transformation. When the two-parameters $a$ and $b$ are turned on, the minimal field redefinitions that meet these requirements are those that connect the DFT components (noted here with an overline) to those of supergravity in the so-called generalized Bergshoeff-de Roo scheme. In such a scheme, the dilaton, vielbein and two-form transform as \cite{MarquesNunez}
\bea
\delta \phi &=& L_\xi \phi \ , \nn\\
\delta e_\mu{}^a &=& L_\xi e_\mu{}^a + e_\mu{}^b \Lambda_b{}^a \ , \label{BdRscheme}\\
\delta b_{\mu \nu} &=& L_\xi b_{\mu \nu} + 2 \partial_{[\mu} \lambda_{\nu]} + \frac a 2 \partial_{[\mu} \Lambda^{a b} \widehat \omega^{(-)}_{\nu]ab} - \frac b 2 \partial_{[\mu} \Lambda^{a b} \widehat \omega^{(+)}_{\nu]ab}  \ . \nn
\eea
Let us now introduce the protagonists in the Green-Schwarz transformation of the two-form. First we define the spin connection 
\be
\omega_{\mu a}{}^b = \partial_{\mu} e_{\nu}{}^b e^\nu{}_a - \Gamma_{\mu \nu}^\rho e_\rho{}^b e^\nu{}_a \ ,  \ \ \ \  \Gamma_{\mu \nu}^{\rho} = \frac 1 2 g^{\rho \sigma} \left( \partial_{\mu} g_{\nu \sigma} + \partial_{\nu} g_{\mu \sigma} - \partial_{\sigma} g_{\mu \nu}\right)\;, \label{spinconnection}
\ee
that transforms as
\be
\delta \omega_{\mu a}{}^b = L_\xi \omega_{\mu a}{}^b + \partial_\mu \Lambda_{a}{}^b + \omega_{\mu a}{}^c \Lambda_c{}^b - \Lambda_a{}^c \omega_{\mu c}{}^b \ .
\ee
We then add torsion to it in two different ways
\bea
\widehat \omega^{(\pm)}_{\mu b c} = \omega_{\mu b c} \pm \frac 1 2 \, \widehat H_{\mu \nu \rho} e^\nu{}_b e^\rho{}_c \ ,
\eea
with the torsion given by
\be
\widehat H_{\mu \nu \rho} = H_{\mu \nu \rho} - \frac 3 2 a \widehat \Omega^{(-)}_{\mu \nu \rho} + \frac 3 2 b \widehat \Omega^{(+)}_{\mu \nu \rho} \ , \ \ \ 
H_{\mu \nu \rho} = 3 \partial_{[\mu} b_{\nu \rho]} \ ,
\ee
where the Chern-Simons three-forms are defined as
\be
\widehat \Omega^{(\pm)}_{\mu \nu \rho} = \widehat \omega^{(\pm)}_{[\mu a}{}^b \partial_{\nu} \widehat \omega^{(\pm)}_{\rho] b}{}^a + \frac 2 3
\widehat \omega^{(\pm)}_{[\mu a}{}^b \widehat \omega^{(\pm)}_{\nu b}{}^c \widehat \omega^{(\pm)}_{\rho] c}{}^a  \ .
\ee
Under diffeomorphisms and Lorentz the Chern-Simons  transform as
\be
\delta \widehat \Omega^{(\pm)}_{\mu \nu \rho} = L_\xi \widehat \Omega^{(\pm)}_{\mu \nu \rho} -
\partial_{[\mu} \left(\partial_{\nu} \Lambda_a{}^b \widehat \omega^{(\pm)}_{\rho] b}{}^a\right) \ ,
\ee
which, combined with the Green-Schwarz transformation of the two-form (\ref{BdRscheme}) renders $\widehat H_{\mu \nu\rho}$ Lorentz invariant. This is then the right three-form curvature tensor to appear in the action. It hiddenly contains an infinite tower of higher derivatives, because it depends on the Chern-Simons terms, which in turn depend on it. So this establishes an infinite recursive relation that allows to expand the corrections order by order. 

We discussed above the minimal and natural all-order completion of what was found in \cite{MarquesNunez} to first order in $\alpha'$, which reproduces exactly the heterotic Green-Schwarz \cite{BdR} extending it to the bi-parametric case. Interestingly, because this fits into a duality covariant picture, T-duality enforces the generalized Green-Schwarz transformation to generate not only the Chern-Simons terms, but also notably the quadratic Riemann interactions present both in bosonic and heterotic supergravity corrections. We will now show in this section that this structure (\ref{BdRscheme}) is preserved by the $\alpha'{}^2$ corrections discussed in this paper, with no further deformations arising.

The starting point is to perform a $GL(D)$ decomposition of $O(D,D)$, by parameterizing all the duality covariant tensors in terms of fields that will later be linked to supergravity. At the moment it is not necessary to impose the strong constraint. The flat and curved $O(D,D)$ invariant metrics are decomposed as follows
\be
\eta_{A B} = \left(\begin{matrix} - g_{\underline {ab}}  & \\ & g_{\overline{ab}} \end{matrix}\right) \ , \ \ \  \eta_{M N} = \left( \begin{matrix}& \delta^\mu_\nu \\ \delta_\mu^\nu &  \end{matrix}\right) \ , 
\ee
where $g$ are Minkowski metrics, that carry different indices because they are acted on separately by the different factors of the double Lorentz group. The generalized fields are parameterized as
\be
e^{-2d} = \sqrt{|\overline {g}|} e^{-2 {\overline \phi}} \ , \ \ \ \ 
E_{M}{}^A = \frac 1 {\sqrt{2}} \left(\begin{matrix} \bar e^\mu{}_{\underline c}\, g^{\underline {ca}} & \bar e^\mu{}_{\overline c}\,  g^{\overline {ca}}
 \\  (\bar b_{\mu \rho} - \bar g_{\mu \rho})\, \bar e^\rho{}_{\underline c} \, g^{\underline {ca}} & (\bar b_{\mu \rho} + \bar g_{\mu \rho})\, \bar e^\rho{}_{\overline c}\, g^{\overline {ca}} \end{matrix}\right) \ , \label{ParamFrame}
\ee
where it is necessary to include a pair of vielbeins, each satisfying
\be
\bar e_\mu{}^{\overline a} = \bar g_{\mu\nu} \bar e^\nu{}_{\overline b} \, g^{\overline {ba}} \ , \ \ \ \  \bar e_\mu{}^{\underline a} = \bar g_{\mu\nu} \bar e^\nu{}_{\underline b} \, g^{\underline {ba}}  \ , \ \ \ \ \bar g_{\mu \nu} = \bar e_\mu{}^{\underline a}\, g_{\underline{ab} }\, \bar e_\nu{}^{\underline b}= \bar e_\mu{}^{\overline a}\, g_{\overline{ab} }\, \bar e_\nu{}^{\overline b}  \ .
\ee
If desired, one can then define the generalized metric by curving its flat version
\be
 {\cal H}_{A B} = \left(\begin{matrix} g_{\underline {ab}}  & \\ & g_{\overline{ab}} \end{matrix}\right) \ , \ \ \ \ {\cal H}_{M N} = \left(\begin{matrix}
                         \bar g^{\mu \nu} & - \bar g^{\mu \rho}  \bar b_{\rho \nu}  \\
                          \bar b_{\mu \rho} \bar g^{\rho \nu} & \bar g_{\mu \nu} -  \bar b_{\mu \rho} \bar g^{\rho \sigma} \bar b_{\sigma \nu}
                       \end{matrix}\right) \ .
\ee
We have included an overline on the dynamical fields because they will ultimately be affected by the generalized Green-Schwarz transformations. These non-standard Lorentz transformations can be removed for the vielbein and the dilaton via field redefinitions, as we will show soon, and then we reserve the notation without an overline for the set of fields that transform as usual with respect to Lorentz symmetries. An equivalent decomposition must apply to generalized coordinates and parameters
\be
\partial_M = \left(\tilde \partial^\mu \,  , \, \partial_\mu\right) \ , \ \ \ \ \xi^M = \left(\bar \xi_\mu \, , \, \xi^\mu \right) \ .
\ee
 Note that the one-form component of the parameter of generalized diffeomorphisms also carries an overline, this is because it will have to be redefined to second order, as we will show here.

We recall the way in which the generalized frame transforms to different orders in $\alpha'$ (or equivalently in powers of $a$ and $b$). We separate the orders from (\ref{VarEup}) and (\ref{VarEdown})
\bea
\delta E_{M}{}^{\underline a} &=& \widehat {\cal L}_\xi E_M{}^{\underline a} + E_M{}^{\underline b} \Lambda_{\underline b}{}^{\underline a} + \Delta^{(1)} E_M{}^{\underline a}  + \Delta^{(2)} E_M{}^{\underline a} + \dots \ ,\\
\delta E_{M}{}^{\overline a} &=& \widehat {\cal L}_\xi E_M{}^{\overline a} + E_M{}^{\overline b} \Lambda_{\overline b}{}^{\overline a} + \Delta^{(1)} E_M{}^{\overline a}  + \Delta^{(2)} E_M{}^{\overline a} + \dots \ ,
\eea
where to first order we have
\bea
\Delta^{(1)} E_M{}^{\underline a} &=& \frac a 2  E_M{}^{\overline b} F_{\overline b \underline {cd}} D^{\underline a} \Lambda^{\underline {c d}} + \frac b 2 E_M{}^{\overline b} D_{\overline b} \Lambda^{\overline {c d}} \, F^{\underline a}{}_{\overline {c d}} \ ,\\
\Delta^{(1)} E_M{}^{\overline a} &=& - \frac a 2 E_M{}^{\underline b} D_{\underline b}\Lambda^{\underline {c d}} F^{\overline a}{}_{\underline {c d}}    - \frac b 2  \, E_M{}^{\underline b} F_{\underline b \overline {c d}} \, D^{\overline a}  \Lambda^{\overline {c d}} \  ,
\eea
and to second 
{\footnotesize \bea
\Delta^{(2)} E_M{}^{\underline a} &=& -\, \frac {a^2} 2 \; E_M{}^{\overline b}\left[
														D^{\underline{a}}D^{\underline c}{\Lambda}^{\underline{e f}}\left(F_{\underline c\overline{d b}} F^{\overline d}{}_{\underline{e f}} + D_{\underline c} F_{\overline b \underline{e f}}\right)	
														\right.
-F_{\overline b \underline{e f}} F_{\overline c \underline d}{}^{\underline f}\left(
F^{\overline c}{}^{\underline{h d}} D^{\underline a}{\Lambda}_{\underline h}{}^{\underline e}
-F^{\overline c}{}^{\underline{h e}} D^{\underline a}{\Lambda}_{\underline h}{}^{\underline d}
\right)\nonumber \\
&&\ \ \ \ \ \ \ \ \ \ \ \ \ \ \  +\;\left. F^{\overline c}{}_{\underline{e f}}\;D^{\underline a}{\Lambda}^{\underline{e}}{}_{\underline g} \left(F_{\overline {b c d}} F^{\overline d \underline{g f}}
-D_{\overline b} F_{\overline c}{}^{\underline{g f}} + 2\; D_{\overline c} F_{\overline b}{}^{ \underline{g f}}
\right)
+F_{\overline b \underline{e f}}D^{\underline a}\left(D^{\overline c}{\Lambda}^{\underline{e d}} F_{\overline c \underline{d}}{}^{\underline f}\right)						
							\right]\nonumber\\&&
 -\frac {a b} 4\, E_M{}^{\overline b}\,\left[D^{\underline a}{\Lambda}^{\underline{ef}}
 \left( F_{\overline{b}\underline{eh}}\,F^{\underline h\overline{cd}} \,F_{\underline f\overline{cd}}- D_{\overline b} F_{\underline e}{}^{\overline{cd}}\;  F_{\underline f\overline{cd}}\right)
+F_{\overline b\underline{ef}}\, D^{\underline a}\left(D^{\underline e}{\Lambda}^{\overline{cd}}\; F^{\underline f}{}_{\overline{cd}}\right) \right.\nn\\&&
\;\;\;\;\;\;\;\;\;\;\;\;\;\;\;\;\;\;
\left.-\, D_{\overline b}{\Lambda}^{\overline{ef}}\left(F^{\underline a}{}_{\overline{eh}}
 F^{\overline h\underline{cd}} F_{\overline f\underline{cd}}
 -D^{\underline a}F_{\overline e}{}^{\underline{cd}} F_{\overline f\underline{cd}}\right) -\,F^{\underline a}{}_{\overline{ef}}\, D_{\overline b}\left( D^{\overline e}{\Lambda}^{\underline{cd}}\, F^{\overline f}{}_{\underline{cd}}\right)\,\right]\nn \\&&
+\frac {b^2} 2\; E_M{}^{\overline b}\left[
D_{\overline{b}}D^{\overline c}{\Lambda}^{\overline{e f}}\left(F_{\overline c\underline d}{}^{\underline a} F^{\underline d}{}_{\overline{e f}} +
D_{\overline c} F^{\underline a}{}_{\overline{e f}}\right)	\right.
-F^{\underline a}{}_{\overline{e f}} F_{\underline c \overline d}{}^{\overline f}\left(
F^{\underline c}{}^{\overline{h d}} D_{\overline b}{\Lambda}_{\overline h}{}^{\overline e}
-F^{\underline c}{}^{\overline{h e}} D_{\overline b}{\Lambda}_{\overline h}{}^{\overline d}
\right)\nonumber \\
&&\ \ \ \ \ \ \ \ \ \ \ \ \ + \left. F^{\underline c}{}_{\overline{e f}}\;D_{\overline b}{\Lambda}^{\overline{e}}{}_{\overline g} \left(F^{\underline a}{}_{\underline{c d}} F^{\underline d \overline{g f}}
-D^{\underline a} F_{\underline c}{}^{\overline{g f}} + 2\; D_{\underline c} F^{\underline a  \overline{g f}}\right)
+F^{\underline a}{}_{\overline{e f}}D_{\overline b}\left(D^{\underline c}{\Lambda}^{\overline{e d}} F_{\underline c \overline{d}}{}^{\overline f}\right)\right] \ ,\nonumber
\eea
\bea
\Delta^{(2)} E_M{}^{\overline a} &=& -\, \frac {b^2} 2\;  E_M{}^{\underline b}\left[
														D^{\overline{a}}D^{\overline c}{\Lambda}^{\overline{e f}}\left(F_{\overline c\underline{d b}} F^{\underline d}{}_{\overline{e f}} + D_{\overline c} F_{\underline b \overline{e f}}\right)	
														\right.
-F_{\underline b \overline{e f}} F_{\underline c \overline d}{}^{\overline f}\left(
F^{\underline c}{}^{\overline{h d}} D^{\overline a}{\Lambda}_{\overline h}{}^{\overline e}
-F^{\underline c}{}^{\overline{h e}} D^{\overline a}{\Lambda}_{\overline h}{}^{\overline d}
\right)\nonumber \\
&&\ \ \ \ \ \ \ \ \ \ \ \ \ \ \  +\;\left. F^{\underline c}{}_{\overline{e f}}\;D^{\overline a}{\Lambda}^{\overline{e}}{}_{\overline g} \left(F_{\underline {b c d}} F^{\underline d \overline{g f}}
-D_{\underline b} F_{\underline c}{}^{\overline{g f}} + 2\; D_{\underline c} F_{\underline b}{}^{ \overline{g f}}
\right)
+F_{\underline b \overline{e f}}D^{\overline a}\left(D^{\underline c}{\Lambda}^{\overline{e d}} F_{\underline c \overline{d}}{}^{\overline f}\right)						
							\right]\nonumber\\&&
 -\frac {a b} 4\, E_M{}^{\underline b}\,\left[D^{\overline a}{\Lambda}^{\overline{ef}}
 \left( F_{\underline{b}\overline{eh}}\,F^{\overline h\underline{cd}} \,F_{\overline f\underline{cd}}- D_{\underline b} F_{\overline e}{}^{\underline{cd}}\;  F_{\overline f\underline{cd}}\right)
+F_{\underline b\overline{ef}}\, D^{\overline a}\left(D^{\overline e}{\Lambda}^{\underline{cd}}\; F^{\overline f}{}_{\underline{cd}}\right) \right.\nn\\&&
\;\;\;\;\;\;\;\;\;\;\;\;\;\;\;\;\;\;
\left.-\, D_{\underline b}{\Lambda}^{\underline{ef}}\left(F^{\overline a}{}_{\underline{eh}}
 F^{\underline h\overline{cd}} F_{\underline f\overline{cd}}
 -D^{\overline a}F_{\underline e}{}^{\overline{cd}} F_{\underline f\overline{cd}}\right) -\,F^{\overline a}{}_{\underline{ef}}\, D_{\underline b}\left( D^{\underline e}{\Lambda}^{\overline{cd}}\, F^{\underline f}{}_{\overline{cd}}\right)\,\right]\nn \\&&
+\frac {a^2} 2 \; E_M{}^{\underline b}\left[
														D_{\underline{b}}D^{\underline c}{\Lambda}^{\underline{e f}}\left(F_{\underline c\overline{d}}{}^{\overline{a}} F^{\overline d}{}_{\underline{e f}} + D_{\underline c} F^{\overline a}{}_{\underline{e f}}\right)	
														\right.
-F^{\overline{a}}{}_{\underline{e f}} F_{\overline c \underline d}{}^{\underline f}\left(
F^{\overline c}{}^{\underline{h d}} D_{\underline{b}}{\Lambda}_{\underline h}{}^{\underline e}
-F^{\overline c}{}^{\underline{h e}} D_{\underline{b}}{\Lambda}_{\underline h}{}^{\underline d}
\right)\nonumber \\
&&\ \ \ \ \ \ \ \ \ \ \ \ \ \ \  +\;\left. F^{\overline c}{}_{\underline{e f}}\;D_{\underline{b}}{\Lambda}^{\underline{e}}{}_{\underline g} \left(F^{\overline a}{}_{\overline{c d}} F^{\overline d \underline{g f}}
-D^{\overline a} F_{\overline c}{}^{\underline{g f}} + 2\; D_{\overline c} F^{ \overline{a}\underline{g f}}
\right)
+F^{\overline a}{}_{\underline{e f}}D_{\underline{b}}\left(D^{\overline c}{\Lambda}^{\underline{e d}} F_{\overline c \underline{d}}{}^{\underline f}\right)						
							\right] \ .\nonumber
\eea}
Note that the transformations of the two components of the generalized frame {\it look} symmetric with respect to the exchange of the projections and the parameters $a$ and $b$. In fact we can make this symmetry manifest 
\bea
\delta E_{M}{}^{\underline a} &=& \widehat {\cal L}_\xi E_M{}^{\underline a} + \left(  E_M{}^{\underline b}\Lambda_{\underline {bc}} +  E_M{}^{\overline b}\Delta_{\overline b \underline c} \right) \eta^{\underline {ca}} \ , \\
\delta E_{M}{}^{\overline a} &=& \widehat {\cal L}_\xi E_M{}^{\overline a} + \left( E_M{}^{\overline b}\Lambda_{\overline {bc}} +  E_M{}^{\underline b}\Delta_{\underline b \overline c } \right)\eta^{\overline {ca}} \ ,\eea
by introducing the following quantity
\bea
\Delta_{\overline b \underline c} &=& - \Delta_{\underline c \overline b} \ = \  E^M{}_{\overline b} \left( \Delta^{(1)} E_M{}^{\underline a}  + \Delta^{(2)} E_M{}^{\underline a} + \dots\right) \eta_{\underline {a c}} \ .
\eea

The idea is to see how the above transformations impact on the components of the generalized fields (\ref{ParamFrame}). There, in order to preserve duality and double Lorentz covariance, the generalized frame had to be parameterized in terms of two vielbeins (both related by a Lorentz transformation). Because we want to establish a connection with supergravity, we must break this symmetry to a single Lorentz transformation and gauge fix the two vielbeins to a single one
\be
\bar e_\mu{}^{ a} = \bar e_\mu{}^{\underline a} \, \delta_{\underline a}{}^{ a} = \bar e_\mu{}^{\overline a} \, \delta_{\overline a}{}^{ a} \ ,
\ee
To this end we have introduced Kronecker deltas to enforce the two Lorentz groups to carry the same set of indices $a, b , c, \dots$, which will be the Lorentz indices in supergravity. The connection to supergravity also requires a specific solution to the strong constraint $\tilde \partial^\mu = 0$, as is well known.

The generalized Green-Schwarz transformation depends on generalized fluxes and flat derivatives. We must then specify how these depend on the supergravity variables
\be
F_{A B C} = 3 D_{[A} E^P{}_{B} E^Q{}_{C]}\, \eta_{P Q} \ , \ \ \ 
F_A = 2 D_A d - \partial_M E^M{}_A \ , \ \ \  D_A = E^M{}_A \partial_M \ .
\ee
We must then specify how these depend on the supergravity fields. The three-form fluxes take the form
\bea
F_{\underline {abc}} &=& - \frac 3 {\sqrt{2}}  \delta^a_{\underline a} \delta^b_{\underline b} \delta^c_{\underline c} \left(\omega_{[abc]} - \frac 1 6 H_{abc}\right) = - \frac 1 {\sqrt{2}}  \delta^a_{\underline a} \delta^b_{\underline b} \delta^c_{\underline c} \left(2 \omega^{(-)}_{[abc]} + \omega^{(+)}_{[abc]} \right) \;,\label{Fddd}\\
F_{\overline {a} \underline{b c}} &=& \frac 1  {\sqrt{2}}  \delta^{a}_{\overline a} \delta^b_{\underline b} \delta^c_{\underline c} \left(\omega_{abc} - \frac 1 2 H_{abc}\right) = \frac 1  {\sqrt{2}}  \delta^{a}_{\overline a} \delta^b_{\underline b} \delta^c_{\underline c} \, \omega^{(-)}_{abc}\;, \label{Fudd}\\
F_{\underline {a} \overline{b c}} &=&  \frac 1 {\sqrt{2}}  \delta^a_{\underline a} \delta^b_{\overline b} \delta^c_{\overline c} \left(\omega_{abc} + \frac 1 2 H_{abc}\right) = \frac 1 {\sqrt{2}}  \delta^a_{\underline a} \delta^b_{\overline b} \delta^c_{\overline c} \, \omega^{(+)}_{abc}\;, \label{Fduu}\\
F_{\overline {abc}} &=& -\frac 3 {\sqrt{2}}  \delta^a_{\overline a} \delta^b_{\overline b} \delta^c_{\overline c} \left(\omega_{[abc]} + \frac 1 6 H_{abc}\right) =  -\frac 1 {\sqrt{2}}  \delta^a_{\overline a} \delta^b_{\overline b} \delta^c_{\overline c} \left(2 \omega^{(+)}_{[abc]} + \omega^{(-)}_{[abc]}\right) \ ,\label{Fuuu}
\eea
while the vectorial fluxes read
\bea
F_{\underline a} &=& \sqrt{2} \delta^a_{\underline a} \left(\omega_{[ad]}{}^d - D_a \overline{\phi}\right) = \frac 1 {\sqrt{2}} \delta^a_{\underline a} \left( \omega^{(+)}_{[ad]}{}^d   + \omega^{(-)}_{[ad]}{}^d - 2 D_a \overline{\phi}\right)\;,\\
F_{\overline a} &=& - \sqrt{2} \delta^a_{\overline a} \left(\omega_{[ad]}{}^d - D_a \overline{\phi}\right) = - \frac 1 {\sqrt{2}} \delta^a_{\overline a} \left( \omega^{(+)}_{[ad]}{}^d   + \omega^{(-)}_{[ad]}{}^d - 2 D_a \overline{\phi}\right) \ .
\eea
Flat derivatives are given by
\be
D_{\underline a} = - \frac 1 {\sqrt{2}} \delta^a_{\underline a} \, D_a  \ , \ \ \ \  D_{\overline a} = \frac 1 {\sqrt{2}} \delta^a_{\overline a} \, D_a \ , \ \ \ \ D_a = \bar e^\mu{}_a \partial_\mu \ ,
\ee
and we see the appearance of the leading order of the spin connections with torsion
\be
\omega^{(\pm)}_{abc} = \omega_{abc} \pm \frac 1 2 H_{a b c} \ , 
\ee
where curved indices are obviously flattened with the vielbein $\bar e$. It is important to emphasize that all these quantities are defined in terms of the over lined component fields $\bar e$, $\bar b$ and $\bar \phi$.

Because we have gauged fixed the vielbein, the double Lorentz symmetry is now broken to its diagonal subgroup, and then the two Lorentz parameters are no longer independent. We must then explore how they are related, and what is the most convenient way to express them in terms of the Lorentz parameter in supergravity.
To this end we first write the two Lorentz invariant metrics in terms of a single one
\be
g^{\underline {ab}} =  \delta^{\underline a}_a \delta^{\underline b}_b   g^{ab} \ , \ \ \ \ \ g^{\overline {ab}} =  \delta^{\overline a}_a \delta^{\overline b}_b   g^{ab} \ ,
\ee
and also express all the Lorentz parameters (including the generalized Green-Schwarz deformation $\Delta$) in terms of the same set of indices
\be
\Lambda_{\overline {ab}} = \delta^a_{\overline a} \delta^b_{\overline b} \overline \Lambda_{ab}  \ , \ \ \ \ \Lambda_{\underline {ab}} = \delta^a_{\underline a} \delta^b_{\underline b} \underline \Lambda_{ab} \ , \ \ \ \
\Delta_{\overline {a} \underline b} = \delta^a_{\overline a} \delta^b_{\underline b} \Delta_{ab} \ . \label{GaugeFixLorentzParam}
\ee
We then get two different transformations for the vielbein from the transformations of the two projections on the generalized frame
\bea
\delta E^{\mu \underline a} \ \ \ \ \ \to \ \ \ \ \ \delta \bar e^\mu{}_{a} &=& \widehat {\cal L}_\xi \bar e^\mu{}_a - \bar e^{\mu b} \left(\underline \Lambda_{b a} + \Delta_{b a}\right)  \ , \\
\delta E^{\mu \overline a} \ \ \ \ \ \to \ \ \ \ \ \delta \bar e^\mu{}_{a} &=& \widehat {\cal L}_\xi \bar e^\mu{}_a + \bar e^{\mu b} \left(\overline \Lambda_{b a} - \Delta_{a b}\right) \ . \label{transfoverlinebien}
\eea
They must obviously coincide
\be
\delta \bar e_\mu{}^a = \widehat {\cal L}_\xi \bar e_\mu{}^a + \bar e_{\mu b} \left(\underline \Lambda^{a b} + \Delta^{a b}\right) = \widehat {\cal L}_\xi \bar e_\mu{}^a + \bar e_{\mu b} \left(\overline \Lambda^{b a} + \Delta^{b a}\right) \ , \label{transfvielbein}\ee
and this imposes the required relation between the double Lorentz parameters
\be
\underline \Lambda_{a b} = - \overline \Lambda_{a b} - 2 \Delta_{[ab]} \ .
\ee
From the transformation of the vielbein we can read that of the metric, and from the generalized frame one can extract in addition the transformation of the two-form
\bea
\delta \bar g_{\mu \nu} &=& \widehat {\cal L}_\xi  \bar g_{\mu \nu} + 2 \bar e_\mu{}^a \bar e_\nu{}^b \Delta_{(ab)} \ , \\
\delta \bar b_{\mu \nu} &=& \widehat {\cal L}_\xi  \bar b_{\mu \nu} - 2 \bar e_\mu{}^a \bar e_\nu{}^b \Delta_{[ab]} \ . \label{transftwoform}
\eea
Finally, regarding generalized diffeomorphisms we find
\bea
\widehat {\cal L}_\xi \bar e_{\mu}{}^{a} = L_\xi e_\mu{}^a \ , \ \ \   \widehat {\cal L}_\xi \bar b_{\mu \nu} = L_\xi \bar b_{\mu \nu} + 2 \partial_{[\mu} \bar \xi_{\nu]} \ . \label{Lieeb}
\eea

We now have the transformations of the over-lined vielbein and two-form in (\ref{transfvielbein}), (\ref{transftwoform}), (\ref{Lieeb}). The plan is to find field and parameter redefinitions that trivialize the anomalous Lorentz transformation of the vielbein, and take that of the two-form to its expected bi-parametric Lorentz Green-Schwarz form. We will name the resulting fields without over-lines $e_\mu{}^a$ and $b_{\mu \nu}$, and demand that they transform as in (\ref{BdRscheme}).

The only explicit derivative expansion in terms of the over-lined fields enters through $\Delta$ 
\be
\Delta_{ab} = \Delta^{(1)}_{a b} + \Delta^{(2)}_{ab} + \dots \;,
\ee
where the supra-label $(n)$ means that it explicitly contains $2n$ derivatives. The first order is
\be
\Delta^{(1)}_{ab} = - \frac{a}{4}\, {D}_{b}{{{\underline \Lambda}}^{c d\, }}\,
\omega^{(-)}_{a c d\, }
+ \frac{b}{4}\, {D}_{a}{{{\overline \Lambda}}^{c d\, }}\,
\omega^{(+)}_{b c d\, } \ ,
\ee
and the second
{\small 
\bea
\Delta^{(2)}_{a b} &=&  a^2 \left[ \frac{1}{8} \right.\, {D}^{c}{{{\underline \Lambda}}^{d\,  e}}\,
{D}_{b}{\omega^{(-)}_{c d\, }\,^{f}}\,  \omega^{(-)}_{a e f}
- \frac{1}{8}\, {D}_{b}{{D}^{c}{{{\underline \Lambda}}^{d\,  e}}\,
}\,  \left( {D}_{c}{\omega^{(-)}_{a d\,  e}}\,  +   \omega^{(+)}_{c
a}\,^{f} \omega^{(-)}_{f d\,  e} +  \omega^{(-)}_{a d\, }\,^{f}
\omega^{(-)}_{c e f}\right) \nn \\ &&
\ \ \ \ \ -  {D}_{b}{{{\underline \Lambda}}^{c d\, }}\,  \left(\frac{1}{8}\right.\,
{D}_{a}{\omega^{(-) e}\,_{c}\,^{f}}\,  \omega^{(-)}_{e d\,  f} -
\frac{1}{4}\, {D}^{e}{\omega^{(-)}_{a c}\,^{f}}\,  \omega^{(-)}_{e
d\,  f} - \frac{1}{12}\, \omega^{(+)}_{a}\,^{e f} \omega^{(-)}_{e
c}\,^{g} \omega^{(-)}_{f d\,  g}  \nn \\ && \ \ \ \ \ \ \ \ \ \ \ \ \ \  + \frac{1}{6}\,
\omega^{(+) e}\,_{a}\,^{f} \omega^{(-)}_{e c}\,^{g}
\omega^{(-)}_{f d\,  g} + \frac{1}{8}\, \omega^{(-)}_{a c}\,^{e}
\omega^{(-) f}\,_{d\, }\,^{g} \omega^{(-)}_{f e g} -
\frac{1}{24}\, \omega^{(-)}_{a}\,^{e f} \omega^{(-)}_{e c}\,^{g}
\omega^{(-)}_{f d\,  g} \nn \\ && \ \ \ \ \ \ \ \ \ \ \ \ \ \  -  \left.\left.\frac{1}{8}\, \omega^{(-)}_{a}\,^{e f}
\omega^{(-) g}\,_{c e} \omega^{(-)}_{g d\,  f} + \frac{1}{12}\,
\omega^{(-) e}\,_{a}\,^{f} \omega^{(-)}_{e c}\,^{g}
\omega^{(-)}_{f d\,  g}\right)\right] \nn \\
& &+ \frac {a b} {16} \, \left[
 \vphantom{\frac{1}{16}}  {D}_{a}\left({{D}^{c}{{{\underline \Lambda}}^{d\,  e}}\,
}\,  \omega^{(-)}\,_{f d  e}\right) \omega^{(+)}_{b c}{}^{f}
 - \, {D}_{b}{{{\underline \Lambda}}^{c d\, }}\,  \left(
{D}_{a}{\omega^{(+)}_{c}\,^{e f}}\,  \omega^{(+)}_{d\,  e f} +
 \omega^{(-)}_{a c}\,^{e} \omega^{(+)}_{d\, }\,^{f
g} \omega^{(+)}_{e f g}\right) \right] \nn \\
&& - \left\{ \vphantom{\frac{1}{16}}  a \leftrightarrow b \, , \, \underline \Lambda \rightarrow \overline\Lambda \, , \, (+) \leftrightarrow (-) \right\}  \ .
\eea}

So we now propose an expansion for the fields and parameters as follows
\bea
\overline \Lambda_{ab} &=& \Lambda_{ab} + \Lambda^{(1)}_{ab} + \Lambda^{(2)}_{ab} + \dots - \Delta^{(1,1)}_{[a b]} - \Delta^{(1,2)}_{[a b]} - \Delta^{(2,2)}_{[a b]}  - \dots \ , \nn\\
\underline \Lambda_{ab} &=& - \Lambda_{ab} - \Lambda^{(1)}_{ab} - \Lambda^{(2)}_{ab} - \dots - \Delta^{(1,1)}_{[a b]} - \Delta^{(1,2)}_{[a b]} - \Delta^{(2,2)}_{[a b]}  - \dots \ , \nn\\
\bar e^\mu{}_a &=& e^{\mu b}  (g_{b a} +  e_{(1)b a} +  e_{(2) b a} + \dots )  \ ,\label{expansions}\\
\bar e_\mu{}^a &=& e_{\mu b} (g^{a b}  -  e_{(1)}^{ab} -  e_{(2)}^{a b} +  e_{(1)}^{a c} e_{(1)c}{}^b  +  \dots ) \ , \nn \\
\bar b_{\mu \nu} &=& e_{\mu}{}^a e_\nu{}^b (b_{a b} +  \, b^{(1)}_{a b} +  \, b^{(2)}_{a b} + \dots ) \ , \ \ \ b_{\mu \nu} \equiv e_{\mu}{}^a e_\nu{}^b b_{a b} \ , \nn \\
\bar \xi_\mu &=& e_\mu{}^a (\lambda_{a} +  \lambda^{(1)}_a +  \lambda^{(2)}_a + \dots) \ , \ \ \ \lambda_{\mu} \equiv e_\mu{}^a \lambda_a \ .\nn
\eea
The fields $e_\mu{}^a$ and $b_{\mu \nu}$ are the supergravity fields, expected to transform as in (\ref{BdRscheme}). The vielbein defines the metric $g_{\mu \nu} = e_\mu{}^a g_{a b} e_\nu{}^b$ and the flat derivatives $\partial_a = e^\mu{}_a \partial_\mu$ in supergravity. The expansions of the Lorentz parameters obey the gauge fixing condition (\ref{GaugeFixLorentzParam}). Let us briefly explain the notation.  The power in $\Delta^{(n)}$ signals its dependence on $a^p b^q$ with $p+q = n$ when written in terms of $\underline \Lambda$, $\overline \Lambda$ and $\bar e$. Instead, the power in $\Lambda^{(n)}$ and $e_{(n)}$ signals its dependence on $a^p b^q$ with $p + q = n$ when written in terms of $\Lambda$ and $e$. The $n$-th power of $\Delta^{(k)}$ when written in terms of $\Lambda$ and $e$ will be noted $\Delta^{(k,n)}$.

It is useful to define an operator that measures the Lorentz non-covariance of the fields
\be
\delta^{\not c} V^a \equiv  (\delta - \widehat {\cal L}_\xi) V^a - V^b \, \Lambda_b{}^a \ .
\ee
Then, for instance we read from (\ref{BdRscheme}) that 
\bea
\delta^{\not c} e_\mu{}^a &=& \delta^{\not c} e^\mu{}_a = 0 \ ,\label{lookfor} \\
\delta^{\not c} b_{\mu \nu} &=& \frac a 2 \partial_{[\mu} \Lambda^{a b} \widehat \omega^{(-)}_{\nu]ab} - \frac b 2 \partial_{[\mu} \Lambda^{a b} \widehat \omega^{(+)}_{\nu]ab} \ . \nn
\eea
We can now insert the expansions (\ref{expansions}) into the transformation of the over-lined vielbein (\ref{transfoverlinebien}), impose (\ref{lookfor}) and decompose order by order:
\bea
{\cal O}(1) \  &\to& \  \left[\delta^{\not c} e_{(1)a b}\right]^{(1)} - \Lambda^{(1)}_{a b} = - \Delta^{(1,1)}_{(a b)} \ , \label{orderbyordereqs}\\
{\cal O}(2) \ &\to& \ \left[\delta^{\not c} e_{(2) ab}\right]^{(2)} - \Lambda^{(2)}_{a b} = - \Delta^{(1,2)}_{(ab)} - \Delta^{(2,2)}_{(a b)} - e_{(1)a}{}^c  \Lambda^{(1)}_{b c} - e_{(1)a}{}^c  \Delta^{(1,1)}_{(bc)} - \left[\delta^{\not c} e_{(1)a b}\right]^{(2)} \ ,\nn\\ &\vdots& \nn
\eea
The LHS of these equations contain the unknowns. Once a given order is solved, then the RHS of the following equation is known, so this must be solved iteratively.
Similar expressions are obtained from the transformation of the two-form (\ref{transftwoform})
\bea
{\cal O}(1)     &\to&  \  \left[\delta^{\not c} b^{(1)}_{a b}\right]^{(1)} = 2 {\cal D}_{[a} \lambda^{(1)}_{b]} - 2 \Delta^{(1,1)}_{[a b]} - \frac a 2 \partial_{[a} \Lambda^{c d} \omega^{(-)}_{b]c d} + \frac b 2 \partial_{[a} \Lambda^{c d} \omega^{(+)}_{b]c d} \ ,\label{orderbyordereqsb}\\
{\cal O}(2)   &\to& \ \left[\delta^{\not c} b^{(2)}_{ab}\right]^{(2)} = 2 {\cal D}_{[a} \lambda^{(2)}_{b]} - 2 \Delta^{(1,2)}_{[a b]}  - 2 \Delta^{(2,2)}_{[a b]} - 2 e_{(1)\, a}^c \Delta^{(1,1)}_{[b c]} + 2 e_{(1)\, b}^c \Delta^{(1,1)}_{[a c]} - \left[\delta^{\not c} b^{(1)}_{a b}\right]^{(2)} \nn\\ &\vdots& \ \ \ \ \ \ \ \ \ \ \ \ \ \ \ \ \ \ \  -\,  \frac 3 8\, a^2\, \partial_{[a} \Lambda^{c d} \Omega^{(-)}_{b] c d} +\,  \frac 3 8\, a b\, \partial_{[a} \Lambda^{c d}\left( \Omega^{(+)}_{b] c d}  -  \Omega^{(-)}_{b] c d} \right)+  \frac 3 8\, b^2\, \partial_{[a} \Lambda^{c d} \Omega^{(+)}_{b] c d} \ . \nn
\eea

The corrections to the Lorentz parameter $\Lambda^{(n)}_{a b}$ are always chosen so as to cancel the antisymmetric part of the RHS in (\ref{orderbyordereqs}), namely
\bea
\Lambda^{(1)}_{a b} &=& 0\ ,\\
\Lambda^{(2)}_{a b} &=& \frac 1 2 e_{(1)[a}{}^c  \Delta^{(1,1)}_{b]c} + \frac 1 2 \Delta^{(1,1)}_{c [b} e_{(1)a]}{}^c \ , \nn\\ &\vdots& \nn
\eea
One these are fixed, we can proceed order by order to find the corrections to the fields. To ${\cal O}(1)$ in (\ref{orderbyordereqs}) and (\ref{orderbyordereqsb}) we find the following redefinitions up to covariant contributions
\bea
e_{(1)a b} &=& -  \frac a 8 \omega^{(-)}_{a c d}  \omega^{(-) c d}_b  -  \frac b 8 \omega^{(+)}_{a c d}  \omega^{(+) c d}_b \label{eRedef1}\\
b^{(1)}_{a b} &=& 0 \ , \ \ \ \ \
\lambda^{(1)}_{a} \,=\, 0 \ .\label{bRedef1}
\eea
They correspond to the minimal redefinitions required to connect with supergravity, sometimes dubbed the {Bergshoeff-de Roo} scheme. Notice that to this order the two-form remains uncorrected and inherits its anomalous Lorentz Green-Schwarz transformation directly from that in the generalized picture. Also, there is no need to redefine the one-form parameter to this order. These expression reproduce exactly the results found in \cite{MarquesNunez}.

The solution to ${\cal O}(2)$ in (\ref{orderbyordereqs}) and (\ref{orderbyordereqsb}), up to covariant contributions, is quite involved but still accessible to an educated guess
{\small \bea
e_{(2)a b} &=& - a^2 \frac{3}{32}\, \omega^{(-)}_{b c d} \Omega^{(-)}_{a}\,^{c d} +
a b \frac{3}{32}\, \omega^{(-)}_{b c d} \Omega^{(+)}_{a}\,^{c d} +
a b \frac{3}{32}\, \omega^{(+)}_{b c d} \Omega^{(-)}_{a}\,^{c d} -
 b^2 \frac{3}{32}\, \omega^{(+)}_{b c d} \Omega^{(+)}_{a}\,^{c d} + \left(a \leftrightarrow b\right)  \nn \\
&& + a^2 \left[ - \frac{1}{32} \right.\, {\partial}^{c}{\omega^{(-)}_{a}\,^{d e}}\,
{\partial}_{c}{\omega^{(-)}_{b d e}}\,
- \frac{1}{32}\, {\partial}_{a}{\omega^{(-) c d e}}\,  \omega^{(-)}_{b
c}\,^{f} \omega^{(-)}_{f d e}
+ \frac{1}{16}\, {\partial}_{a}{\omega^{(-) c d e}}\,  \omega^{(-)}_{b
d}\,^{f} \omega^{(-)}_{c e f} \nn \\
&& \ \ \ \  \ \ \ \   - \frac{1}{8}\, {\partial}^{c}{\omega^{(-)}_{a}\,^{d e}}\,  \omega^{(-)}_{b
d}\,^{f} \omega^{(-)}_{c e f}
- \frac{1}{16}\, {\partial}^{c}{\omega^{(-)}_{a}\,^{d e}}\,  \omega^{(-)}_{c
b}\,^{f} \omega^{(-)}_{f d e}
+ \frac{1}{16}\, {\partial}^{c}{\omega^{(-)}_{a}\,^{d e}}\,  \omega^{(+)}_{b
c f} \omega^{(-) f}\,_{d e} \nn \\
&&  \ \ \ \  \ \ \ \  - \frac{1}{16}\, \omega^{(-)}_{a}\,^{c d} \omega^{(-)}_{b c}\,^{e}
\omega^{(-) f}\,_{d}\,^{g} \omega^{(-)}_{f e g}
+ \frac{3}{256}\, \omega^{(-)}_{a}\,^{c d} \omega^{(-)}_{b}\,^{e f}
\omega^{(-) g}\,_{c d} \omega^{(-)}_{g e f}\nn \\
&&  \ \ \ \  \ \ \ \
+ \frac{1}{16}\, \omega^{(-)}_{a}\,^{c d} \omega^{(-)}_{b}\,^{e f}
\omega^{(-) g}\,_{c e} \omega^{(-)}_{g d f} - \frac{1}{8}\, \omega^{(-)}_{a}\,^{c d} \omega^{(-) e}\,_{b}\,^{f}
\omega^{(-)}_{e c}\,^{g} \omega^{(-)}_{f d g} \nn \\ &&  \ \ \ \   \ \ \ \
- \frac{1}{32}\, \omega^{(-) c}\,_{a}\,^{d} \omega^{(-)}_{c b}\,^{e}
\omega^{(-)}_{d}\,^{f g} \omega^{(-)}_{e f g}
+ \frac{1}{8}\, \omega^{(+)}_{b c d} \omega^{(-)}_{a}\,^{e f}
\omega^{(-) c}\,_{e}\,^{g} \omega^{(-) d}\,_{f g} \nn \\
&&  \ \ \ \  \ \ \ \  \left. + \frac{1}{16}\, \omega^{(+)}_{b c d} \omega^{(-) c}\,_{a}\,^{e}
\omega^{(-)}_{e}\,^{f g} \omega^{(-) d}\,_{f g}
- \frac{1}{32}\, \omega^{(+)}_{a c}\,^{d} \omega^{(+)}_{b e d}
\omega^{(-) c f g} \omega^{(-) e}\,_{f g} + \left(a \leftrightarrow b\right) \right] \nn
\\
&& + a b \left[\frac{1}{32} \right.\, {\partial}^{c}{{H}^{d e f}}\,  {H}_{a d e} {\omega}_{b c f}
+ \frac{1}{16}\, {\partial}^{c}{{\omega}^{d e f}}\,  {\omega}_{a c d} {\omega}_{b e f}
+ \frac{3}{256}\, {H}_{a}\,^{c d} {H}_{c d}\,^{e} {\omega}_{b}\,^{f g}
{\omega}_{e f g}  \label{eRedef2}\\
&&  \ \ \ \  \ \ \ \  - \frac{1}{32}\, {H}_{a}\,^{c d} {H}_{c d}\,^{e} {\omega}_{b}\,^{f g}
{\omega}_{f e g}
+ \frac{1}{16}\, {H}_{a}\,^{c d} {H}_{c}\,^{e f} {\omega}_{b e}\,^{g}
{\omega}_{g d f}
- \frac{3}{512}\, {H}_{a}\,^{c d} {H}_{b}\,^{e f} {\omega}^{g}\,_{c d}
{\omega}_{g e f} \nn \\
&&  \ \ \ \  \ \ \ \ - \frac{3}{256}\, {H}_{a}\,^{c d} {H}^{e f g} {\omega}_{b e f} {\omega}_{g c d}
- \frac{3}{512}\, {H}^{c d e} {H}_{c}\,^{f g} {\omega}_{a d e} {\omega}_{b f g}
- \frac{1}{64}\, {H}^{c d e} {H}_{c}\,^{f g} {\omega}_{a d f} {\omega}_{b e g}
\nn \\ &&  \ \ \ \  \ \ \ \  + \frac{1}{16}\, {\omega}_{a}\,^{c d} {\omega}_{b}\,^{e f} {\omega}_{c
d}\,^{g} {\omega}_{g e f}
- \frac{1}{16}\, {\omega}_{a}\,^{c d} {\omega}_{b}\,^{e f} {\omega}_{c
e}\,^{g} {\omega}_{d f g}
\nn \\  && \ \ \ \  \ \ \ \ \left. + \frac{3}{128}\, {\omega}_{a}\,^{c d} {\omega}_{b}\,^{e f}
{\omega}^{g}\,_{c d} {\omega}_{g e f} + \left( a \leftrightarrow b \right) \right] \nn \\
&& + b^2 \left[- \frac{1}{32} \right.\, {\partial}^{c}{\omega^{(+)}_{a}\,^{d e}}\,
{\partial}_{c}{\omega^{(+)}_{b d e}}\,
- \frac{1}{32}\, {\partial}_{a}{\omega^{(+) c d e}}\,  \omega^{(+)}_{b
c}\,^{f} \omega^{(+)}_{f d e}
+ \frac{1}{16}\, {\partial}_{a}{\omega^{(+) c d e}}\,  \omega^{(+)}_{b
d}\,^{f} \omega^{(+)}_{c e f} \nn \\
&& \ \ \ \  \ \ \ \   - \frac{1}{8}\, {\partial}^{c}{\omega^{(+)}_{a}\,^{d e}}\,  \omega^{(+)}_{b
d}\,^{f} \omega^{(+)}_{c e f}
- \frac{1}{16}\, {\partial}^{c}{\omega^{(+)}_{a}\,^{d e}}\,  \omega^{(+)}_{c
b}\,^{f} \omega^{(+)}_{f d e}
+ \frac{1}{16}\, {\partial}^{c}{\omega^{(+)}_{a}\,^{d e}}\,  \omega^{(-)}_{b
c f} \omega^{(+) f}\,_{d e} \nn \\
&&  \ \ \ \  \ \ \ \  - \frac{1}{16}\, \omega^{(+)}_{a}\,^{c d} \omega^{(+)}_{b c}\,^{e}
\omega^{(+) f}\,_{d}\,^{g} \omega^{(+)}_{f e g}
+ \frac{3}{256}\, \omega^{(+)}_{a}\,^{c d} \omega^{(+)}_{b}\,^{e f}
\omega^{(+) g}\,_{c d} \omega^{(+)}_{g e f}\nn \\
&&  \ \ \ \  \ \ \ \
+ \frac{1}{16}\, \omega^{(+)}_{a}\,^{c d} \omega^{(+)}_{b}\,^{e f}
\omega^{(+) g}\,_{c e} \omega^{(+)}_{g d f} - \frac{1}{8}\, \omega^{(+)}_{a}\,^{c d} \omega^{(+) e}\,_{b}\,^{f}
\omega^{(+)}_{e c}\,^{g} \omega^{(+)}_{f d g} \nn \\ &&  \ \ \ \   \ \ \ \
- \frac{1}{32}\, \omega^{(+) c}\,_{a}\,^{d} \omega^{(+)}_{c b}\,^{e}
\omega^{(+)}_{d}\,^{f g} \omega^{(+)}_{e f g}
+ \frac{1}{8}\, \omega^{(-)}_{b c d} \omega^{(+)}_{a}\,^{e f}
\omega^{(+) c}\,_{e}\,^{g} \omega^{(+) d}\,_{f g} \nn \\
&&  \ \ \ \  \ \ \ \  \left. + \frac{1}{16}\, \omega^{(-)}_{b c d} \omega^{(+) c}\,_{a}\,^{e}
\omega^{(+)}_{e}\,^{f g} \omega^{(+) d}\,_{f g}
- \frac{1}{32}\, \omega^{(-)}_{a c}\,^{d} \omega^{(-)}_{b e d}
\omega^{(+) c f g} \omega^{(+) e}\,_{f g} + \left(a \leftrightarrow b\right) \right] \;,\nn
\eea
\bea
b^{(2)}_{a b} &=& a^2 \left[ \frac{1}{4}\right.\, {\partial}_{a}{\omega^{(-) c d e}}\,
{\partial}_{c}{\omega^{(-)}_{b d e}}\,
 - \frac{1}{4}\,
{\partial}_{a}{\omega^{(-) c d e}}\,  \omega^{(-) f}\,_{d e}
\omega^{(+)}_{b c f}
+ \frac{1}{8}\, {\partial}_{a}{\omega^{(-) c d
e}}\,  \omega^{(-)}_{b c}\,^{f} \omega^{(-)}_{f d e}   \nn \\ && \ \ \ \ \ \ \ \
+ \frac{1}{4}\,
{\partial}_{a}{\omega^{(-) c d e}}\,  \omega^{(-)}_{c b}\,^{f}
\omega^{(-)}_{f d e}
+ \left. \frac{1}{4}\, {\partial}_{a}{\omega^{(-) c d
e}}\,  \omega^{(-)}_{b d}\,^{f} \omega^{(-)}_{c e f} \right]  \nn \\ &&
+ a b \left[ - \frac{1}{8}\right.\, {\partial}^{c}{{H}^{d e f}}\,  {\omega}_{a c
d} {\omega}_{b e f}
+ \frac{1}{4}\, {\partial}^{c}{{\omega}^{d e f}}\,
 {H}_{a e f} {\omega}_{b c d} - \frac{1}{16}\, {H}_{a}\,^{c d}
{H}_{c}\,^{e f} {H}_{d e}\,^{g} {\omega}_{b f g}  \nn \\ &&  \ \ \ \ \ \ \ \
- \frac{1}{8}\,
{H}^{c d e} {\omega}_{a c d} {\omega}_{b}\,^{f g} {\omega}_{f e g}
-
\frac{1}{4}\, {H}^{c d e} {\omega}_{a c}\,^{f} {\omega}_{b d}\,^{g}
{\omega}_{f e g}    \label{bRedef2}\\ &&  \ \ \ \ \ \ \ \
+ \frac{1}{4}\, {H}_{a}\,^{c d} {\omega}_{b}\,^{e f}
{\omega}^{g}\,_{c d} {\omega}_{e f g}
- \left.\frac{1}{4}\, {H}_{a}\,^{c d}
{\omega}_{b}\,^{e f} {\omega}_{e c}\,^{g} {\omega}_{f d g}\right]  \nn \\ &&
- b^2 \left[\frac{1}{4}\right.\, {\partial}_{a}{\omega^{(+) c d e}}\,
{\partial}_{c}{\omega^{(+)}_{b d e}}\,
 - \frac{1}{4}\,
{\partial}_{a}{\omega^{(+) c d e}}\,  \omega^{(+) f}\,_{d e}
\omega^{(-)}_{b c f}
+ \frac{1}{8}\, {\partial}_{a}{\omega^{(+) c d
e}}\,  \omega^{(+)}_{b c}\,^{f} \omega^{(+)}_{f d e}   \nn \\ && \ \ \ \ \ \ \ \
+ \frac{1}{4}\,
{\partial}_{a}{\omega^{(+) c d e}}\,  \omega^{(+)}_{c b}\,^{f}
\omega^{(+)}_{f d e}
+ \left. \frac{1}{4}\, {\partial}_{a}{\omega^{(+) c d
e}}\,  \omega^{(+)}_{b d}\,^{f} \omega^{(+)}_{c e f}\right] \;, \nn \\
\lambda^{(2)}_{a} &=& - \frac {a^2} 8 \partial^{b} \Lambda^{c d} \partial_a \omega^{(-)}_{b c d} + \frac {b^2} 8 \partial^{b} \Lambda^{c d} \partial_a \omega^{(+)}_{b c d}\;.
\eea}
We see to this order that the structure of the expected transformations (\ref{BdRscheme}) can be maintained, but now at the expense of redefining the one-form parameter. 

Let us finally comment on the dilaton field. It can be read from this expression
\bea
d = \bar \phi - \frac 1 4 \log |\bar g| = \phi - \frac 1 4 \log |g| \ .
\eea
The field $d$ transforms as usual, so $\bar \phi$ receives corrections in the Lorentz transformations but transforms as usual under Buscher rules (together with the T-duality covariant fields $\bar g$ and $\bar b$). On the other hand, $\phi$ is Lorentz invariant, but its Buscher rules receive higher derivative corrections.

\section{Summary and Outlook}

We extended the results in \cite{GenBdR} to the bi-parametric family of T-duality deformations of DFT introduced in \cite{MarquesNunez}. The strategy relies on an duality covariant generalization of the Bergshoeff-de Roo (BdR) identification between gauge (independent) and gravity (composite) dof, originally designed to implement higher order supersymmetry in heterotic supergravity \cite{BdR}. On the one hand this identification relates interactions with different amount of derivatives, and on the other, due to its duality covariance it also relates same order interactions beyond those obtained from the original identification \cite{BdR}. As an example, it not only enforces the expected Lorentz Chern-Simons terms, but also the quadratic Riemann interactions \cite{MarquesNunez}, and presumably the full tower of higher derivatives implicitly contained in them.  From the Point of view of DFT, the identification deforms the double Lorentz symmetry. This deformation was dubbed the generalized Green-Schwarz transformation, computed in \cite{MarquesNunez} to first order, in \cite{GenBdR} to second order in the mono-parametric case, and here to second order in the bi-parametric case. We also introduced the second order invariant bi-parametric action. 

\noindent Let us briefly provide a guide to the main original results in the paper:
\begin{itemize}
    \item The bi-parametric identification was established in (\ref{EABC}-\ref{LockingBi}). It is exact, and generates the generalized Green-Schwarz transformation (\ref{exactGGS}) in DFT. 
    \item We develop the perturbative expansion in powers of $\alpha'$ to second order. That of the generalized Green-Schwarz transformation can be found in (\ref{VarEup}-\ref{VarEdown}) and the invariant action in Section \ref{SectBi-parDFT}.
    \item We show in Section \ref{SecSugra} that this deformation reproduces the Green-Schwarz transformation of the two-form (\ref{BdRscheme}) in the so-called Bergshoeff-de Roo scheme of supergravity. For this it is necessary to realize the non-covariant field redefintions (\ref{eRedef2}-\ref{bRedef2}).
\end{itemize}

\noindent There are a number of questions that arise, and many open problems that remain:
\begin{itemize}
    \item The generalized Green-Schwarz transformation is infinitesimal, as it depends linearly on the Lorentz parameters. From the point of view of the identification, it arises from infinitesimal generalized diffeomorphisms in the extended space. It is then natural to ask what the finite version of these transformations is. For generalized diffeomorphisms, these issues were extensively discussed in \cite{Finite}. To first order in $\alpha'$ the finite form of the generalized Green-Schwarz transformation was originally derived in \cite{YB}. The computation strongly relies on the imposition of the strong constraint. The same result was reproduced in a  double language and very nicely related to Born geometry in  \cite{Hassler}. 
    
    \item Recently, the first order generalized Green-Schwarz transformation of \cite{MarquesNunez} was exploited in \cite{Hassler,BorsatoWulff,CodinaMarques} to study how higher derivatives deform the action of generalized T-dualities. These include the standard Abelian ones, and also non-Abelian and Poisson-Lie, possibly among further generalizations. The idea is that within DFT there are two important Lorentz gauges, one that allows an immediate contact with supergravity, and another one in which generalized dualities act linearly on the backgrounds. Studying the effect of generalized dualities on supergravity backgrounds then requires the composition of Lorentz and $O(D,D)$ transformations. In this paper we provide all the necessary tools to pursue this line of research to second order. For this to be possible, the first step should be to find the corresponding finite form of the Green-Schwarz transformation presented here. It should be noted that the action and the equations of motion can be written only in terms of flat derivatives and generalized fluxes, a fact that is important to guarantee that the action of generalized dualities works like a solution generating technique.
    
    \item The first order corrections that emerge from the bi-parametric deformation were shown in \cite{MarquesNunez} to contain Lorentz Chern-Simons corrections to the three-form curvature \cite{GreenSchwarz} and quadratic Riemann interactions \cite{MetsaevTseytlin}. We expect that the results here yield cubic Riemann plus Gauss-Bonnet terms both for the bosonic \cite{CubicBosonic} and HSZ cases \cite{HSZ,HSZcont}, but no cubic Riemann interaction for the heterotic string \cite{BdR}. To higher orders, we expect that the quartic Riemann interactions fall in two categories. Those  that are captured by this setup (the ones present in \cite{BdR}) and those proportional to $\zeta(3)$ \cite{GrossSloan}, which presumably require new deformations or the existence of a new invariant in DFT with eight derivatives. All these speculations remain to be confirmed.
    
    \item Our results can presumably be used to extract non-perturbative aspects for this tower of corrections. At the moment how to do this, and what sort of information one should aim at, remains unclear to us. An interesting aspect of this construction is that it allows a systematic procedure to extract order by order corrections in a perturbative expansion. The counterpoint is that the expressions that emerge from such an expansion get harder to deal with as the orders increase. An example of this is the second order action, which we showed here as a existence proof, but whose length makes it hard to work with. A smarter embedding of the double space into the extended space could simplify the outcome. It is known for instance that the first order action found in \cite{MarquesNunez} and later rewritten in terms of generalized fluxes in \cite{OddStory} can be drastically simplified using Bianchi Identities, as shown in \cite{Wulff}. Another source of simplification is to truncate the theory before implementing the iterative approach, $e.g.$ we could directly use this method with a time-dependent cosmological background as the starting point, in the line of \cite{CosmoDFT}. 
    
    \item Regarding the identification, we insist that at the moment it lacks a precise mathematical structure. The rules that make it work are clear to us, but the underlying infinite dimensional group structure calls for a better understanding.

    \item Our results can contribute to many recent works that study the role of T-duality for higher derivatives in the context of black-holes \cite{BlackHoles} and other solutions of cosmological relevance \cite{CosmoDFT, CosmoDFT2}.

\end{itemize}

\section*{Acknowledgments} We acknowledge F. Hassler and A. Gitsis for comments on the manuscript (see footnote \ref{typo}) . Our work is supported by CONICET. WB receives financial support from ANPCyT-FONCyT (PICT-2015-1525) and the Grant PIP-UE B\'usqueda de nueva f\'isica.

\section*{Appendix}

\begin{appendix}

\section{$O(D,D)$ decomposition of the extended generalized fluxes} \label{parameterization}

We display here the exact decomposition of the extended generalized fluxes in terms of those in the double setup



The remaining projections are simply obtained by switching everywhere 
\begin{eqnarray}
(\underline{a},\underline{b},\underline{c},...;\ut\mu,\ut\nu,\ut\rho,...;\, g_1)\longleftrightarrow (\overline{a},\overline{b},\overline{c},...;\tilde\mu,\tilde\nu,\tilde\rho,...;\, g_2)\,.\nonumber
\end{eqnarray}
where $\Omega_{\ut\mu\tilde\nu}:=-\Omega_{\tilde\nu\ut\mu}$ .

\section{The identification in the scalar sector}

A naive counting of dof in the heterotic case seems to be in agreement with the expectations. The generalized frame in the extended space has $D(D + k)$ independent dof and the generalized BdR identification (\ref{LockingMono}) consist of $D k$ relations, whose effect is to leave as unique physical $D^2$ dof those of the double generalized vielbein. Although this simple counting works, one has to bear in mind that the identifications are far from linear in the sense that we are not directly linking gauge with gravity dof (by this we mean the full NSNS sector). Instead, we are identifying gauge with gravity + gauge dof, and it is only after working out explicitly the derivative expansion through the iterative process described in previous sections, that the iteration converges in such a way that the gauge dof are finally replaced by the gravitational ones in the double space. The explicit computation at ${\cal O}(\alpha'{}^2)$ confirms an agreement with the naive counting analysis.

The situation is more involved in the bi-parametric case. The generalized frame in the extended space starts now with $(D+k)^2$ independent dof, which means that the generalized BdR identification needs to fix $2Dk$ (vectorial) + $k^2$ (scalar) dof this time, in order to leave only the $D^2$ dof captured by the double generalized frame. An apparent inconsistency emerges after counting the $2 D k$ (from ${\cal E}_{\tilde\mu}{}^{\underline{a}},\, {\cal E}_{\ut\mu}{}^{\overline{a}}$) + $2 k^2$ (from ${\cal E}_{\tilde\mu}{}^{\underline{\alpha}},\, {\cal E}_{\ut\mu}{}^{\overline{\alpha}}$) generalized BdR identifications (\ref{LockingBi}). A potential conflict then arises in the scalar sector, which is absent in the mono-parametric generalized BdR identification.

We now show how this tension is resolved to second order in $\alpha'$, but first  let us show explicitly where the apparent over-constraints appear. We have  parameterized the scalar sector in (\ref{FrameParameterizationA}) and (\ref{FrameParameterizationB}) in such a way that 
${\cal E}_{\tilde\mu}{}^{\underline{\alpha}} =  - (\Box^{\frac 12})_{\tilde \mu}{}^{\tilde \rho}\Omega_{\tilde\rho}{}^{\ut\nu}\,\underline{e}_{\ut\nu}{}^{\underline\alpha}$ and ${\cal E}_{\ut\mu}{}^{\overline{\alpha}} = (\Box^{\frac 1 2})_{\ut \mu}{}^{\ut \rho} \Omega^{\tilde\nu}{}_{\ut\rho}\,\overline{e}_{\tilde\nu}{}^{\overline\alpha}$.  There are then two possible ways to read $\Omega$
\begin{eqnarray}
\Omega^{\tilde\mu}{}_{\ut\nu}\; &=&\; -\,(\Box^{-\frac12})^{\tilde\mu}{}_{\tilde\rho} \; {\cal{E}}^{\tilde\rho}{}_{\underline\alpha} \; \underline{e}_{\ut\nu}{}^{\underline\alpha}\,, \label{Omega1}\\
\Omega^{\tilde\mu}{}_{\ut\nu}\; &=&\; (\Box^{-\frac12})_{\ut\nu}{}^{\ut\sigma} \; {\cal{E}}_{\ut\sigma}{}^{\overline\alpha}  \; \overline{e}^{\tilde\mu}{}_{\overline\alpha} \ .\label{Omega2}
\end{eqnarray}
This rises no conflict from the point of view of the extended space because the extended generalized frame is a constrained field, so its component are related. It is precisely the fact that it is $\cal G$-valued that relates its components in this form. What turns on the alarms is the generalized BdR identification (\ref{LockingBi}), as it identifies the two scalar directions in different ways ${\cal E}_{\tilde\mu}{}^{\underline{\alpha}} \leftrightarrow {\cal F}^{\underline\alpha}{}_{\overline{\cal BC}}$ and ${\cal E}_{\ut\mu}{}^{\overline{\alpha}} \leftrightarrow {\cal F}^{\overline\alpha}{}_{\underline{\cal BC}}$.   Implementing the identification (\ref{LockingBi}) into (\ref{Omega1})-(\ref{Omega2}), leads on the one hand to 
\begin{eqnarray}
\Omega^{\tilde\mu}{}_{\ut\nu}\; &=&\frac{1}{g_2\, X_{R_2}} \,\left(\delta^{\overline{\cal AB} }_{\overline{\cal CD} }-\frac{1}{ g_2^2\, X_{R_2}} {\cal F}^{\underline{e}\overline{\cal AB} }\, {\cal F}_{\underline{e} \overline{\cal CD} } \right)^{-\frac12}  {\cal F}_{\underline{\alpha}}{}^{\overline{\cal CD} }  \; \underline{e}_{\ut\nu}{}^{\underline\alpha}\, (t^{\tilde\mu})_{\overline{\cal AB}}\,,\label{OmegaA}
\end{eqnarray}
and on the other to
\begin{eqnarray}
\Omega^{\tilde\mu}{}_{\ut\nu}\; &=&\frac{1}{g_1\, X_{R_1}} \,\left(\delta^{\underline{\cal CD} }_{\underline{\cal AB} }+\frac{1}{ g_1^2\, X_{R_1}} {\cal F}_{\overline{e}\underline{\cal AB} }\, {\cal F}^{\overline{e} \underline{\cal CD} } \right)^{-\frac12}  {\cal F}^{\overline{\alpha}}{}_{\underline{\cal CD} }  \; \overline{e}^{\tilde\mu}{}_{\overline\alpha}\, (t_{\ut\nu})^{\underline{\cal AB}}\,.\label{OmegaB}
\end{eqnarray}
The first identity depends on ${\cal F}_{\underline{\cal A}\overline{\cal{BC}}}$ while the second on ${\cal F}_{\overline{\cal A}\underline{\cal{BC}}}$, and there is no evident reason why these two expressions should agree.

 Let us now explain how this is resolved. The generalized BdR identification truncates the extended setup, rising new relations on the extended fluxes. As an example, in the extended setup the vectorial components of the generalized frame ${\cal E}_{\tilde \mu}{}^{\underline a}$ and ${\cal E}_{\ut \mu}{}^{\overline a}$ are generic, but after the identification they become related with generalized fluxes, which satisfy Bianchi Identities. Hence, the question is whether the truncated (\ref{OmegaA}) coincides with the truncated (\ref{OmegaB}). This is very ambitious, as it requires an explicit realization of the generators $(t_{\hat\mu})_{\cal A}{}^{\cal B}$, which is beyond the scope of this paper. Instead, we will only explore this at leading order in a derivative expansion, which turns out to be enough to compute second order corrections to the action.  The reason is that neither the generalized Green Schwarz transformation nor the DFT deformed action contain free internal (gauge) indices.
 This means that scalars are always contracted with vectors ($\sim g^{-1}$) or with other scalars ($\sim g^{-2}$), which implies on the one hand that the leading contribution of terms with $\Omega$ is ${\cal O}(\alpha'{}^2)$ and on the other hand that only the leading ($i.e.$ ${\cal O}(\alpha')$) part of $\Omega$ contributes at this order. 
 
Let us multiply both expressions (\ref{Omega1}) and (\ref{Omega2}) with the generators $(t_{\tilde \mu})^{\overline{\cal A B}}$ and $(t^{\ut \nu})_{\underline{\cal C D}}$, and name them $X$ and $Y$ respectively  
\begin{eqnarray}
X^{\overline{\cal AB}}{}_{\underline{\cal CD}}&:=&-\,(\Box^{-\frac12})^{\tilde\mu}{}_{\tilde\rho} \; {\cal{E}}^{\tilde\rho}{}_{\underline\alpha} \; \underline{e}_{\ut\nu}{}^{\underline\alpha}\;(t_{\tilde\mu})^{\overline{\cal AB}}\,(t^{\ut\nu})_{\underline{\cal CD}}\, ,\\
Y^{\overline{\cal AB}}{}_{\underline{\cal CD}} &:=&\; (\Box^{-\frac12})_{\ut\nu}{}^{\ut\sigma} \; {\cal{E}}_{\ut\sigma}{}^{\overline\alpha}  \; \overline{e}^{\tilde\mu}{}_{\overline\alpha}\;(t_{\tilde\mu})^{\overline{\cal AB}}\,(t^{\ut\nu})_{\underline{\cal CD}}\, .
\end{eqnarray}
The plan is to explore the equality of these after the generalized BdR identification is imposed, to leading order. A rapid treatment first leads to
\begin{eqnarray}
X^{\overline{\cal AB}}{}_{\underline{\cal CD}} =-\frac{1}{g_2} {\cal F}^{\underline\alpha \overline{\cal AB}}\; \underline{e}^{\ut\nu}{}_{\underline\alpha}\;(t_{\ut\nu})_{\underline{\cal CD}}+{\cal O}(\alpha'{}^2)
\;,\;\;\;\;\;
Y^{\overline{\cal AB}}{}_{\underline{\cal CD}}=\frac{1}{g_1} {\cal F}^{\overline\alpha}{}_{\underline{\cal CD}}\; \overline{e}^{\tilde\mu}{}_{\overline\alpha}\;(t_{\tilde\mu})^{\overline{\cal AB}}+{\cal O}(\alpha'{}^2)\,.\;\;\;\;
\end{eqnarray}
Let us analyze each component individually

\begin{itemize}
\item $X^{\overline{\cal AB}}{}_{\underline{c\delta}}-Y^{\overline{\cal AB}}{}_{\underline{c\delta}}$ and $X^{\overline{a \beta}}{}_{\underline{\cal{C D}}}-Y^{\overline{a \beta}}{}_{\underline{\cal{C D}}}$ vanish at ${\cal O}(g^{-2})$. 

That is automatically satisfied as pairs of indices with mixed $h$ and $H$ components are always of higher order: $X^{\overline{a b}}{}_{\underline{c\delta}}, X^{\overline{a \beta}}{}_{\underline{c d}}, Y^{\overline{a b}}{}_{\underline{c\delta}}, Y^{\overline{a \beta}}{}_{\underline{c d}}\sim {\cal O}(g^{-3})$ and  $X^{\overline{a \beta}}{}_{\underline{c\delta}}, X^{\overline{a \beta}}{}_{\underline{c \delta}}, Y^{\overline{a \beta}}{}_{\underline{c\delta}},$ $Y^{\overline{a \beta}}{}_{\underline{c \delta}}\sim {\cal O}(g^{-4})$. 

\item $X^{\overline{ab}}{}_{\underline{cd}} -  Y^{\overline{ab}}{}_{\underline{cd}} = {\cal O}(g^{-4})$ holds due to Bianchi Identities.

Indeed, after (\ref{F1alpbcGF}) and the generalized BdR identification one readily finds
\begin{eqnarray}
Y^{\overline{ab}}{}_{\underline{cd}}=\frac{1}{g_1\,g_2}\left(F^{\underline e \overline{ab}} F_{\underline{e c d}}- 2\, F_{\underline c}{}^{\overline{e}[\overline{a}}\,F_{\underline{d} \overline{e}}{}^{\overline{b}]}  - 2 D_{[\underline{c}}F_{\underline d ]}{}^{\overline{ab}} \right)+{\cal O}(g^{-4}) \, .\label{Xabcd}
\end{eqnarray} 
Repeating the same for $X^{\overline{ab}}{}_{\underline{cd}}$, one easily verifies that 
\begin{eqnarray}
X_{\overline{ab} \underline{cd}}-Y_{\overline{ab} \underline{cd}}=\frac{4}{g_1 \, g_2}\left( D_{[\overline a}F_{\overline b \underline{c d}]}- \frac34 F_{[\overline{ab}}{}^{\underline{e}} F_{\underline{cd}] \underline{e}}- \frac34 F_{[\overline{ab}}{}^{\overline{e}} F_{\underline{cd}] \overline{e}}\right)\, +{\cal O}(g^{-4})\;,
\end{eqnarray}
vanishes at leading order because of the Bianchi identities for generalized fluxes in the double space. 

\item $X^{\overline{\alpha\beta}}{}_{\underline{\gamma\delta}} -  Y^{\overline{\alpha\beta}}{}_{\underline{\gamma\delta}}=0$ does not impose a condition on $\Omega$, but a condition on the generators. 

Indeed, from the  flux decomposition in Appendix A, one obtains 
\begin{eqnarray}
X^{\overline{\alpha\beta}}{}_{\underline{\delta\gamma}}&=&\Omega^{\tilde\mu}{}_{\ut\nu} (t_{\tilde \mu})^{\overline{\alpha\beta}} (\tau_{\ut \nu})_{\underline{\delta\gamma}}+{\cal O}(g^{-4})\;,\cr
Y^{\overline{\alpha\beta}}{}_{\underline{\delta\gamma}}&=&\Omega^{\tilde\mu}{}_{\ut\nu} (\tau_{\tilde \mu})^{\overline{\alpha\beta}} (t_{\ut \nu})_{\underline{\delta\gamma}}+{\cal O}(g^{-4})\;,
\end{eqnarray}
where $\tau$ denote the generators in the Adjoint representation
\begin{eqnarray}
(\tau_{\ut\mu})_{\underline\alpha}{}^{\underline\beta}\,=\,-\,f_{\ut\mu\ut\nu}{}^{\ut\rho}\; \underline{e}^{\ut\nu}{}_{\underline\alpha}\, \underline{e}_{\ut\rho}{}^{\underline\beta}  \;,\;\;\;\;\;\;\;\;\;\;\;\;\;\;\;\;
(\tau_{\tilde\mu})_{\overline\alpha}{}^{\overline\beta}\,=\,-\,f_{\tilde\mu\tilde\nu}{}^{\tilde\rho}\; \overline{e}^{\tilde\nu}{}_{\overline\alpha}\, \overline{e}_{\tilde\rho}{}^{\overline\beta} \;.
\end{eqnarray} 
Hence, the validity of $X^{\overline{\alpha\beta}}{}_{\underline{\gamma\delta}} -  Y^{\overline{\alpha\beta}}{}_{\underline{\gamma\delta}}={\cal O}(g^{-4})$ is simply a consequence of requiring that the internal components of the generators (upon contraction) agree at leading order with the adjoint representation. This can be alternatively verified from comparison of the generalized BdR decomposition and the leading terms of the flux decomposition (\ref{F1aalpbetGF}), (\ref{F1alpbetgamGF}) 
\begin{eqnarray}
{\cal F}_{\overline{\cal A}\underline{\beta}}{}^{\underline{\gamma}}=
\,g_1\,f_{\ut\mu\ut\nu}{}^{\ut\rho}\; \underline{e}^{\ut\nu}{}_{\underline\alpha}\; \underline{e}_{\ut\rho}{}^{\underline\beta}\;\; {\cal E}^{\ut\mu}{}_{\overline{\cal A}} + \dots\,,\label{AdjAnsatz}
\end{eqnarray}
and the analogous expression for ${\cal F}_{\underline{\cal A}\overline{\beta}}{}^{\overline{\gamma}}$.

\item $X^{\overline{\alpha\beta}}{}_{\underline{cd}} - Y^{\overline{\alpha\beta}}{}_{\underline{cd}}=0$ is a true constraint that implicitly fixes some $\Omega$ dof
\begin{eqnarray}
X^{\overline{\alpha \beta}}{}_{\underline{c d}}&=& \Omega^{\tilde\mu}{}_{\ut\nu}\,(\tau_{\tilde\mu})^{\overline{\alpha \beta}}\,(t^{\ut\nu})_{\underline{c d}}\, +{\cal O}(g^{-4})\;,
\end{eqnarray}
while
\begin{eqnarray}
Y^{\overline{\alpha\beta}}{}_{\underline{cd}}&=&\frac{1}{g_1\,g_2}\left({\cal F}^{\underline e \overline{\alpha\beta}} F_{\underline{e c d}}- 2\, {\cal F}_{\underline c}{}^{\overline{\gamma}[\overline{\alpha}}\,{\cal F}_{\underline{d} \overline{\gamma}}{}^{\overline{\beta}]}  - 2 D_{[\underline{c}}{\cal F}_{\underline d ]}{}^{\overline{\alpha\beta}} \right)+{\cal O}(g^{-4})\;.\label{Yalpbetcd}
\end{eqnarray} 

\item $X^{\overline{ab}}{}_{\underline{\gamma\delta}} -  Y^{\overline{ab}}{}_{\underline{\gamma\delta}}=0$ also fixes some $\Omega$ dof through
\begin{eqnarray}
Y^{\overline{ab}}{}_{\underline{\gamma\delta}}&=& \Omega^{\tilde\mu}{}_{\ut\nu}\,(t_{\tilde\mu})^{\overline{ab}}\,(\tau^{\ut\nu})_{\underline{\gamma\delta}}\, +{\cal O}(g^{-4})\;,
\end{eqnarray}
while
\begin{eqnarray}
X^{\overline{ab}}{}_{\underline{\gamma\delta}}&=&
-\frac{1}{g_1\,g_2}\left({\cal F}_{\overline e \underline{\gamma\delta}} F^{\overline{e a b}} - 2\, {\cal F}^{\overline a}{}_{\underline{\beta}[\underline{\gamma}}\,{\cal F}^{\overline{b} \underline{\beta}}{}_{\underline{\delta}]}  - 2 D^{[\overline{a}}{\cal F}^{\overline b ]}{}_{\underline{\gamma\delta}} \right)+{\cal O}(g^{-4})\;.\label{Xalpbetcd}
\end{eqnarray} 
 
So we find once again two different relations for $\Omega$. Since we do not have an explicit realization of the generators, at this stage it is unclear if these are equivalent up to generalized Bianchi Identities or if they are constraints on different components of $\Omega$. 

\end{itemize}

Let us emphasize that the issues raised above are a consequence of trying to analyze $\Omega$ uncontracted. As mentioned, the scalar matrix always appears with its indices totally contracted, some examples being: $\Omega^{\tilde\mu \ut\nu}\,\Omega^{\tilde\rho \ut\sigma}\,\Omega^{\tilde\tau \ut\lambda}\,g_1\, f_{\tilde\mu \tilde\rho\tilde\tau}\,g_2\, f_{\ut\nu \ut\sigma\ut\lambda}$ , ${\cal E}^{\tilde\mu \underline a}\, \Omega^{\tilde\nu\ut\sigma}\,\left( D_{\underline a}\Omega^{\tilde\rho}{}_{\ut\sigma}\right) \;g_2\;f_{\tilde\mu\tilde\nu\tilde\rho}$ , ${\cal E}^{\ut\mu \overline a}\;{\cal E}^{\ut\nu}{}_{\overline a}\; \Omega^{\tilde\tau\ut\rho}\;\Omega_{\tilde\tau}{}^{\ut\sigma}  \,g_1^2\,f_{\ut\mu\ut\rho\ut\lambda}\;f_{\ut\nu\ut\sigma}{}^{\ut\lambda}$ , $\Omega^{\tilde\mu}{}_{\ut\nu}\;{\cal E}^{\ut\nu \overline a}\;{\cal E}^{\tilde\rho \underline b}\;{\cal E}^{\tilde\sigma \underline c} \;g_2\,f_{\tilde\mu\tilde\rho\tilde\sigma}$ , ${\cal E}^{\ut\mu\overline c}\,{\cal E}_{\tilde{\nu}}{}^{\overline d}\,\Omega^{\tilde{\nu}}{}_{\ut\mu}$. The process to replace scalar dof in term of the $2D$ dof of DFT follows similar steps like those we already took for the vector dof at the end of Section \ref{sectionMono}. It requires the use of cyclic relations whose effect is a renormalization of the coefficients leading to $a$ and $b$ parameters. 

Two remarks are in order: 
\begin{itemize}
\item The first is that, even if we do not have an all-order proof of the validity for the mechanism responsible for the elimination of the gauge dof and the subsequent formation of $a$ and $b$ parameters, this mechanism empirically works for hundred of terms independently, strongly suggesting that it plausibly holds at higher orders.
\item Despite the success in to getting rid of the vector and scalar dof, there is a subtle point in the latter because in principle there are two way to do this depending on the choice (\ref{Omega1}) or (\ref{Omega2}). All the terms containing scalar fields were shown to be equivalent independently of that choice. 
\end{itemize}  

It is worth  illustrating how conclusive identities can be obtained when $\Omega$ appears contracted. For concreteness we consider the contraction ${\cal E}^{\ut\mu\overline c}\,{\cal E}_{\tilde{\nu}}{}^{\overline d}\,\Omega^{\tilde{\nu}}{}_{\ut\mu}$
\begin{eqnarray}
{\cal E}^{\ut\mu\overline c}\;{\cal E}_{\tilde{\nu}}{}^{\overline d}\;\Omega^{\tilde{\nu}}{}_{\ut\mu}
&=&
\left(\frac{1}{X_{R_1}g_1}\right)\,\left(\frac{1}{X_{R_2}g_2^2}\right){\cal F}^{\overline c}{}_{\underline {\cal EF}}\; (t^{\ut\mu})^{\underline {\cal EF}} \;{\cal F}^{\underline d}{}_{\overline{gh}} \; {\cal F}_{\underline \alpha}{}^{\overline{gh}}\; \underline{e}_{\ut\mu}{}^{\underline \alpha}\cr
&+& 
\left(\frac{1}{X_{R_1}g_1}\right)\,\left(\frac{1}{X_{R_2}g_2^2}\right){\cal F}^{\overline c}{}_{\underline {\cal EF}} \;(t^{\ut\mu})^{\underline {\cal EF}} \;{\cal F}^{\underline d}{}_{\overline{\beta\gamma}}\;  {\cal F}_{\underline \alpha}{}^{\overline{\beta\gamma}}\; \underline{e}_{\ut\mu}{}^{\underline \alpha}\;+\;{\cal O}(\alpha'{}^3)\cr
&=&\left(\frac{1}{g_1^2(X_{R_1}-1)}\right)\,\left(\frac{1}{X_{R_2}g_2^2}\right)F^{\underline d}{}_{\overline{gh}}\;
\left( \vphantom{\frac12}\right.-F^{\overline a\underline{ef}} \;F^{\overline c}{}_{\underline{ef}} \;F_{\overline a}{}^{\overline{gh}}\cr
&&\;\;\;\;\;\;\;\;\;\;\;\;\;\;\;\; +\;2\; F_{\overline{g}}{}^{\underline{ea}} \;F_{\overline{h}\underline a}{}^{\underline{f}} \;
F^{\overline{c}}{}_{\underline{ef}} \;+\; 2\left(D^{\overline g} F^{\overline h\underline{ef}}\right)\; F^{\overline c}{}_{\underline{ef}}  \left.\vphantom{\frac12}\right)\cr
&+&\frac{1}{X_{R_2}} {\cal E}^{\ut\mu\overline c}\;{\cal E}_{\tilde{\nu}}{}^{\overline d}\;\Omega^{\tilde{\nu}}{}_{\ut\mu}
\;\;+\;{\cal O}(\alpha'{}^3)\;.\label{EEOmega}
\end{eqnarray}
Then we conclude 
\begin{eqnarray}
{\cal E}^{\ut\mu\overline c}\;{\cal E}_{\tilde{\nu}}{}^{\overline d}\;\Omega^{\tilde{\nu}}{}_{\ut\mu}
=\frac{a\;b}{4}\;F^{\underline d}{}_{\overline{gh}}\; F^{\overline c}{}_{\underline{ef}}\;
\left( \vphantom{\frac12}\right.-F^{\overline a\underline{ef}}\;F_{\overline a}{}^{\overline{gh}}\; +\;2\; F^{\overline{g} \underline{ea}} \;F^{\overline{h}\underline{f}}{}_{\underline a} \;+\; 2\,D^{\overline g} F^{\overline h\underline{ef}} \left.\vphantom{\frac12}\right)\;+\;{\cal O}(\alpha'{}^3)\,.\cr\label{OmegaTrick}
\end{eqnarray}
In the first line of (\ref{EEOmega}) we used the ansatz (\ref{EABC}), (\ref{LockingBi}) and (\ref{Omega1}). In the third line we used the parameterization of the Appendix \ref{parameterization}. On the other hand one can repeat the same procedure but using (\ref{Omega2}) instead. One arrives at 
\begin{eqnarray}
{\cal E}^{\ut\mu\overline c}\;{\cal E}_{\tilde{\nu}}{}^{\overline d}\;\Omega^{\tilde{\nu}}{}_{\ut\mu}
=\frac{a\;b}{4}\;F^{\underline d}{}_{\overline{gh}}\; F^{\overline c}{}_{\underline{ef}}\;
\left( \vphantom{\frac12}\right.F^{\underline{a e f}}\;F_{\underline a}{}^{\overline{gh}}\; -\;2\; F^{\underline{e} \overline{g a}} \;F^{\underline{f}\overline{h}}{}_{\overline a} \;-\; 2\,D^{\underline e} F^{\underline f\overline{g h}}  \left.\vphantom{\frac12}\right)\;+\;{\cal O}(\alpha'{}^3)\;,\cr\label{OmegaTrick2}
\end{eqnarray}
and we conclude that both alternative expressions agree up to Bianchi Identities. Notice that this term depends on both parameters $a$ and $b$. This is a general property of all terms whose origin can be traced back to the scalar dof. The reason is obvious, all these terms should disappear if we turn off one of the parameters as in the mono-parametric case there are no scalars dof.
\end{appendix}

\end{document}